\documentclass[%
aip,
amsmath,amssymb,
reprint,%
author-year,%
]{revtex4-1}

\usepackage{graphicx}
\usepackage{dcolumn}
\usepackage{bm}

\usepackage[utf8]{inputenc}
\usepackage[T1]{fontenc}
\usepackage{mathptmx}
\usepackage{placeins}  
\usepackage{bibentry}

\begin{document}

\preprint{AIP/123-QED}

\title[Materials loss measurements using superconducting microwave resonators]{Materials loss measurements using superconducting microwave resonators}

\author{C.R.H. McRae}
\email{coreyrae.mcrae@colorado.edu}
\affiliation{
Department of Physics, University of Colorado, Boulder, Colorado 80309, USA}
\affiliation{ 
National Institute of Standards and Technology, Boulder, Colorado 80305, USA
}%
\affiliation{ 
JILA, University of Colorado, Boulder, Colorado 80309, USA
}%
\affiliation{ 
Boulder Cryogenic Quantum Testbed, University of Colorado, Boulder, Colorado 80309, USA
}%
\author{H. Wang}%
\affiliation{
Department of Physics, University of Colorado, Boulder, Colorado 80309, USA}
\affiliation{ 
National Institute of Standards and Technology, Boulder, Colorado 80305, USA
}%
\affiliation{ 
JILA, University of Colorado, Boulder, Colorado 80309, USA
}%
\affiliation{ 
Boulder Cryogenic Quantum Testbed, University of Colorado, Boulder, Colorado 80309, USA
}%
\author{J. Gao}%
\affiliation{
Department of Physics, University of Colorado, Boulder, Colorado 80309, USA}
\affiliation{ 
National Institute of Standards and Technology, Boulder, Colorado 80305, USA
}%
\author{M.R. Vissers}%
\affiliation{ 
National Institute of Standards and Technology, Boulder, Colorado 80305, USA
}%
\author{T. Brecht}%
\affiliation{ 
HRL Laboratories,
Malibu, California 90265, USA
}%
\author{A. Dunsworth}
\affiliation{
Google Inc.,
Mountain View, California 94043, USA}%
\author{D.P. Pappas}%
\affiliation{ 
National Institute of Standards and Technology, Boulder, Colorado 80305, USA
}%
\author{J. Mutus}
\affiliation{
Google Inc.,
Mountain View, California 94043, USA}
\affiliation{ 
Boulder Cryogenic Quantum Testbed, University of Colorado, Boulder, Colorado 80309, USA
}%

\date{\today}

\begin{abstract}
The performance of superconducting circuits for quantum computing is limited by materials losses. In particular, coherence times are typically bounded by two-level system (TLS) losses at single photon powers and millikelvin temperatures. The identification of low loss fabrication techniques, materials, and thin film dielectrics is critical to achieving scalable architectures for superconducting quantum computing. Superconducting microwave resonators provide a convenient qubit proxy for assessing performance and studying TLS loss and other mechanisms relevant to superconducting circuits such as non-equilibrium quasiparticles and magnetic flux vortices. In this review article, we provide an overview of considerations for designing accurate resonator experiments to characterize loss, including applicable types of loss, cryogenic setup, device design, and methods for extracting material and interface losses, summarizing techniques that have been evolving for over two decades. Results from measurements of a wide variety of materials and processes are also summarized. Lastly, we present recommendations for the reporting of loss data from superconducting microwave resonators to facilitate materials comparisons across the field.
\end{abstract}

\maketitle

\section{\label{sec:intro}Introduction}

Superconducting quantum computers consist of two basic components: the qubit, which is used to manipulate quantum information, and the superconducting microwave resonator, a multipurpose component used for qubit protection and readout \citep{Blais2004,Wallraff2004} as well as qubit-qubit coupling \citep{Mariantoni2011,Sillanpaa2007} and quantum memory \citep{Hofheinz2009,Reagor_10ms3D}.

At millikelvin temperatures and single photon powers, loss in both qubits and resonators is thought to be dominated by coupling to two-level systems (TLS) that may be present in device materials. Microwave energy in the device can couple to an ensemble of defects in nearby amorphous material, causing energy decay \citep{Martinis2005}. Although the origin of TLS loss is still up for debate \citep{Muller2019}, it is known that TLS loss is generated largely in bulk dielectrics and at the interfaces between materials \citep{GaoThesis,Calusine2018}. Resonators are useful test structures for measuring TLS loss due to their simplicity in design, fabrication and measurement, and their sensitivity to loss at low temperatures and photon number. Resonators can also be measured over a wider range of photon occupation and temperatures than qubits, allowing for exploration of defects over a large experimental parameter regime. While some loss mechanisms are specific to qubit implementations and Josephson junction circuit elements, non-junction-related loss mechanisms can be investigated with resonators. 

Accurate resonator measurements allow us to engineer interfaces and circuit designs to identify regions that contribute to loss as well as to identify low-loss materials in order to decrease the qubit footprint and enable multilayer architectures. Resonator measurements can also provide information about losses due to microwave and cryogenic experimental issues, including magnetic \citep{Song2009,Chiaro2016} and radiative \citep{Sage2011} losses.

In this review, we introduce the general properties of superconducting microwave resonators (Sec.~\ref{sec:beginnerstuff} and summarize in detail the best practices of resonator measurements. These include the evaluation of applicable losses mechanisms (Sec.~\ref{sec:losses}), cryogenic and electronic experimental set-up (Sec.~\ref{sec:setup}), and experimental decisions involved in device design and fabrication (Sec.~\ref{sec:methods}). Finally, we present a summary of existing TLS loss measurements (Sec.~\ref{sec:experiments}) and suggest additional parameters to report in future studies. This provides an overview of metrics that are comparable across various material sets and device implementations.

\section{\label{sec:beginnerstuff}The Superconducting Microwave Resonator}

\begin{figure}
\includegraphics[width=3.35in]{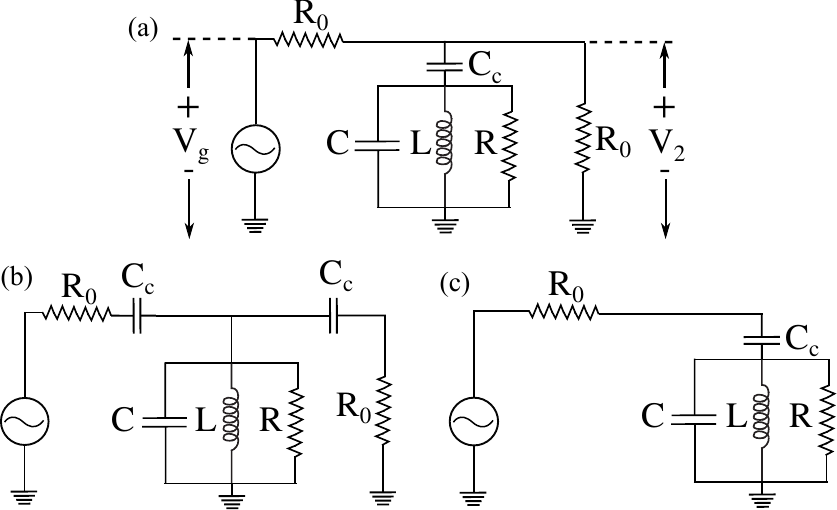}
\caption{\label{fig:circuits} Circuit diagrams of three measurement orientations of parallel RLC resonators. (a) Resonator in hanger mode, capacitively coupled to a feedline for measurement in transmission. $V_g$ is the source voltage, $R_0$ is the source and load impedance which is often equal to the transmission characteristic impedance $Z_0$ for impedance matching, and $V_{2}$ is the load voltage. The arrows on the voltages indicate the direction of the microwaves being measured. (b) Resonator in transmission mode, capacitively coupled to an input and an output feedline. (c) Resonator in reflection mode, capacitively coupled to the end of a feedline.}
\end{figure}

A basic resonator circuit, Fig.~\ref{fig:circuits}, consists of an inductive component with inductance $L$ coupled to a capacitive component with capacitance $C$. Energy stored in the resonator oscillates between the inductor (stored as current) and the capacitor (stored as charge) with a resonance frequency $f_0$ determined by
\begin{equation}
    f_0 = \frac{1}{2 \pi \sqrt{L C}},
\end{equation}
neglecting any external coupling. To take advantage of commercial microwave test and measurement equipment as well as commercial RF components, superconducting microwave resonators for quantum computing are generally designed to have resonance frequencies between 4 and 8 GHz. The resonator circuits commonly consist of thin film superconducting metals and thin film dielectrics, with typical thicknesses in the range of hundreds of nanometers, patterned on a crystalline dielectric substrate hundreds of micrometers thick \citep{Zmuidzinas2012}.

\begin{figure}
\includegraphics[width=\linewidth]{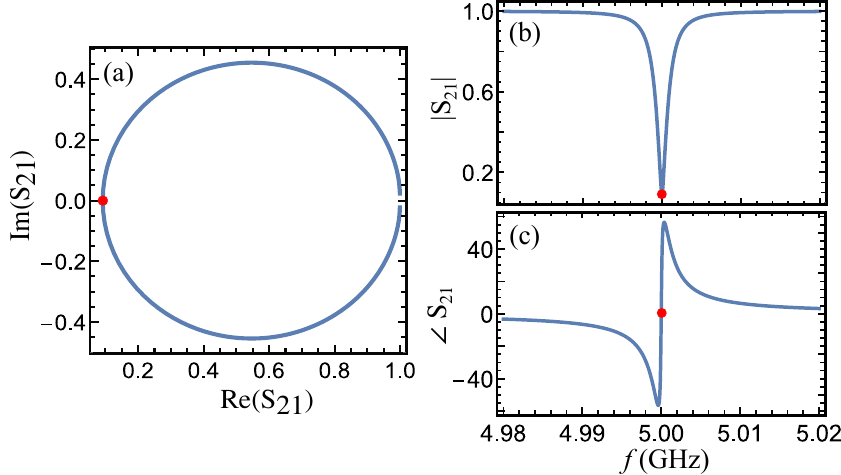}
\caption{\label{fig:S21plot} Simulated $S_{21}$ response of a 5 GHz resonator~[Fig.~\ref{fig:circuits}~(a)] in hanger mode with $Q_i=22,000$ and $Q_c=2,200$, with the red dot indicating resonance frequency, on (a) the complex plane, (b) amplitude as a function of frequency, and (c) phase as a function of frequency. The resonator impedance $Z_{in}$ has similar Lorentzian behavior.}
\end{figure}

In practice, the resonator, like all superconducting circuits, is coupled to energy dissipation channels, causing loss and reducing its ability to store energy. For superconducting microwave resonators, loss is low compared to normal metal resonators due to the low dissipation nature of the superconducting metal. We can combine all the resonator losses into an effective resistance $R$. 
The resonator also has input impedance $Z_{in}$, defined as:
\begin{equation}
    Z_{in} = (\frac{1}{R} + \frac{1}{i \omega L} + i \omega C)^{-1},
\end{equation}
$Z_{in}$ can be plotted as a function of frequency and behaves as a Lorentzian, similar to that in Fig.~\ref{fig:S21plot}(a)-(c). 

To quantify resonator performance, we define a dimensionless quantity, the quality factor, as
\begin{equation}
Q = \frac{2 \pi f_0 W}{P_{\mathrm{R}}},
\end{equation}
with total time-averaged energy $W$ and power loss due to the resistor defined as
\begin{equation}
P_{\mathrm{R}} = \frac{|V|^2}{2R},
\end{equation}
where $V$ is the voltage on the resistor $R$ \citep{pozar} for a parallel RLC resonator~(Fig.~\ref{fig:circuits}). In general terms, the quality factor describes the ratio of the total energy stored near resonance and the total energy loss in the system. Qualitatively, as  $Q$ increases, the resonance line-shape becomes sharper.

\begin{figure}
\includegraphics[width=85mm]{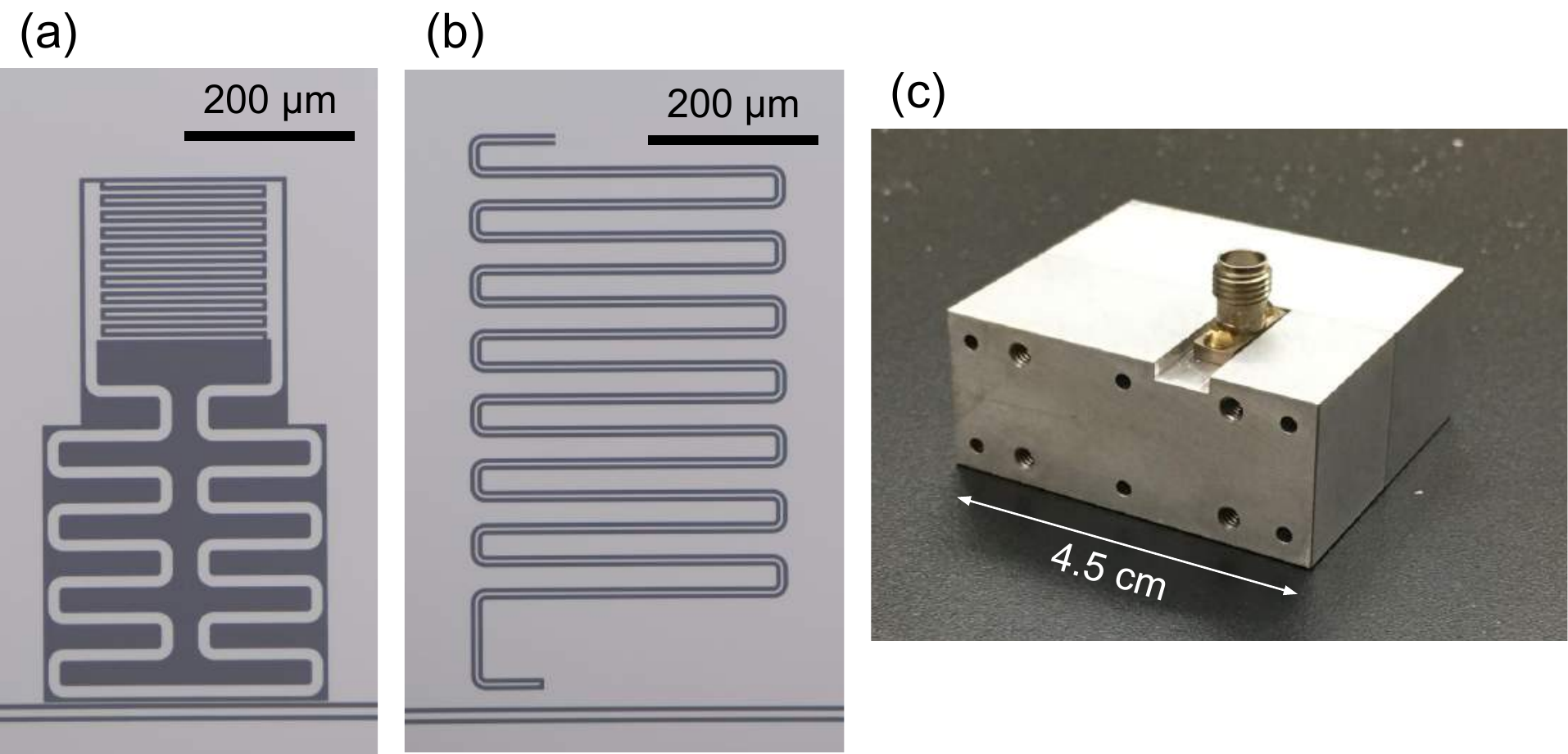}
\caption{\label{fig:resonator_micrograph}Three common resonator implementations. (a) Al on Si lumped element resonator with a meandered inductor (below) connected to an interdigitated capacitor (above) and inductively coupled to a feedline. (b) Al on Si $\lambda$/4 CPW resonator capacitively coupled to a feedline. (c) 3D resonant cavity made of high purity Al.}
\end{figure}

A resonator is normally connected to the measurement circuit through capacitive and/or inductive coupling, which contributes to the resonator quality factor. Thus, the total quality factor $Q$ can be divided into an internal and coupling component, as
\begin{equation}
    \frac{1}{Q} = \frac{1}{Q_i} + \frac{1}{Q_c},
\end{equation}
where $Q_c$ is the coupling quality factor and is defined as the rate at which energy stored in the resonator escapes into the larger circuit, and $Q_i$ is the internal quality factor and is defined as the rate at which energy is lost to parasitic effects of the environment. The loading will shift the resonance frequency  downwards, which is normally negligible as long the coupling capacitance and inductance is small compared to the C and L values of the resonator.

$Q_c$ is a design parameter as it is determined by specifying the admittance to ground through coupling capacitance and mutual inductance between the resonator and transmission feedline. Conversely, $Q_i$ is determined by the intrinsic losses within the resonator itself, and can thus be used as a performance metric of the resonator.

The inverse of resonator $Q_i$ is sometimes represented as the loss tangent, as
\begin{equation}
\frac{1}{Q_i} = \tan{\delta} \simeq \delta
\end{equation}
for loss $\delta$ and $\tan{\delta} < 10^{-2}$. Loss is a convenient metric for distinguishing between multiple contributions to performance, as loss contributions add linearly. $Q_i$, $\tan{\delta}$, and $\delta$ are all used to compare resonator performance in different contexts, where $\tan{\delta}$ and $\delta$ are generally used to denote performance when loss mechanisms are being analyzed, and $Q_i$ is used to compare general performance, especially for high-performance devices.

Although $\tan{\delta}$ conventionally refers to dielectric loss, $\delta$ will be used in this review to refer to any loss, regardless of dissipation channel.

Resonators are typically characterized with a vector network analyzer (VNA). To measure a resonator with a VNA, microwave signals are sent into the circuit input port (port 1) through a feedline coupled to the resonator, and the response is measured through the output port (port 2). The amplitude and phase of the signal are sampled, achieving an $S_{21}$ measurement. The complex $S_{21}$ can be visualized by breaking it into a magnitude and phase component, or into a real and imaginary component, as shown in Fig.~\ref{fig:S21plot}. One can extract $Q_i$, $Q_c$ and $f_0$ from this data by fitting to a Lorentzian model over several bandwidths surrounding the resonance frequency.

Resonator circuits can be designed in several form factors. Fig.~\ref{fig:resonator_micrograph} shows examples of three common resonator geometries. Lumped element resonators [Fig.~\ref{fig:resonator_micrograph}(a)] consist of discrete inductive and capacitive components connected together. Distributed element resonators like coplanar waveguide (CPW) [Fig.~\ref{fig:resonator_micrograph}(b)] and microstrip resonators do not have discrete capacitors and inductors, but rather consist of a transmission line segment with some capacitance and inductance per unit length. These are usually meandered in order to minimize their on-chip footprint. 3D cavity resonators [Fig.~\ref{fig:resonator_micrograph}(c)] differ from the types above as they are not formed by integrated circuits on a chip but rather as empty volumes within superconducting walls, the geometrical boundary conditions of which determine resonant modes. Both superconducting qubits and input/output circuitry can be coupled to the 3D cavity modes via radiative antennae. By completely eliminating bulk dielectric loss and storing a large fraction of the energy in vacuum, 3D cavities can attain the highest quality factors of these three types of resonators. While machined blocks of superconducting metal are bulky, micromachined 3D cavities embedded within stacks of bonded silicon wafers have reached quality factors as high as 300 million \citep{Lei2020}, and show promise for use in multilayer microwave integration 
\citep{brecht2016multilayer,Rosenberg2019}.
Additionally, microwave resonators that do not fit neatly into these three categories include those featuring planar multilayer structures such as vacuum-gap transmission lines \citep{Cicak2010,Lewis2017,Minev2013}.

Currently, in the field of superconducting quantum computing, CPW resonators are most commonly implemented due to their ease of design, single-layer lithographic fabrication, and ability to circumvent lossy dielectrics. CPWs consist of a center conductor several micrometers wide separated from ground planes on either side by a gap of a few micrometers. The conductor and ground planes are composed of superconducting metal and are positioned on a dielectric substrate.

\section{\label{sec:losses}Losses in Superconducting Microwave Resonators}

Superconducting resonators, while low loss, nevertheless suffer from various forms of intrinsic loss. Characterizing the origins of that loss is the subject of much ongoing research. The dependence of resonator loss on experimental conditions such as applied power and temperature can be used to allocate the loss to several mechanisms.

Resonator loss mechanisms that are most important to quantum circuits include two-level systems (TLS), radiation, magnetic vortices, and quasiparticles originating from stray infrared (IR) light and microwave-induced pair-breaking. In practice, superconducting microwave resonators generally exhibit total losses in the $10^{-5}$ to $10^{-6}$ range. In addition to being sources of dissipation and degradation of resonator performance, these loss mechanisms are also sources of decoherence in qubit applications as well as sources of noise impacting superconducting detector applications. For these reasons, considerable effort has been made over the past two decades to understand resonator loss mechanisms and reduce their negative effects. 

\begin{table}
\centering
\begin{ruledtabular}
\begin{tabular}{cccccc}
    Loss Type & L or C & Power & Temperature & Frequency & Geometry \\
    \hline
    TLS & C & - & - & + & -\\
    qp-thermal & L & No & + & + &  \\
    qp-IR & L & No & No &   &  \\
    qp-$\mathrm{\mu}$w & L & + & + & + &   \\
    vortex & L & No & + & + & +\\
    radiation & L, C & No & No & + & + \\
    par. modes & L, C & No & No & Yes & \\
\end{tabular}
\end{ruledtabular}
\caption{Summary of resonator loss mechanisms and their properties. L or C denotes whether the loss is associated with capacitive or inductive resonator components. Columns 3 to 6 denote positive (+), negative (-), no (No), or unknown (blank) correlation between each loss type and experimental parameter. Yes indicates that a correlation exists. TLS: two-level system loss. qp-IR: quasiparticle loss due to stray infrared light. qp-$\mathrm{\mu}$w: quasiparticle loss due to microwave-induced pair-breaking. par. modes: parasitic modes. Geometry refers to the conductor/gap widths of CPW and IDC for TLS and radiation loss, and the conductor width of the inductor for vortex loss.}
\label{tab:losssummary}
\end{table}

Table~\ref{tab:losssummary} gives a summary of common loss types with their important properties. As can be seen from this table, only TLS loss is associated with the capacitance of the resonator while the other loss mechanisms are associated with the inductance. This is because TLS have electrical dipole moments and couple to the electric field, while the other losses are coupled or related to the current in the resonator. 

\subsection{TLS Loss}
Amorphous solids exhibit very different thermal, acoustic, and dielectric properties from crystalline solids at low temperatures, which can be modeled by the two-level system (TLS) model introduced by \cite{Phillips} and \cite{Anderson1972}. Although the microscopic nature of TLS is still unclear \citep{Muller2019}, it is thought that one or a group of atoms in a disordered solid can tunnel between two sites, giving rise to tunneling quantum states with a broad spectrum of transition energy in amorphous solids. A TLS can be excited by absorbing a photon (from a resonator or qubit) and relaxes by emitting a phonon into the bath, causing resonator loss and qubit decoherence \citep{Martinis2005}. Furthermore, because of the randomness of the absorption and emission, TLS also cause fluctuations in dielectric constant (often referred to as two-level fluctuators or TLFs) and introduce excess phase noise to the resonators\citep{Gao_2Dnoise}.

As first shown by \citep{Martinis2005}, TLS in deposited dielectrics such as $\mathrm{SiO_{2}}$ and $\mathrm{Si_3N_{4}}$ are a significant source of loss and decoherence in superconducting quantum circuits. Gao showed that even for a superconducting resonator on crystalline Si without a deposited amorphous dielectric, TLS on surfaces and interfaces still dominate resonator intrinsic loss \citep{Gao2008b} and is responsible for excess frequency noise in these resonators \citep{Gao2007}. Since then, TLS-induced loss, decoherence, and noise have been active areas of research.

\begin{figure}[t]
    \centering
    \includegraphics[width=75mm]{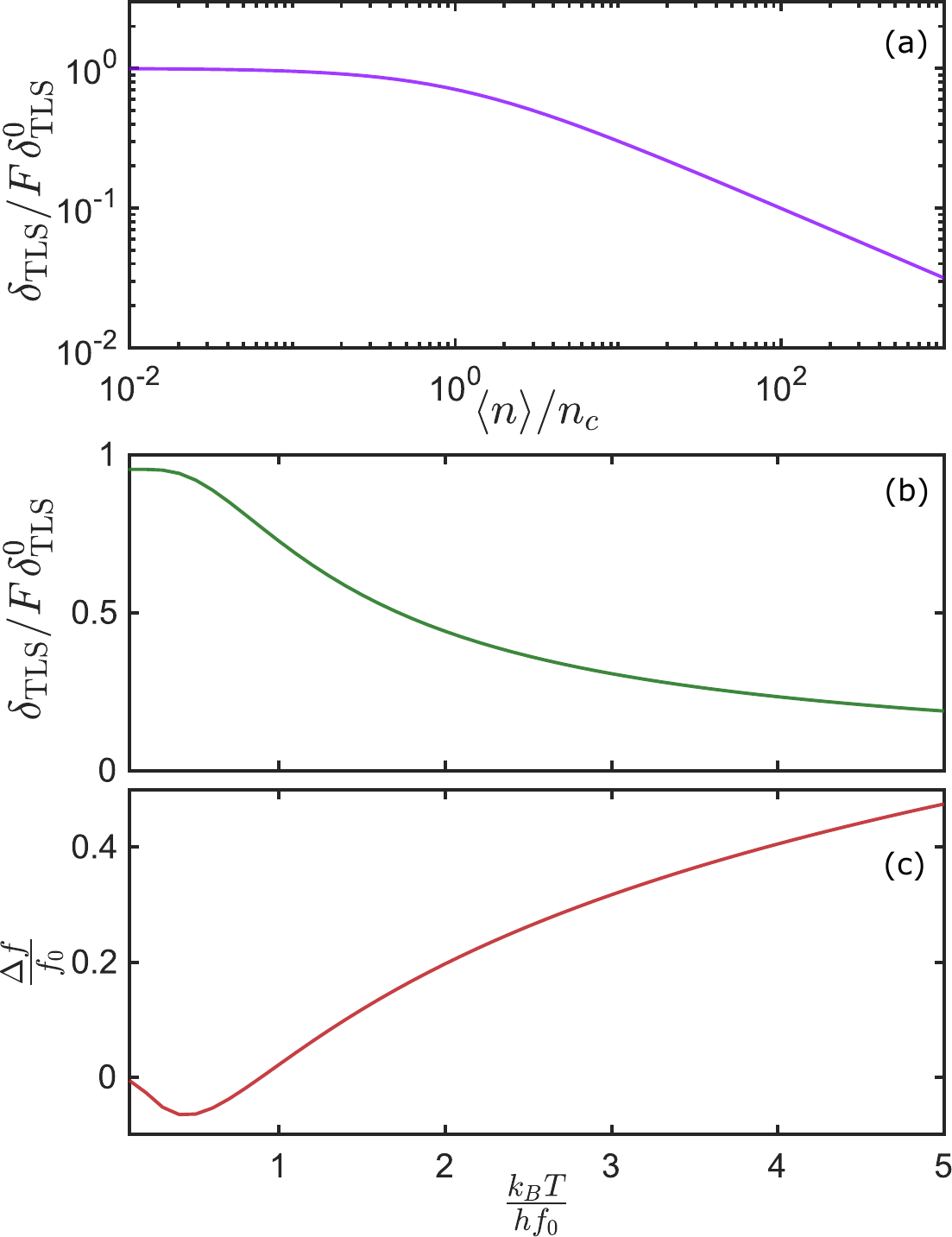}
    \caption{Theoretical TLS loss curves demonstrating TLS behavior outlined in Eqs.~\ref{eqn:TLSloss} and \ref{eqn:TLSfreqshift}. TLS resonator loss $\delta_{TLS}$ normalized by $F \delta^0_\mathrm{TLS}$ as a function of normalized (a) average photon number, and (b) temperature. (c) Temperature dependence of frequency shift due to TLS.}
    \label{fig:TLSTheoCurves}
\end{figure}

According to the TLS model, TLS induce a power- and temperature-dependent resonator loss $\delta_{TLS}$ and temperature-dependent resonance fractional frequency shift $\Delta f/f_0$ given by~\citep{GaoThesis}
\begin{equation}
\label{eqn:TLSloss}
\delta_{TLS} = F \delta_{TLS}^0 \frac{\tanh(\frac{\hbar \omega}{ 2 k_B T})}{\sqrt{1 + \frac{\langle n \rangle}{n_c}}},
\end{equation}
\begin{equation}
\label{eqn:TLSfreqshift}
\frac{\Delta f}{f_0} = \frac{F \delta_{TLS}^0}{\pi} \left[\mathrm{Re}\left(\Psi\left(\frac{1}{2}+\frac{1}{2 \pi i}\frac{\hbar \omega}{k T}\right)\right) - \log{\frac{\hbar \omega}{2\pi k T}}\right],
\end{equation}
where $\omega$ is the angular frequency, $\delta^0_{TLS}$ is the intrinsic TLS loss, $\langle n \rangle$ is the average photon number in the resonator, $n_c$ is the characteristic photon number of TLS saturation, $\Delta f=f-f_0$ is the difference between measured resonance frequency $f = \omega / 2 \pi$ and TLS-free resonance frequency $f_0$, $\mathrm{Re}(\Psi)$ is the real part of the complex digamma function, and $F$ is the filling factor of the TLS material, defined as the fraction of the resonator's total electrical energy stored in the TLS material. TLS are present in parts of the resonator (such as interfaces and dielectrics) and not others (such as superconductors and air), which reduces the effective loss due to these systems. 

When more than one TLS medium is present, as is the case in the vast majority of physical circuits, $F \delta_{TLS}^0$ in Eq.~\ref{eqn:TLSloss} and Eq.~\ref{eqn:TLSfreqshift} are replaced by a sum of contributions from each TLS material $\sum_{n} F_{n} \delta_{TLS,n}^0$ where
\begin{equation}
    F_{n} = \frac{U_n}{U_{tot}} = \frac{\int_{V_n} \frac{1}{2}\epsilon_n \lvert E^2 \rvert dV}{\int_V \frac{1}{2} \epsilon \lvert E^2 \rvert dV}
\end{equation}
where $U_{tot}$ is the total electric energy stored in resonator, $U_n,~F_n,~V_n,~\epsilon_n$ are the electrical energy, filling factor, volume and dielectric constant of the nth TLS medium.

The intrinsic TLS loss $\delta_{TLS}^0$ is an important dielectric property that characterizes the microwave dissipation in the material. It is physically set by the density and electric dipole moment of the tunneling states in the material, and is therefore an intrinsic property of the material. Values of $\delta_{TLS}^0$ can be compared directly between different dielectric materials.

It is clear from Eqs.~\ref{eqn:TLSloss}-\ref{eqn:TLSfreqshift} that TLS-induced loss and frequency shifts are dependent on geometry, through the filling factor $F$, as well as power, temperature, and frequency. Theoretical curves are plotted in Fig.~\ref{fig:TLSTheoCurves}. These curves demonstrate that TLS-induced loss is highest at low power and temperature. It decreases as power and temperature increase which is called TLS saturation effect. In this saturation regime, the population difference between the excited and ground states of the TLS is reduced due to transitions induced by a strong microwave drive or thermal phonons, leading to reduced net absorption of microwave photons. We may define the average number of photons in the resonator at the onset of TLS saturation as $<n_c>$. Most quantum computing experiments operate close to single photon regime with $<n>\lesssim <n_c>$. According to Eq.~\ref{eqn:TLSloss}, in the single photon power and millikelvin temperature regime of quantum information processing applications, the TLS loss takes its maximum value, approaching~$\delta_{TLS}^0$.

As shown in the equations above, $\delta_{TLS}^0$ can be determined either by fitting measured loss $1/Q_i$ as a function of power at low temperature to Eq.~\ref{eqn:TLSloss}, or by fitting the measured resonance frequency shift $\Delta f/f_0$ as a function of temperature to Eq.~\ref{eqn:TLSfreqshift}, with example fits shown in Fig.~\ref{fig:experimental_scurve}.

The standard theory of TLS in glasses assumes each TLS has a random energy minima and barrier height and hence there is an uniform energy distribution of TLS \citep{Anderson1972,Phillips,GaoThesis}, and studies of CPWs have not seen a frequency dependence of TLS density \citep{Lindstrom_2009_TLS}, even up to mm-wave frequencies \citep{gao2009_150GHz}. In conflict with this, \cite{Skacel2015} shows a frequency-dependent TLS distribution and loss in microstrip resonators, with low power loss increasing with frequency.

\subsection{Thermal Quasiparticle Loss}

\begin{figure}
    \centering
    \includegraphics[width=75mm]{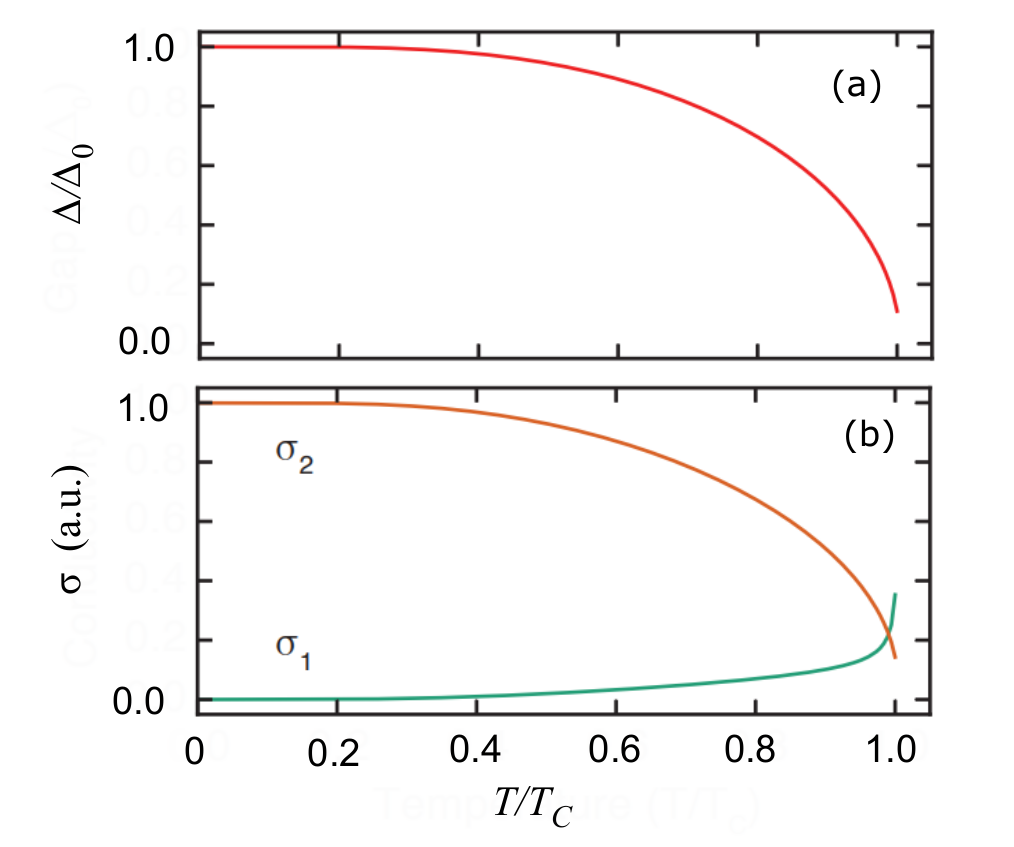}
    \caption{Temperature dependence of surface impedance in BCS superconductor. (a) The energy gap between free quasiparticles and bound Cooper pairs decreases at a significant fraction of $T_C$. (b) The real and imaginary parts of the AC surface conductivity, $\sigma_1 + i\sigma_2$. Figure adapted with permission from \cite{ReagorThesis}.
    }
    \label{fig:QPThermal}
\end{figure}

A superconducting metal possesses a complex surface conductivity  $\sigma_1 + i\sigma_2$ that is both a function of frequency and temperature. This behavior, and the resulting  thermal quasiparticle loss, is reviewed in \cite{Zmuidzinas2012} and illustrated here in Fig.~\ref{fig:QPThermal}. The AC surface conductivity manifests experimentally as a complex surface impedance $Z_{S} = R_{S} + iX_{S}$, composed of a surface resistance $R_{S}$ and a surface reactance $X_{S} = \omega \mu \lambda$, where $\mu$ is the magnetic permeability and $\lambda$ is the penetration depth. Surface quality of the material is defined by $Q_{S} = \frac{X_{S}}{R_{S}}$, which contributes to resonator loss scaled by the ratio of magnetic energy stored in the superconductor to the total magnetic energy: 
\begin{equation}
     \frac{\alpha}{Q_{S}}= \frac{R_{S}}{\omega\mu_{0}\lambda} \frac{\lambda \int_{S} \lvert H^2 \rvert dV}{\int_V \lvert H^2 \rvert dV}.
\end{equation}
 $\alpha$, defined by the second fraction on the right, is known equivalently as the conductor participation ratio or the kinetic inductance fraction, and it is analogous to the dielectric filling factor above.
 
 Thermally excited quasiparticles are intrinsic to the superconductor itself. At higher temperatures, quasiparticle density increases and the superconducting energy gap $\Delta$ is reduced. This changes both $R_{S}$ and $\lambda$ in the equation above. The change in surface impedance has the effect of shifting the frequency down and reducing the quality factor according to
 \begin{equation}
 \label{eq:surface_impedance}
     \frac{1}{Q} + 2i\frac{\delta f}{f} = \frac{\alpha}{\omega \mu \lambda}( R_{S} + i \delta X_{S} ).
 \end{equation}
Temperature sweep data can be compared to numerical integration of Mattis-Bardeen's formulas for AC conductivity of a BCS superconductor \citep{Mattis1958,Barends2008,Reagor_10ms3D}, as in Fig.~\ref{fig:Quasiparticle_temp_expt}. One can place bounds on $R_{S}$ of materials while remaining agnostic about the underlying physical mechanism (i.e. exact physics of quasiparticle dynamics).

The loss due to thermally excited quasiparticles is reduced exponentially as the bath temperature approaches zero. According to Mattis-Bardeen theory \citep{Mattis1958}, loss reaches $\delta < 10^{-7}$ for $T < T_c/10$, where $T_c$ is the superconducting critical temperature, which is a condition easily satisfied at the base temperature of a dilution refrigerator (DR) or an adiabatic demagnetization refrigerator (ADR). 

\subsection{Quasiparticle Loss Due to Stray Infrared Light and Ionizing Particles}
As mentioned above, thermally excited quasiparticles contribute negligibly to resonator loss in the quantum computing low temperature regime. Quantum experiments, however, find a non-equilibrium quasiparticle population that leads to loss and degradation of the quantum circuit \citep{Serniak2018, Zmuidzinas2012}. The source of these "excess" quasiparticles has been identified as pair breaking from stray radiation. \cite{Barends2011} and \cite{Corcoles2011} unambiguously show that resonator $Q_i$ and qubit energy relaxation times can be significantly degraded by excess quasiparticles from stray IR radiation, as well as demonstrate how this influence can be fully removed by isolating the devices from the radiative environment with multistage shielding. \cite{kreikebaum2016optimization} also show a systematic study of the shielding effect and report similar results. More recently, as the qubit coherence time continues to improve and puts more and more stringent requirement on excess quasiparticles, the non-equilibrium quasiparticles generated by ionization particles from cosmic ray and environmental radioactivity has become a new concern \citep{cardani2020, vepslinen2020, Karatsu2019}. Operating quantum devices with additional radiation shielding and in an underground facility has been proposed. These mitigation strategies are an active area of research.    

\subsection{Quasiparticle Loss Due to Microwave-induced Pair-breaking}
The microwave tone used to read out the resonator, with a frequency well below the gap frequency of the superconductor ($\hbar \omega \ll 2\Delta$), can indirectly break pairs and generate excess quasiparticles in the resonator. In this process, a strong microwave tone can pump a quasiparticle from an energy level of $E \gtrsim \Delta$ to a ladder of higher energy levels $E + n\hbar \omega$ (n=1,2,3...). When such a quasiparticle reaches a high enough energy level and relaxes back to an energy level slightly above the gap, a pair-breaking phonon can be emitted which produces more quasiparticles. This quasiparticle multiplication process has been theoretically studied \citep{Goldie2012} and experimentally confirmed \citep{DeVisser2014}. Quasiparticle relaxation in superconductors has been studied as a function of temperature \citep{Barends2008}. Additionally, qubits have been used as detectors to study non-equilibrium quasiparticle dynamics in superconductors \citep{Wang2014}.

\subsection{Vortex Loss}
Vortices of trapped magnetic flux form islands of normal metal in an otherwise continuous superconducting film, and thus can dissipate power when current is present at the core of the vortex. A systematic study of microwave response of vortices was carried out by \cite{Song2009} on thin Re and Al films where they found the trapped flux can lead to substantial reductions in the internal quality factor. They have showed that cooling Al and Re resonators of 12 um wide and 150 nm thick through $T_c$ in a residual magnetic field comparable to the earth field can trap vortices in the resonators and introduce excess fractional frequency shift and excess loss on the order of $10^{-5}-10^{-4}$. It is also well known that wider superconducting traces are more susceptible to vortex trapping and $B_{th} \approx \Phi_0/w^2$ (where $\Phi_0$ is the magnetic flux quanta and $w$ is the strip width) gives an good estimate of the threshold magnetic field for vortex trapping \citep{Song2009,Stan2004}. 

\subsection{Radiative Loss}
Strictly speaking, radiative loss refers a resonator losing its energy by radiating into free space. This loss generally increases with frequency and device dimension \citep{Sage2011}. Embedding the device in a high-Q cavity can effectively eliminate the radiation loss to free space, as is done in 3D qubit architecture which has achieved impressive long energy relaxation times \citep{Paik_PhysRevLett.107.240501}. Because most planar superconducting devices are mounted in a device box or package, their radiative loss can be generalized to include energy dissipation far from the device's intended extent. For example, a resonator's field may induce currents in the lossy metal walls of the box, or propagate into lossy dielectrics elsewhere in the package. One way to reduce radiative loss is to coat the backside of the chip with superconducting films, effectively reducing the mode's extent into the rest of the package \citep{Goetz2016}. For example, \cite{Sandberg2013} has suggested the radiation loss limited qubit $T_1$ time can be improved by two orders of magnitude by backside coating. A second mitigation strategy involves constructing the device box of superconducting material and avoiding dielectrics where possible.

\subsection{Parasitic Modes}

In addition to the previously described loss mechanisms, a resonator's $Q_i$ can be reduced through unwanted hybridization with nearby lower-Q modes, including slotline modes, box modes, chip modes, and even other on-chip devices \citep{Sheldon2017,Houck2008}. Such a reduction in $Q_i$ is referred to as parasitic loss. Two modes occupying the same volume share energy to the extent that they have mutually overlapping fields. The amount of hybridization, and thus the loss imparted to the higher-Q mode, is frequency dependent, as coupling decreases with increasing mode detuning. Strategies to avoid parasitic loss include increasing the detuning and reducing the field overlap between the resonator mode and any other modes present, as well as increasing the $Q_i$ of all modes present.

Parasitic slotline modes are a type of mode supported by a CPW's ground planes when they are not electrically connected or are uneven in width, leading to a voltage differential across the conductor. They can couple strongly to CPW resonators due to large field overlap, and they can have low $Q_i$ due to their larger spacial extent into dielectric and packaging. Slotline modes can be eliminated by adding an electrical connection between the ground planes using wirebonds \citep{Wenner2011}, airbridges \citep{Abuwasib2013,chen2014_airbridge,dunsworth2018_airbridge}, vias \citep{Vahidpour2017}, or bump bonding \citep{Lei2020,Foxen2018,Rosenberg2017}.

Box modes are modes sustained within the free space in the sample box \citep{Wenner2011}. The frequencies of these modes depend on the geometry of the box. Box mode frequencies can be intentionally raised by modifying the geometry or boundary conditions of the cavity. For example, adding an evacuated space below the chip \citep{Wenner2011} or pogo pins \citep{Bejanin2016,Mcconkey2017,Bronn2018} raises the box mode frequencies. Similarly, chip modes are sustained within the substrate, and can be raised in frequency by the addition of through-substrate vias.

\subsection{Total Loss}

\begin{figure}[t]
    \centering
    \includegraphics[width=85mm]{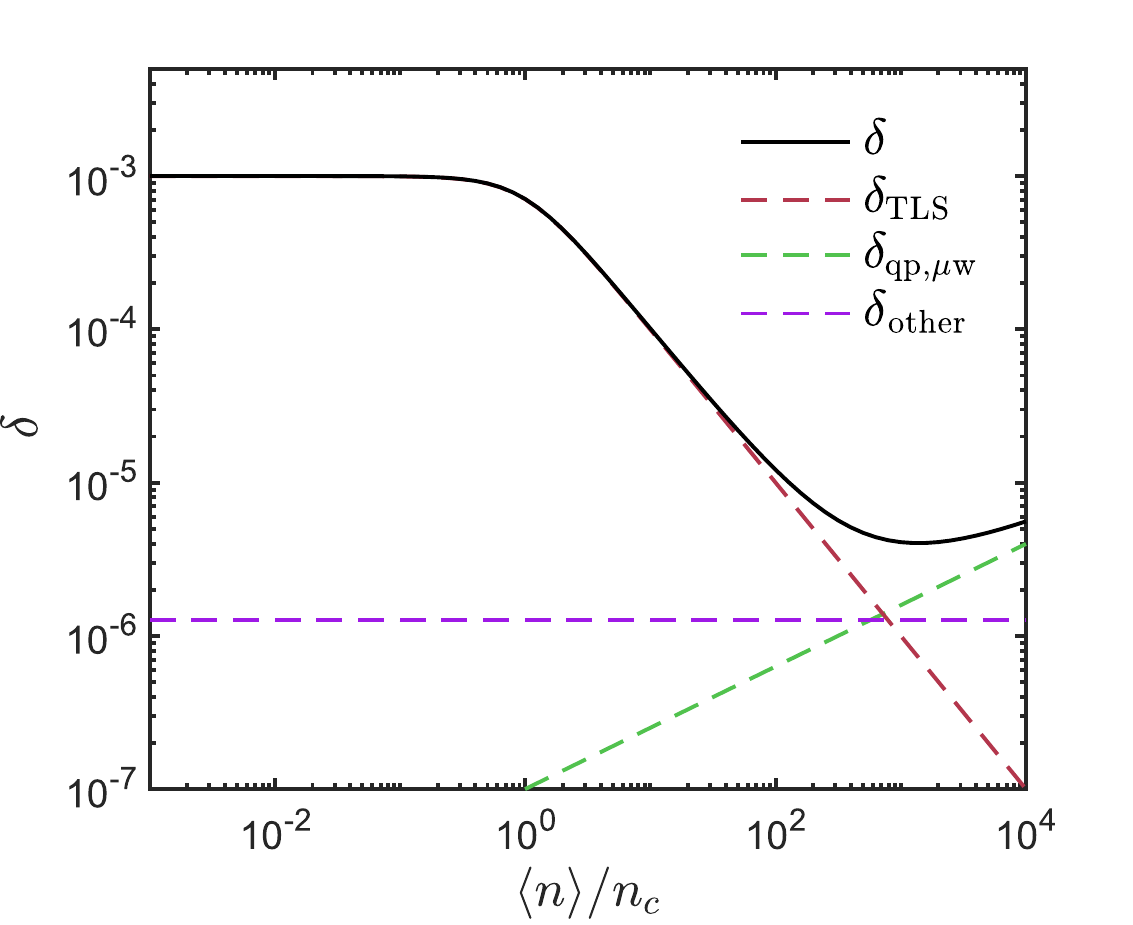}
     \caption{Theory curves demonstrating total loss (solid line) and primary loss mechanisms (dashed lines) in a superconducting microwave resonator. In this resonator model, $F \delta^0_{\mathrm{TLS}} = 10^{-3}$, $\delta_\mathrm{qp,\mu w} = 10^{-7}(\langle n \rangle / n_c)^{0.4}$, and $\delta_{\mathrm{other}} = 1.5\times10^{-6}$. The resonator, modelled using Eq.~\ref{eq:totallossinfig}, is dominated by TLS loss $\delta_{TLS}$ (red) in the single photon regime, where $\langle n \rangle / n_c\lesssim1$. At higher power, microwave-induced quasiparticle loss $\delta_\mathrm{qp,\mu w}$ (green) becomes an important contributor (a simple power law is assumed here for illustration purposes). Power-independent loss $\delta_{\mathrm{other}}$ (purple) from various sources is given as a total sum.}
    \label{fig:losssummary}
\end{figure}

Fig.~\ref{fig:losssummary} shows the power dependence as well as relative contributions from each loss channel in a superconducting resonator, where the total loss is given as
\begin{equation}
    \delta = \delta_{TLS} + \delta_\mathrm{qp,\mu w} + \delta_\mathrm{other}
\label{eq:totallossinfig}
\end{equation}
where $\delta_\mathrm{other}$ is the sum of all power-independent losses acting on the resonator, including loss from radiation, quasiparticles generated from infrared or thermal sources, vortices, and parasitic modes. These power-independent losses can usually be suppressed to $\delta_\mathrm{other} \lesssim~10^{-6}$ by the use of mitigation strategies (see Sec.~\ref{sec:setup} and Sec.~\ref{sec:methods}). Fig.~\ref{fig:losssummary} shows that TLS loss dominates the total loss in the single photon regime where quantum computers operate, while the microwave-induced quasiparticle loss dominates at high power where microwave kinetic inductance detectors [MKIDs \citep{Zmuidzinas2012}] operate. There are few exceptions where TLS is not the dominant source of loss at low power. These exceptions include 3D cavities, which have very small TLS loss and the metal surface quality factor $Q_{S}$ may play an important role \citep{Reagor_10ms3D}, and planar resonators associated with GaAs where the total loss can be dominated by piezoelectric loss \citep{Scigliuzzo2020}. 

\subsection{Qubit Loss Mechanisms}

Superconducting qubits are sensitive to all of the above loss mechanisms, in addition to losses specific to Josephson junctions \citep{catelani2011relaxation}. Qubit coherence is commonly described by energy relaxation time constant $T_1$, which can be compared to $Q_i$ as
\begin{equation}
    T_1 = Q_i / (2 \pi f_{10})
\end{equation}
where $f_{10}$ is the frequency of the qubit state.  The exact circuit used for qubit implementation greatly affects the participation of the these loss mechanisms. Direct comparison experiments have shown striking correlations between TLS loss in resonators and qubits \citep{Dunsworth2017,Quintana2014}. Due to their requirement for single-photon operating power, qubits cannot measure the power dependence of loss. However, unlike resonators, qubits can easily implement frequency tuning via flux biasing Josephson junctions, which is another useful technique for differentiating loss channels. A summary of frequency dependencies of loss mechanisms can be found in \cite{dunsworth2018thesis}. Using frequency tuning, qubits have been shown to interact coherently with resonant TLS defects, leading to energy level splittings, as well as incoherently, as shown by loss peaks in frequency spectra \citep{Martinis2005,barends2013coherent}. These strongly coupled TLS are a key error mechanism in superconducting qubits.

Many loss mechanisms that plague qubits stem from additional control requirements for performing high fidelity quantum computation. While isolated qubits have been shown to have $T_1 \sim 170 \ \mu$s ($Q_i \sim 4.2\times 10^6$ at 4 GHz), nearing that of their resonator counterparts \citep{dial2016bulk_surf}, systems of qubits tend to have shorter $T_1 \ \sim 15 \ \mu$s ($Q_i \sim 0.6 \times 10^6$ at 6.5 GHz) \citep{arute2019quantum}. This trade-off is a manifestation of optimizing for large system fidelity rather than individual qubit lifetime. Each additional control line or on-chip circuit element increases the complexity and speed of quantum operations at the cost of increasing loss. The ratio of operation time to coherence time $T_{gate}/T_1$ gives a heuristic for energy relaxation induced incoherent error rates in quantum processors.  While these additional loss mechanisms can become significant in systems, they are designed, and can be optimized for system performance.  Materials loss is still among the dominant losses in qubit devices, especially with additional materials and processing required for qubit devices, and thus materials analysis is vital to pushing the cutting edge of quantum computation.

\subsection{Noise}

The random nature of TLS-resonator photon exchange leads to effective fluctuations in the dielectric constant. These fluctuations manifest as phase or frequency noise, and are equivalent to resonance frequency jitter \citep{Zmuidzinas2012}. TLS noise shows the same power, temperature, and geometry dependence (see Table~\ref{tab:losssummary}) as TLS loss \citep{Gao2007,GaoThesis}. As shown by \cite{DeGraaf2017}, a reduction of TLS effectively reduces TLS-induced resonator loss and noise as well as qubit decoherence. The simultaneous measurement of multiple TLS effects, including loss, frequency shift, noise, and qubit decoherence time, may yield richer information, more reliable data, and a more comprehensive picture of TLS.


Noise due to TLS fluctuators is the dominant noise source for MKIDs. Therefore, understanding the TLS physics, searching for material with lower TLS loss, and developing techniques to reduce the negative effects have become a common interest to and an important joint effort for the superconducting quantum computing and detector research communities.

\section{\label{sec:setup}Measurement Setup}

\begin{figure}
\includegraphics[width=85mm]{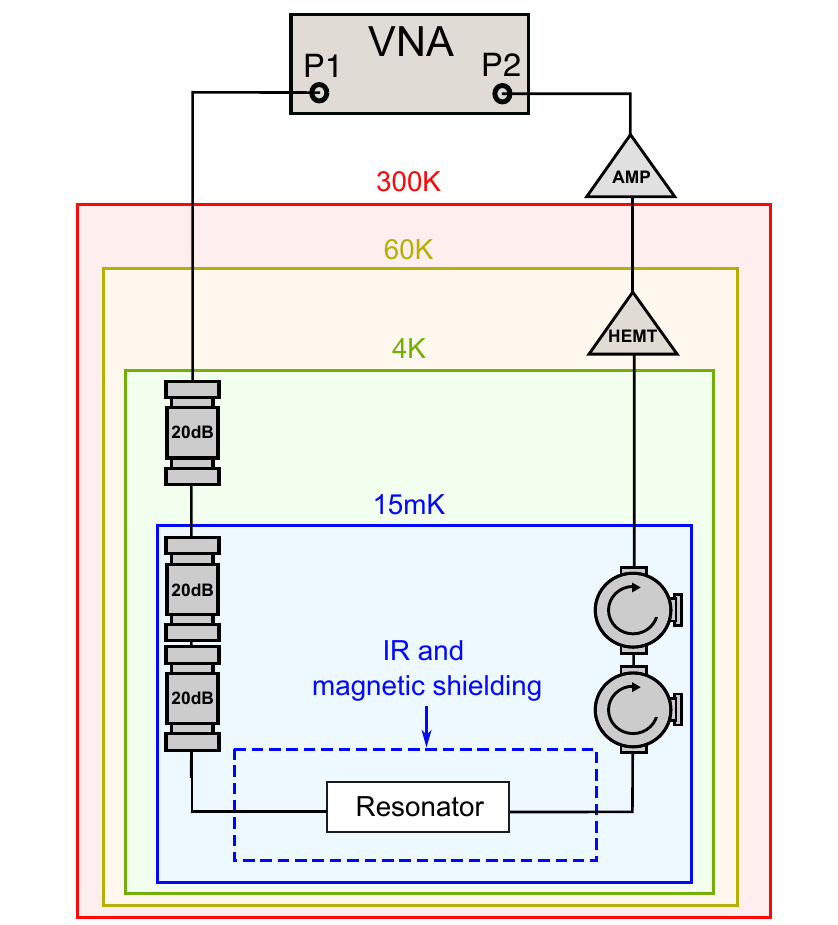}
\caption{\label{fig:cryosetup}Standard wiring diagram of a measurement setup for a superconducting resonator in a dilution refrigerator (DR). A vector network analyzer (VNA) is used for resonator excitation and readout. On the input line (left), attenuators at each temperature stage are used for thermalization and minimizing thermal noise. On the output (right), 40~dB double stage isolators can be implemented for blocking radiation from the output line. A high mobility electron transistor (HEMT) at 4~K and an amplifier at room temperature are used to amplify the output signal.}
\end{figure}

A dilution refrigerator (DR) is commonly used to cool devices to millikelvin operating temperatures. Fig.~\ref{fig:cryosetup} shows a typical DR wiring diagram for resonator control and readout between millikelvin and ambient temperatures. In order to minimize noise and sufficiently thermalize input cables at 4~K and 15~mK stages, a series of cryogenic attenuators, sometimes combined with filters, are connected along the input line. Amplifiers and isolators are connected to the output in order to allow signal readout through a vector network analyzer (VNA) \citep{kreikebaum2016optimization}. An adiabatic demagnetization refrigerator (ADR) can be set up similarly for this purpose, but the presence of magnetic fields may detract from the reliability of the measurement. 

Due to the extreme environment, small measurement signal, and susceptibility to environmental coupling, superconducting microwave resonators can be deceptively difficult to measure accurately. Impedance mismatches in the measurement largely come from components inside the DR, which are not commonly included in calibration and can affect measurement accuracy. Microwave setup, measurement mode, thermalization, and number of photons inside the resonator all play significant roles in measurement outcomes. Magnetic shielding and IR absorption and filtering can be implemented to minimize losses due to trapped vortices and RF induced quasiparticles, as discussed in Sec.~\ref{sec:losses}. 

Sample packaging can also affect measurement outcomes. Unless carefully designed, box modes can be present that couple to the resonator. Normal metals situated close to the device such as in silver paint can lead to irreproducible high power losses. Al, Cu, and gold-coated Cu are all commonly used packaging materials and exhibit different magnetic shielding and thermalization behavior. Refer to \cite{Lienhard2019} and \cite{Rosenberg2019} for detailed discussions.


\subsection{Microwave Components and Wiring}

Stainless steel coaxial cables are widely used in superconducting quantum computing experiments, especially to connect the room temperature plate of the refrigerator to the 4~K plate, due to their low loss, moderate thermal conductivity, and low passive thermal loads. From 4~K down to 15~mK, cables made of superconductors such as NbTi are commonly used to reduce loss and increase thermal isolation. At the base temperature plate, cryogenic coaxial semi-rigid cables are used for connecting different components and devices. Besides the cable materials, thermalizing microwave lines can be achieved with cavity attenuators \citep{Wang2019} and specialized microwave attenuators \citep{Yeh2017}.

In order to reduce the room temperature thermal noise to at least two orders of magnitude lower than the single photon power level, the input line needs roughly 60~dB of added attenuation \citep{krinner2019engineering}, which is typically achieved by placing attenuators at different temperature stages, as in Fig.~\ref{fig:cryosetup}. 
While the input signal is heavily attenuated, the output line has at least two stages of amplification, the first being a 40~dB high electron mobility transistor (HEMT) amplifier located at the 4~K stage, where $T_{noise}$~$\sim$~2.2~K, about an order of magnitude higher than vacuum noise at 6 GHz. This is followed by a 30~dB amplifier at room temperature to amplify the output signal sufficiently for VNA measurement. Based on the Friis formula \citep{kraus1986}, the noise factor of the first amplifier sets the noise floor of the amplification chain. Thus, although HEMT is a high gain low noise amplifier, a quantum limited amplifier like a Josephson parametric amplifer~(JPA) \citep{gao2011strongly,white2015traveling,castellanos2007widely} or Josephson parametric converter~(JPC) \citep{bergeal2010} can be used to increase the signal to noise ratio~(SNR). 

On the output, the same level of attenuation as the input is required. Since the HEMT can block radiation from above the 4~K plate, a double-stage isolator with total isolation of nearly 40~dB is commonly mounted to the 15~mK stage. As these isolators contain magnets, careful placement is essential in order to avoid magnetic loss.


\subsection{Impedance Matching and Calibration}

SMA connectors, coaxial cables, wirebonds, and on-chip transmission lines often have impedances deviating from $50~\Omega$. A mismatch between these components and the $50~\Omega$ VNA source impedance creates unwanted reflections, causing ripples in the transmission baseline and an asymmetric resonance lineshape, and impeding $Q_i$ and $Q_c$ extraction. These reflections can also lead to overestimation of power at the resonator. Calibration of the microwave environment can correct these errors. The commonly used two-port thru-reflect-line (TRL) and short-open-load-thru (SOLT) calibration methods \citep{SOLTvsTRL} require known standards at cryogenic temperatures which are not easily obtained. In addition, the non-reciprocal microwave environment increases the calibration difficulty: directional couplers or circulators and microwave switches are needed to separate reflection signals from calibration elements. Thus, it is common to perform bench top calibration using the VNA manufacturer calibration kit and assume coaxial cables and microwave components perform similarly at room temperature and when cooled. However, the electrical and mechanical properties of the materials of cables and other components at low temperature will change, and thus cause impedance mismatches when applying the room temperature calibration. \cite{ranzani2013two} demonstrates a method for two-port calibration at base temperature which implements a calculation of the error matrix elements and uses extra setup hardware. A millikelvin calibration method performed by the VNA itself is an existing goal in the superconducting quantum computing field. Since $S_{21}$ is the crucial parameter in these experiments, one-port calibration alone could be effective enough for calibrating the measurement setup chain.

\subsection{Measurement Mode and Fitting}\label{subsec:GeoandFitting}

A resonator can be measured in three commonly used modes: hanger~[Fig.~\ref{fig:circuits}(a)] \citep{Gao2008b}, transmission~[Fig.~\ref{fig:circuits}(b)] \citep{petersan1998measurement}, and reflection~[Fig.~\ref{fig:circuits}(c)] \citep{castellanos2007widely}. The hanger mode is realized by coupling a resonator to a transmission feedline which is connected to the measurement environment directly. This mode results in a Lorentzian lineshape with a dip in magnitude at the resonance frequency and a phase shift lower than $\pi$. As illustrated in Fig.~\ref{fig:circuits}(a), the total complex $S_{21}(f)$ can be represented by the ratio of the output voltage $V_2$ to the source voltage $V_g$ as $S_{21}$=2$V_{2}/V_g$, or in terms of quality factor, as
\begin{equation}\label{eqn:hanger}
  S_{21}(f) =1-\frac{\frac{Q}{Q_c}}{1+2 i Q \frac{f-f_{0}}{f_{0}}},
\end{equation}
where $f_{0}$ is the resonance frequency. One advantage of hanger mode is its natural frequency multiplexing: devices can be designed such that multiple resonators share a microwave input and output line on a single package or chip.

Transmission mode is commonly used for directly measuring half-wave ($\lambda/2$) CPW resonators \citep{goppl2008coplanar, OConnell2008}. It results in a Lorentzian line shape with a peak in magnitude at the resonance frequency and a $\pi$ phase shift \citep{pozar}. The total complex $S_{21}(f)$ can be represented as
\begin{equation}\label{eqn:transmission}
  S_{21}(f) = \frac{\frac{Q}{Q_c}}{1+2 i Q \frac{f-f_{0}}{f_{0}}},
\end{equation}
which is a mirror image of a hanger-type resonance. A drawback of inline transmission resonator measurements is that calibrating unity transmission is difficult because cable loss and resonator loss become indistinguishable \citep{Probst2015}, thus reducing quality factor accuracy.

The reflection mode is realized by coupling the resonator to the input microwave transmission directly through a capacitor or inductor and then measuring the reflection. Since the reflected signal cannot be detected directly at the input port, a non-reciprocal device must be adopted. By using a circulator or a directional coupler to separate the incoming signal to the resonator and the reflection from the resonator, the transmission $S_{21}$ can represent the reflection to the input and can be described by \citep{manuelthesis}
\begin{equation}\label{eqn:reflection}
  S_{21}(f) =1-\frac{2\frac{Q}{Q_c}}{1+2 i Q \frac{f-f_{0}}{f_{0}}},
\end{equation}
differing from the hanger case by a factor of 2 in signal. The line shape is also a Lorenztian dip in magnitude, but the phase shift can be $2\pi$, $\pi$ or less than $\pi$, depending the ratio between $Q_i$ and $Q_c$. 

Eqs.~\ref{eqn:hanger}-\ref{eqn:reflection} assume perfect impedance matching to the environment and a flat, featureless background, which are assumptions rarely satisfied by measured $S_{21}$ data. In order to extract $Q_c$, $Q_i$ and $f_0$ from measurement, some background normalization or removal method, as well as a correction of the rotation angle $\phi$, must be implemented. $\phi$ describes the rotation of the resonance circle in an Im($S_{21}$) v.s. Re($S_{21}$) plot, as in Fig.~\ref{fig:S21plot}~(a), with respect to the off-resonance point (1,0) as well as the asymmetry of the resonance lineshape in an $S_{21}$ v.s. $f$ plot [Fig.~\ref{fig:S21plot}~(b) and (c)] \citep{GaoThesis,Khalil2012}.

A summary of common fitting models and methods for hanger mode measurements are outlined below. To the authors' knowledge, there exist no methods for correcting rotation caused by impedance mismatch in transmission and reflection measurement modes, leading to inaccuracies not present in hanger mode fitting.

The $\phi$~rotation method ($\phi$RM) \citep{GaoThesis} was the first fitting method to take impedance mismatch into account, with the model
\begin{equation}\label{eqn:phiRM}
  S_{21}(f) =1-\frac{\frac{Q}{Q_c} e^{i\phi}}{1+2 i Q \frac{f-f_{0}}{f_{0}}},
\end{equation}
where $e^{i\phi}$ is used to account for circuit asymmetry. A circle fit, rotation and translation to origin, and phase angle fit are implemented to determine an initial guess of parameters, followed by a direct fit to Eq.~\ref{eqn:phiRM}.

The diameter correction method (DCM) \citep{Khalil2012} modifies the $\phi$RM as
\begin{equation}\label{eqn:DCM}
    S_{21}(f) =1-\frac{Q / \hat {Q}_c}{1+2 i Q \frac{f-f_{0}}{f_{0}}},
\end{equation}
where $\hat {Q}_c = Q_c e^{-i\phi}$ is the complex coupled quality factor. By accounting for the origin of the rotation as a component of $Q$, a correction of $1/(\cos(\phi)-1)$ is applied to the $Q_i$ value, which increases in magnitude with increasing circuit asymmetry.

The inverse $S_{21}$ method (INV) \citep{Megrant2012} uses the model \begin{equation}\label{eqn:INV}
    S_{21}^{-1}(f) = 1+\frac{Q_i}{Q^{\ast}_c}e^{i \phi}\frac{1}{1+2 i Q_i \frac{f-f_{0}}{f_{0}}},
\end{equation}
where $Q_c^* = (Z_0/|Z|) Q_c$ is the rescaled coupling quality factor, 
$Z_0$ is the cable impedance, and $Z = |Z|e^{i\phi}$ is half the inverse sum of the environment impedances on the input and output sides of the resonator. Similarly to the DCM model, the coupled quality factor is used to absorb the impedance mismatch in order to return $Q_i$ accurately. However, unlike DCM, INV cannot be used to determine $Q_c$ without environment impedance, which is very difficult or impossible to determine. 

In practice, DCM and INV give largely similar values for $Q_i$, though an exhaustive comparison has yet to be performed, while $\phi$RM systematically overestimates $Q_i$ as a function of $\phi$. Python fitting code exists for all fitting methods mentioned above,\footnote{https://github.com/Boulder-Cryogenic-Quantum-Testbed/measurement/} as well as the lesser-known closest pole and zero method (CPZM) \citep{Deng2013}, and allows for straightforward comparison between these methods.

In measured $S_{21}$ data, total signal transmission is shifted from $0$~dB by the total attenuation, losses and amplification in the measurement setup. As well, a nonlinear background exists due to external reflections, which cause ripples in transmission, and parasitic modes, which cause broad resonance features. Normalization is required in order to fit the data into the models above. 

When $Q$ is high enough, the resonance linewidth is much smaller than the background features and a linear normalization can be used by performing a linear fit to off-resonant points on either side of the resonance. Alternatively, a high-temperature background removal can be performed \citep{GaoThesis} by heating the circuit to $T_c/2 < T < T_c$, where thermally-populated quasiparticles decrease $Q_i$ to the point that a resonance is no longer visible, and using the resulting spectrum as a background to be subtracted from the resonance data.

The effects of normalization on reported $Q_i$ and $Q_c$ are not well-understood, so the use of a traceable protocol is essential for fit reproducibility. 

\subsection{Temperature and Thermalization}



In order to avoid thermally-populated quasiparticle loss~\citep{Mattis1958}, resonator measurements are normally performed at temperatures lower than one tenth of the superconducting transition temperatures $T_c$ of the electrode material. Common superconductors used in microwave circuits have $T_c$ between 1.2 and 18~K. DR systems are ideal for resonator measurements as their standard base temperature is less than 20~mK and can operate at base temperature indefinitely. A low measurement temperature is also critical for reporting accurate TLS loss, as TLS loss decreases as temperature increases in the low temperature regime (as shown in Fig.~\ref{fig:TLSTheoCurves}). Insufficient sample thermalization keeps the resonator at a higher temperature than the cold plate of the refrigerator and artificially lowers the loss extracted from measurement. Thus, to avoid systematically under-reporting loss, thermalization the resonator is crucially important.
 
Resonator thermalization methods include using oxygen-free high thermal conductivity copper to act as an ideal cold resistor and thermal link for sample thermalization, and shielding the sample from black-body radiation from higher temperature stages. On-chip refrigeration can also be performed \citep{Partanen2016} at the cost of both increasing the amount of processing the chip is exposed to as well as adding normal metal to the sample, which itself can introduce loss. However, there is ambiguity in the literature about the degree to which chips are thermalized in conventional work. As reported by \cite{geerlings2012}, when mounted on the mixing chamber plate of a dilution refrigerator with temperature of $\textrm{15~mK}$, the effective sample temperature could be as high as $\textrm{50~mK}$.
 
\subsection{Vortex Loss Mitigation}

Vortex loss has been shown to cause up to 10$\%$ of measured loss, and can be suppressed by magnetic field shielding \citep{nsanzineza2016}. Because magnetic shielding is effective and can be easily implemented \citep{kreikebaum2016optimization}, it is commonly utilized in loss measurements. The use of vortex-pinning hole arrays \citep{Chiaro2016,Stan2004} is another widely-used method, requiring at least 6~$\mathrm{\mu}$m of separation between the vortex-pinning hole arrays and the CPW center trace \citep{bothner2011improving,Chiaro2016} in order to avoid increasing TLS participation. Trenching a slot on the center strip of CPW resonator is another straightforward method for reducing loss due to trapped vortices \citep{song2009reducing}. Single vortices trapped at current anti-nodes of CPW resonators can reduce quasiparticle loss \citep{nsanzineza2014trapping}. Although these vortex trapping techniques can help achieve higher $Q_i$ values, implementing them adds an extra level of complexity of fabrication and measurement.

\subsection{Measurement Variability}

Resonator loss variability has been observed for a single resonator over different cooldowns \citep{Calusine2018}, for nominally identical resonators on different chips \citep{Ohya2014}, and for a single resonator over time \citep{neill2013fluctuations}. All of these types of resonator loss variability warrant more in-depth study.

Repeated measurements of the exact same resonator have been shown to yield different $Q_i$ values at both low and high power, though the variation at high power is more pronounced (at a difference of up to two times or more), but show no change in intrinsic TLS loss \citep{Calusine2018}. This cooldown-to-cooldown variability can be attributed to changes in power-independent losses, and emphasizes the benefit of performing a TLS loss extraction rather than reporting the low power loss.

Resonators exposed to the same processing conditions on the same chip can fall within a range of $Q_i$ and $Q_c$ values. \cite{Ohya2014} reports $Q_c$ values varying by more than an order of magnitude, and $Q_i$ varying by up to an order of magnitude. It is possible to explain the change of $Q_c$ by small variations in microwave set-up and environment between measurements. Variations in $Q_i$ indicate that there are some subtleties in fabrication that are difficult to control. Thus, the loss measurement of a single resonator could have associated uncertainty of up to an order of magnitude.

Time fluctuations of $1/Q_i$ due to TLS loss have been seen to scale with the magnitude of the loss and fluctuate at $\sim 20\%$ of the mean \citep{neill2013fluctuations}. Further characterization of this phenomenon is warranted.

\section{\label{sec:methods}Methods for Materials Loss Measurements Using Resonators}

\begin{figure}
\includegraphics[width=85mm]{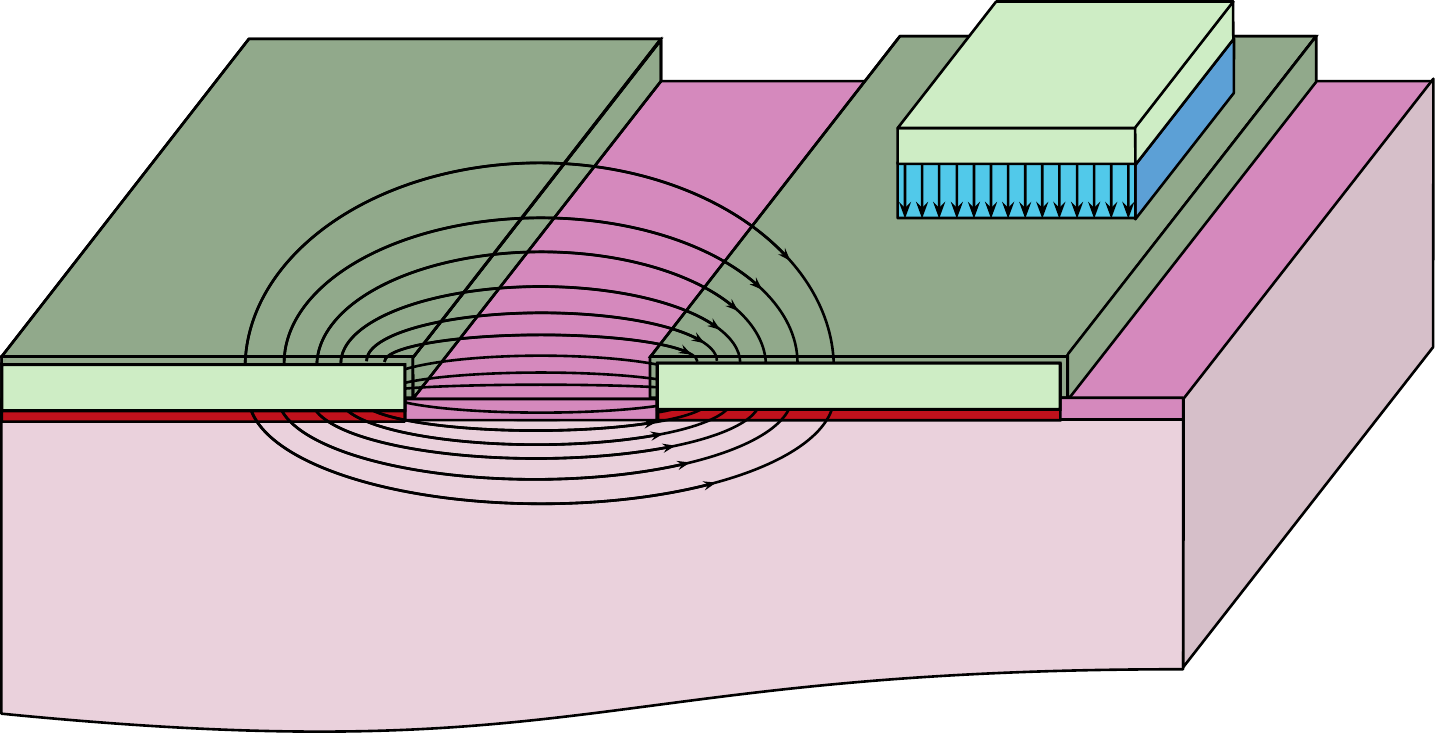}
\caption{\label{fig:lossschematic}Regions of TLS loss in a planar superconducting circuit with electric field lines shown in black.  Light green: superconducting metal (no TLS contribution). Dark green: Metal-air (MA) interface. Light pink: Dielectric substrate. Dark pink: Substrate-air (SA) interface. Red: Metal-substrate (MS) interface. Light blue: thin film dielectric in a parallel-plate configuration.}
\end{figure}

The term ``materials loss'' encompasses all forms of intrinsic loss induced in a superconducting microwave resonator due to its constituent materials and interfaces. The main source of materials loss in planar devices is TLS loss, but other types such as piezoelectric loss are applicable in rare instances.

Materials loss is present in three general regions, as shown in Fig.~\ref{fig:lossschematic}: in bulk dielectrics such as substrates, in thin film dielectrics such as Josephson junctions and deposited dielectrics, and in surfaces and interfaces between materials. In high performance superconducting microwave circuits for quantum information, surfaces and interfaces can have the highest loss, up to $10^{-2}$ \citep{Wenner2011}, while intrinsic crystalline bulk dielectric substrates can have the lowest, $\le 10^{-7}$, \citep{Calusine2018} and deposited dielectrics vary between $10^{-2}$ and $10^{-5}$ \citep{OConnell2008}.

3D cavity resonators obviate both radiative and bulk dielectric losses by maximizing the participation of vacuum. They are susceptible only to loss in the cavity wall's surface oxide and conductor loss extending into the wall a distance of the London penetration depth \citep{Romanenko2017,Reagor_10ms3D}. Cavities have also been used intentionally to measure bulk dielectric loss \citep{Pogue2012,Krupka1999}.

The low power loss $\delta_{\mathrm{LP}}$, the resonator-induced intrinsic TLS loss $F \delta^0_{\mathrm{TLS}}$ or simply the material's intrinsic TLS loss $\delta^0_{\mathrm{TLS}}$ are the most common metrics used to compare losses. The TLS loss is extracted from a power or temperature sweep using Eq.~\ref{eqn:TLSloss} or Eq.~\ref{eqn:TLSfreqshift}, while $\delta_{\mathrm{LP}}$ is the loss at some ``low enough'' power, typically on the order of a single photon occupation in the resonator, where it is expected that TLS loss contribution is dominant. This is often determined empirically when loss plateaus at low power.

Due to their sensitivity to TLS, resonators are seemingly ideal devices for measuring the intrinsic losses of dielectric materials and extracting the TLS loss of materials. However, measuring the TLS loss of a single material can prove difficult. Every interface, substrate, and deposited dielectric contributes to the filling factor, i.e., the total TLS loss in proportion to the amount of capacitive energy stored in it (see Sec.~\ref{sec:losses} for more information). Therefore, the device must be carefully designed, and loss extraction analysis methods are often required, in order to determine an accurate and precise TLS loss value for a single region or material of interest. Dielectric loss has been measured using a wide variety of device structures including CPW resonators, distributed-element microstrip resonators, lumped element resonators with parallel plate capacitors, and lumped element resonators with interdigitated capacitors.

\subsection{Form Factor}

Changes in the design of planar, multilayer, distributed, or lumped element resonators can be used to optimize the resonator's sensitivity to bulk or surface dielectric losses. Since TLS loss is capacitive \citep{Vissers2012}, the design of the resonator's capacitor determines the TLS regions to which it is sensitive. Planar devices are sensitive to surface dielectric losses, whether they are CPW \citep{Gao2008a} or IDC \citep{Deng2014,Vissers2012} designs. While they are also compatible with measurements of a wide variety of subsequently-added dielectrics, only a fraction of the electric field will be stored in the added material, implying that the planar devices will have a low sensitivity to the bulk TLS loss of these deposited dielectrics. Since CPWs are also used in high performance superconducting quantum circuits for algorithm implementation, these resonators can be used as a proxy and measured in order to improve the quantum devices themselves, in addition to their use as a direct TLS measurement technique.

On the other hand, multilayer devices involving deposited dielectrics and multiple wiring layers (such as microstrip or parallel plate lumped element resonators) can concentrate fields in the thin film or bulk dielectric of interest. Multilayer resonators also have a lower participation of native surface oxides, which are difficult to control during processing and often require simulation to disentangle their loss from the loss of interest. However, these devices necessitate more fabrication steps and can thus introduce unwanted interface losses that are difficult to distinguish from the loss of interest.

It has been assumed that lumped element inductors introduce a negligible fraction of capacitance into lumped element resonators \citep{OConnell2008} thus allowing their TLS loss contribution to be ignored. However, recent study shows that depending on the magnitude of loss in the capacitor compared to the generally high magnitude of interface loss, this assumption does not necessarily hold \citep{McRae2020}.

In 3D cavity resonators, losses due to surface oxide and conductor loss can be effectively scaled as the ratio of surface area to volume, depending on the cavity geometry.  Alternatively, a piece of bulk dielectric material of interest can be inserted inside a 3D cavity for loss study \citep{LeFloch2014}. In cases where the resonant mode is supported by currents traversing a joint in the cavity's construction, the conductor loss at such joints has been studied and reduced by employing In bonds \citep{Lei2020}.

\subsection{Resonator Coupling}

A resonator is coupled to a feedline by creating a capacitive or inductive link. The fraction of the resonator's total capacitance or inductance respectively that is used for this circuit element should be carefully tuned to ensure a close match between $Q_i$ and $Q_c$ at the temperature and power of interest in order to measure the resonator's loss accurately. For $Q_c \ll Q_i$, $Q_c$ dominates the signal vs.~frequency lineshape and the accuracy of measured $Q_i$ is decreased. For $Q_c \gg Q_i$, the resonator becomes difficult to measure due to reduced signal \citep{GaoThesis}. Thus, in order to be able to accurately measure resonator $Q_i$, we must either first estimate the order of magnitude of $Q_i$ or measure a series of resonators with varied $Q_c$ values and use the data from the resonator with the closest match between $Q_i$ and $Q_c$.

Theoretically, if an accurate equivalent circuit model of a resonator measurement setup is known, the coupling loss can be calculated by $1/Q_c=\mathrm{Re}[Y(\omega_0)]Z_r$, where $Y(\omega_0)$ is the total admittance to ground through the coupling components at resonance frequency and $Z_r=\sqrt{\frac{L}{C}}$ is the characteristic impedance of the resonator \citep{schuster2007,goppl2008coplanar}. Performing microwave simulations of the design with lossless materials (i.e., $Q = Q_c$) and estimating the total $Q$ by measuring the bandwidth of the resonance dip or peak is a standard way to determine $Q_c$ for a given design prior to fabrication and measurement. Other simulation methods have also been developed that greatly improve the speed of measurement by examining the impedance of a virtual internal port \citep{wisbey2014_Qc}. While these ideal simulations can predict $Q_c$, box modes, standing waves along the feedline, and stray couplings can cause variation in actual devices as seen in large arrays of resonators \citep{Mazin_LE_MKIDArray}.

\subsection{Filling Factor}\label{sec:filling} 

Regardless of the form factor of the resonator chosen, the experimenter must be careful to allocate the contribution of materials to the overall loss measured by the resonator. The resonator-induced intrinsic TLS loss, $F \delta^0_{TLS}$ depends on both the intrinsic TLS loss, $\delta^0_{TLS}$, of the material as well as what fraction of the capacitive energy is stored in the material of interest, its dielectric participation ratio or filling factor $F$. Each material in the device, including substrate, native oxides, interfaces, and deposited dielectrics, will have its own filling factor.

The filling factor of a dielectric can approach 100 $\% $ in the case of a thick parallel plate capacitor, or be extremely small, $< 10^{-9}$ as in the surface oxide of a 3D cavity \citep{Reagor_10ms3D}. Depending on the goal of the experimental study, design choices can be made to either decrease the total loss of a component by decreasing its associated $F$, or by maximizing the electric field in the material of interest in order to increase $F$ and make the experiment more sensitive to loss in that material.

In planar circuits such as CPWs, the filling factor depends on the thickness of the dielectric as well as the geometry of the electrodes.  In a CPW, the width of the center trace and gap between center trace and ground plane have a significant effect on the measured loss of the resonator \citep{Gao2008b,Wenner2011}. Higher performance devices can be achieved by increasing conductor width, while higher sensitivity to TLS loss can be achieved with narrower conductors  \citep{Sage2011,geerlings2012,wang2015,Calusine2018, Megrant2012}. Similar effects can be seen in lumped element resonators, where increasing the distance between capacitors leads to reduced participation of the surface oxides and bulk TLS \citep{dial2016bulk_surf}.

In addition to resonators, it is also possible to take advantage of the geometry dependence to reduce TLS loss in qubits. Indeed, \cite{Chang2013} has unambiguously shown improved coherence times as the finger/gap of the interdigitated capacitor (IDC) increases from 2 to 30 $\mu m$ for both Al and TiN qubits. Although quite effective, moving to larger geometries increases the footprint of qubit as well as the radiation loss, requiring optimizations to be made.

The removal of dielectric material is another method for decreasing TLS loss. This can be done by etching excess substrate in the gaps of the CPW, trenching the substrate \citep{Barends2008,Vissers2012,sandberg2012etch,Calusine2018,Bruno2015}, while keeping in mind that the removal of substrate reduces the effective dielectric constant. This decrease in capacitance per unit size changes the impedance and requires larger devices to maintain the same $f_0$. In addition, the relocation of the substrate's native oxide from the edges and corners of the CPW conductors has an outsized effect since the electric field diverges near the the corners of the electrodes \citep{Wenner2011}, significantly reducing the filling factor of this surface.

Detailed simulations using 3D finite element solvers are often necessary to precisely calculate the amount of TLS loss, particularly of low filling factor or low loss materials. Finite element models have been used to calculate the filling factor of different resonator geometries \citep{Gao2008a,wang2015}, deposited dielectrics \citep{OConnell2008}, metal and substrate surfaces in CPW resonators \citep{Wenner2011}, the effect of trenching the Si gap in CPW resonators \citep{Barends2008,sandberg2012etch,vissers2012reduced,Calusine2018}, and the effect of different etches on interface loss \citep{sandberg2012etch,Calusine2018, Nersisyan2019}. 3D resonators have also been simulated \citep{Romanenko2017} in order to determine the effects of surface loss of the cavity.

Several rules of thumb exist that give intuition into filling factor behavior. Roughly, scaling up the CPW or IDC conductor width and gap by a factor of 2 decreases the interface contribution by a factor of 2 as well \citep{Gao2008a,Wenner2011}. Trenching of the gap reduces the contribution of the bulk substrate and any substrate surface dielectric loss linearly, for depths less than or equal to the CPW conductor width \citep{sandberg2012etch}, though for deep aspect ratio or isotropic etches \citep{Bruno2015,Calusine2018}, this simple relationship is no longer accurate.

\subsection{Resonator Fabrication}

The total TLS loss of even the simplest resonator depends on all materials in the circuit, as shown in Fig.~\ref{fig:lossschematic}: substrate bulk and surface, superconductor surface oxide, deposited dielectrics, and interfaces between materials, which are affected by processing used to fabricate the circuit.

Since the performance of a device with multi-step fabrication can be affected by all the steps in the fabrication process \citep{Nersisyan2019}, resonators included as \emph{in situ} witnesses (i.e.  resonators fabricated from just a single step in the process) are an important diagnostic tool for process monitoring and iterative improvements to performance. Alternatively, a large number of wafers or devices could be fabricated from a wide variety of processes \citep{Nersisyan2019}. These resonators can aid in analysis of loss added by each constituent step.

\subsection{Power and Temperature Sweeps}

While the applications of some microwave resonators, e.g. photon detectors or MKIDs, are at high photon numbers, quantum information applications require operation at single photon occupation. Power sweeps down to the single photon regime are necessary to measure TLS loss and predict how the devices would perform in qubit circuits. It can be very difficult to determine the actual power within the resonator, but estimation methods exist. The average number of photons in the resonator can be estimated from the power at the input of the resonator, applied power $P_{\mathrm{app}}$, as \citep{GaoThesis,Bruno2015,Burnett2018}:
\begin{equation}
    < n > = \frac{< E_{int} >}{\hbar \omega_0} = \frac{2}{\hbar \omega_0^2} \frac{Z_0}{Z_r} \frac{Q^2}{Q_c} P_{\mathrm{app}},
\end{equation}
where $Z_0$ is the characteristic impedance of the microwave environment to which the resonator is coupled, $Z_r$ is the resonator characteristic impedance, and $< E_{int} >$ is the average energy stored within the resonator.

To understand the dominant loss mechanism in a resonator material system, it takes much more than measuring loss at the lowest cryostat temperature at single photon powers. Measuring the temperature and power dependence of loss can be used to elucidate a great deal about the dominant loss mechanism present in the device under test. Both power and temperature sweeps can be used to fit for TLS loss in a superconducting microwave resonator. By measuring loss as a function of power at a temperature well below where the TLS are unsaturated, $T_{bath} \ll \frac{\hbar\omega}{2k_{B}}$, the TLS model in Eq.~\ref{eqn:TLSloss} can be used to determine $\delta_{TLS}^0$. Additionally, measuring loss as a function of temperature at low powers and fitting to the same equation yields similar results. Ideally, fits to the saturation of both temperature and power dependent loss are measured to find self-consistent results. Deviations could be an indication that an assumption about the temperature or power at the device are not accurate \citep{Barends2011,Chang2013}.

In addition to dependence on loss, the frequency of the resonator itself can shift depending on microscopic loss effects among the ensemble of TLS. In \cite{GaoThesis}, a method is shown to measure resonance frequency shift as a function of temperature at any power and fit to Eq.~\ref{eqn:TLSfreqshift} to determine $\delta_{TLS}^0$. However, \cite{Pappas2011} states that measuring loss as a function of power or temperature determines the TLS loss in a narrow frequency band near resonance, while measuring resonance frequency as a function of temperature determines wide-band TLS loss. Since the off-resonance TLS remain unsaturated, the measurement can be performed at powers far above the single photon limit required by other measurements, reducing measurement time.  

\begin{figure}
\includegraphics[width=85mm]{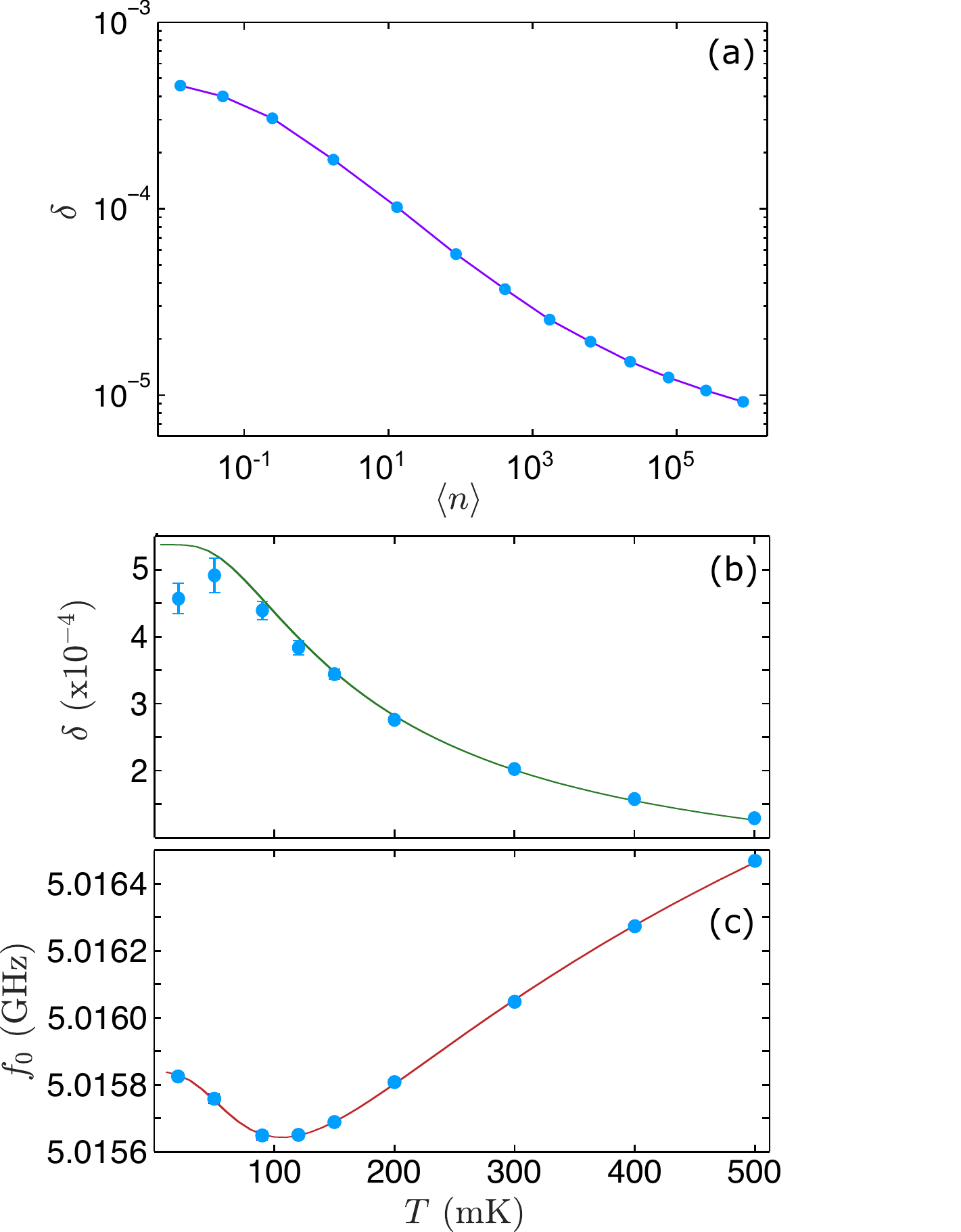}
\caption{Power and temperature dependence of the loss of a Nb on $\mathrm{Al_2O_3}$ microstrip resonator with an interlayer of sputtered $\mathrm{SiO_2}$ \citep{GaoThesis}. Loss $\delta$ as a function of (a) mean photon number $\mathrm{\langle n \rangle}$ at 20 mK, and (b) temperature $T$ at $\mathrm{\langle n \rangle \sim 1}$. (c) Resonance frequency $f_0$ as a function of temperature. Data (points) is fit to models in Eq.~\ref{eqn:TLSloss} and Eq.~\ref{eqn:TLSfreqshift} (lines). Figure adapted with permission from Gao, J., \emph{The physics of superconducting microwave resonators}, Ph.D. thesis, California Institute of Technology (2008).
\label{fig:experimental_scurve}}
\end{figure}

\begin{figure}
\includegraphics[width=85mm]{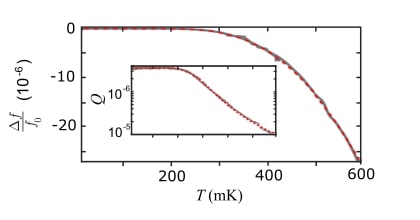}
\caption{Frequency shift and quality factor (inset, same temperature range) as a function of temperature of an Al resonator dominated by conductor loss. At temperatures a significant fraction of $T_{c}$ (here 1.18 K), quasiparticles contribute to loss in the superconducting film. Theory curves (red) are from a fit to the real and imaginary parts of Eq.~\ref{eq:surface_impedance} and include a numerical integration of Mattis-Bardeen theory of AC surface conductivity. TLS loss contributes negligibly in this case, evidenced by these monotonically decreasing curves, and by power independence over many orders of magnitude in photon number. Adapted from Minev, Z., Pop, I., and  Devoret, M., “Planar superconducting whispering gallery mode resonators,” Appl. Phys. Lett. \textbf{103}, 142604 (2013), with the permission of AIP Publishing.
\label{fig:Quasiparticle_temp_expt}}
\end{figure}

Accurate extraction of TLS loss from temperature sweep data is only possible when the frequency shift is dominated by TLS around $T_{bath} \sim \frac{\hbar\omega}{2k_{B}}$. This condition is met by Nb resonators with the characteristic hook shape seen in Fig.~\ref{fig:TLSTheoCurves} and Fig.~\ref{fig:experimental_scurve}. However, for lower $T_c$ materials such as Al, a frequency shift in the opposite direction can arise from large thermal quasiparticle population in the superconductor. From Eq.~\ref{eq:surface_impedance}, increased temperature leads to decrease in frequency and quality factor, as shown in Fig.~\ref{fig:Quasiparticle_temp_expt}. When both effects are present, such as in \cite{Chang2013}, the anomalous temperature-dependent shifts will produce fits with unphysical TLS loss extraction results. Therefore, superconductors of low $T_c$ and high kinetic inductance are not suitable for using temperature sweeps to study TLS loss. On the other hand, there exist regimes in which TLS loss is sufficiently small compared to conductor loss that temperature sweeps reveal much more about properties of the superconductor than TLS loss of nearby materials \citep{Reagor_10ms3D, Minev2013}.

\section{\label{sec:experiments}Materials Loss Experiments and Analysis}

\begin{figure*}
\includegraphics[width=140mm]{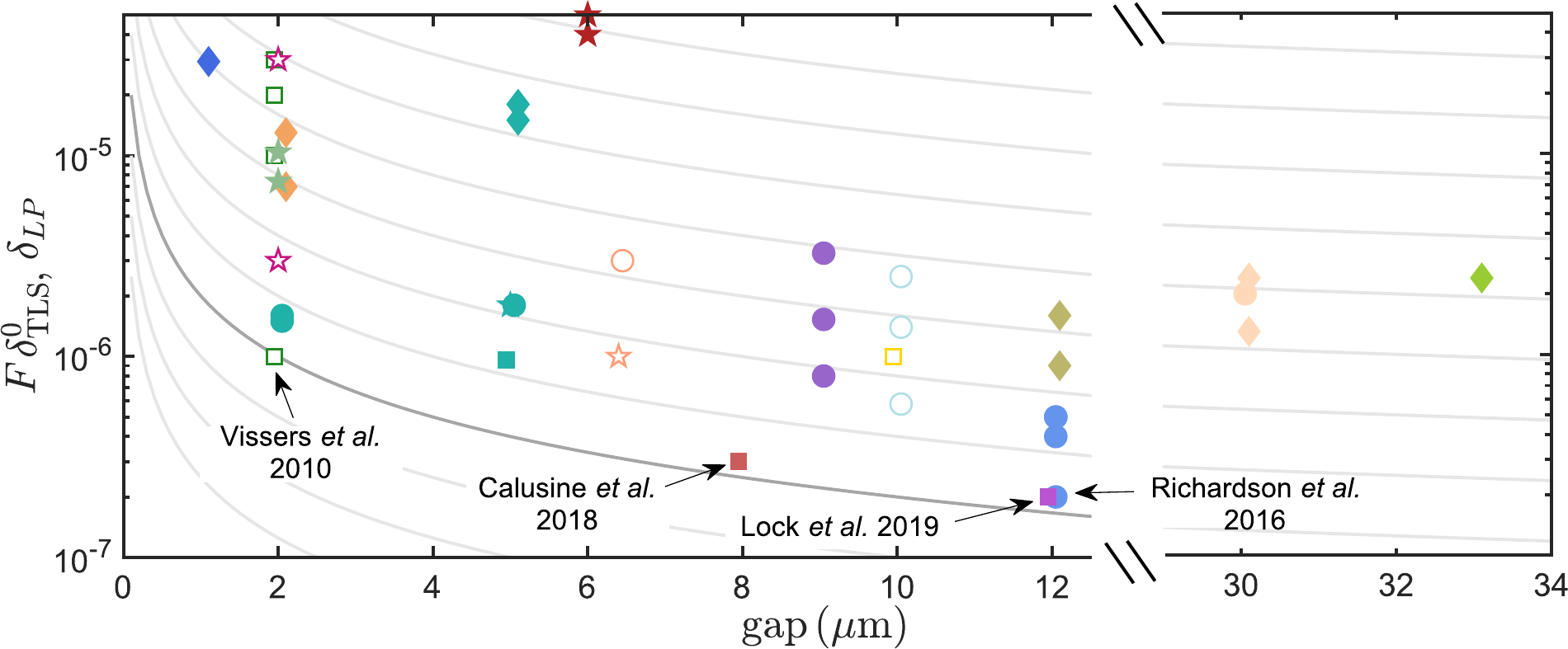}
\caption{Summary of thin film superconductor loss measurements of hero devices. CPW resonator loss as a function of width of the gap between conductor and ground plane for the highest performance device of each type. Solid markers: $F \delta_{\mathrm{TLS}}^0$ values. Open markers: $\delta_{\mathrm{LP}}$ values. Superconducting microwave resonators patterned from Al (circles), TiN (squares), Nb (diamonds), and other superconductors (stars) are shown. Marker color denotes reference in which each measurement is reported, with associated colors and references shown in Table~\ref{tab:metalstable}. Grey curves are guides to the eye showing a rough linear correlation between gap and $F$, with dark grey curve denoting the current performance limit of on-chip superconducting microwave resonators.
\label{fig:review_scatter}}
\end{figure*}

\begin{table*}
\caption{\label{tab:dielectricstable}Table of dielectric material measurements. Columns (from left to right): Material of interest (Material), reference where measurement is reported (Reference), method of dielectric deposition (Dep.),  design of measured device (Geom.), reported low power loss $\delta_{LP}$, and resonator-induced intrinsic TLS loss $F \delta_{TLS}^0$. CVD: chemical vapor deposition. Therm. ox.: Thermal oxidation. PECVD: Plasma-enhanced CVD. ICP CVD: Inductively-coupled plasma CVD. MBE: Molecular beam epitaxy. ECR-PECVD: Electron cyclotron resonance PECVD. LE: lumped element resonator design. LE PPC: Lumped element resonator with parallel plate capacitor. IDC: Lumped element resonator with interdigitated capacitor. Note: Values cannot be directly compared due to differences in TLS filling factor, resonance frequency, $Q_{i}/Q_{c}$ matching, fabrication, magnetic shielding, IR filtering and isolation, thermalization, and data fitting.}
\begin{ruledtabular}
\begin{tabular}{cccccc}
Material & Reference & Dep. & Geom. & $\delta_{LP}$ & $F \delta_{TLS}^0$\\
 & & & & [$\times 10^{-5}$] & [$\times 10^{-5}$] \\
\hline
a-$\mathrm{SiO_2}$ & \cite{Martinis2005} & CVD  & LE PPC & 500 & \\
a-$\mathrm{SiN_{x}}$ & \cite{Martinis2005} & CVD & LE PPC & 20 & \\
a-Si:H & \cite{OConnell2008} & NA & LE PPC, CPW & 1-13 & \\
$\mathrm{SiN_x}$ & \cite{OConnell2008} & NA & LE PPC, CPW & 10-20 & \\
$\mathrm{SiO_2}$ & \cite{OConnell2008} & Therm. ox. & CPW & 30-33 & \\
$\mathrm{Si}$ & \cite{OConnell2008} & Sputtered & CPW & 50-60 & \\
$\mathrm{AlN}$ & \cite{OConnell2008} & NA & CPW & 110-180 & \\
$\mathrm{SiO_2}$ & \cite{OConnell2008} & PECVD & CPW & 270-290 & \\
$\mathrm{MgO}$ & \cite{OConnell2008} & NA & CPW & 500-800 & \\
a-$\mathrm{SiO_2}$ & \cite{Cicak2010} & ECR-PECVD & LE PPC & 600 & \\
a-$\mathrm{SiN}$ & \cite{Cicak2010} & ECR-PECVD & LE PPC & 40-50 & \\
a-$\mathrm{Si}$ & \cite{Cicak2010} & Sputter & LE PPC & 150-200 & \\
$\mathrm{Nb_2O_5}$ & \cite{kaiser2010_nb2O5} & Anodic ox. & LE PPC & 100-400 & \\
$\mathrm{SiO}$ & \cite{kaiser2010_nb2O5} & Therm. evap. & LE PPC & 20-50 & \\
$\mathrm{SiN_x}$ & \cite{kaiser2010_nb2O5} & PECVD & LE PPC & 10-30 & \\
a-SiN & \cite{Paik2010} & ICP CVD & LE PPC & 2.5-120 & \\
$\mathrm{AlO_x}$ & \cite{Pappas2011} & Therm. ox. & CPW & & $F$ $\times 100$ \\
$\mathrm{Al_2O_3}$ & \cite{weides2011_epiAl2O3} & MBE & LE PPC & 6 & \\
$\mathrm{HfO_2}$ & \cite{Burnett2013} & Sputter & LE IDC & & 1.5-2.5 \\
$\mathrm{Al_2O_3}$ & \cite{Burnett2013} & Sputter & LE IDC & & 2.0-2.5 \\
$\mathrm{Al_{2}O_{3}}$ & \cite{Cho2013} & PLD & LE PPC & 3-5 & \\
$\mathrm{SiO_{x}}$ & \cite{li2013_siox} & ECR-PECVD & Microstrip & 100-700 & \\
$\mathrm{AlO_x}$ & Deng \emph{et al.} (\citeyear{Deng2014}) & Plasma ox. & LE overlap & 140-180 & \\
$\mathrm{SiN}$ & \cite{duff2016_SiN} & ICP-PECVD & Microstrip &  & 78 \\
$\mathrm{SiO_2}$ & \cite{Goetz2016} & Therm. ox. & CPW &  & 0.34-0.89 \\
a-Si & \cite{Lecocq2017} & PECVD & LE PPC & 15-50 & \\
$\mathrm{SiN_{x}}$ & \cite{Sarabi2016} & PECVD & LE PPC & 78 & \\
$\mathrm{AlO_x}$ & \cite{Brehm2017} & Anodic ox. & CPW + PPC & & 4-22 \\
$\mathrm{B_4C}$ & \cite{wisbey2019_boron} & Sputter & CPW & & 10-15 \\
$\mathrm{BN}$ & \cite{wisbey2019_boron} & Sputter & CPW & & 6 \\
$\mathrm{Al_2O_3}$ & \cite{McRae2020} & Sputter & LE PPC & & 100 \\
$\mathrm{HSQ}$ & \cite{Niepce_2020} & Spin-on-glass & CPW &  800 & \\
\end{tabular}
\end{ruledtabular}
\end{table*}

\begin{table*}
\caption{\label{tab:metalstable}Table of loss measurements of superconducting metals. Columns (from left to right): symbol denoting the measurement in Table~\ref{fig:review_scatter} (unlabelled), superconductor in measured device (SC), reference where measurement is reported (Reference), method of metal deposition (Dep.), surface treatments applied and etch type used to define resonators (Surf./Etch), substrate on which metal was deposited (Substrate), design of measured device (Geom.), reported low power loss $\delta_{LP}$, resonator-induced intrinsic TLS loss $F \delta_{TLS}^0$, and width $w$ of conductor and gap $g$ between conductor and ground plane. Therm. evap.: Thermal evaporation. MBE: Molecular beam epitaxy. PVD: Plasma vapor deposition. RIE: Reactive ion etch. H-pass.: hydrogen-passivated. $\mathrm{\lambda/2}$ or $\mathrm{\lambda/4}$: CPW resonator of length stated. LE IDC: Lumped element resonator with interdigitated capacitor. NA: information not available. Note: Values cannot be directly compared due to significant experimental differences.}
\begin{ruledtabular}
\begin{tabular}{cccccccccc}
& SC & Reference & Dep. & Surf./Etch & Substrate & Geom. & $\delta_{LP}$ & $F \delta^0_{TLS}$ & w/g \\ 
& & & & & & & [$\times 10^{-6}$]& [$\times 10^{-6}$] & [$\mathrm{\mu m}$/$\mathrm{\mu m}$] \\
\hline

\includegraphics[width=2.5mm]{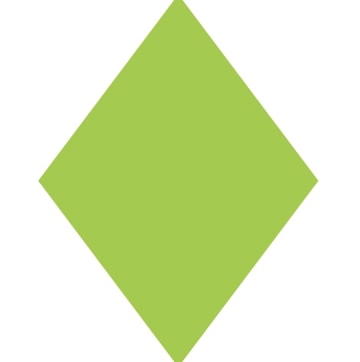}&$\mathrm{Nb}$ & \cite{Gao2008b} & NA & NA & $\mathrm{Al_{2}O_{3}}$ &$\mathrm{\lambda/4}$ & & 2.4-29.8 & 3/2-50/33 \\ 
\includegraphics[width=2.5mm]{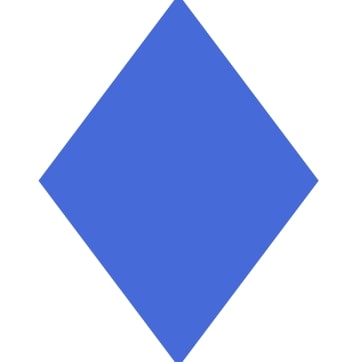}&$\mathrm{Nb}$ & \cite{Kumar2008} & NA & NA, RIE & $\mathrm{Si}$ & $\mathrm{\lambda/4}$ & & 29.4 & 5/1 \\ 
&$\mathrm{Al}$ & \cite{OConnell2008} & NA & NA & Si & $\mathrm{\lambda/2}$ & <5-12 & \\
&$\mathrm{Re}$ & \cite{OConnell2008} & NA & NA & $\mathrm{Al_{2}O_{3}}$ & $\mathrm{\lambda/2}$ & <6-10 & \\
&$\mathrm{Al}$ & \cite{OConnell2008} & NA & NA & $\mathrm{Al_{2}O_{3}}$ & $\mathrm{\lambda/2}$ & <9-21 & \\
&$\mathrm{Al}$ & \cite{OConnell2008} & NA & NA & $\mathrm{Al_{2}O_{3}}$ & LE IDC & 41-47 & \\
\includegraphics[width=2.5mm]{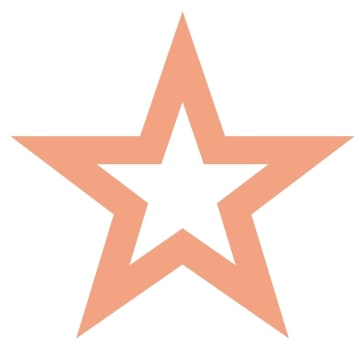} & $\mathrm{Re}$ & \cite{Wang2009} & E-beam & NA, RIE & $\mathrm{Al_{2}O_{3}}$ & $\mathrm{\lambda/4}$ & 1-3 & & 16/6.4-5/2 \\ 
\includegraphics[width=2.5mm]{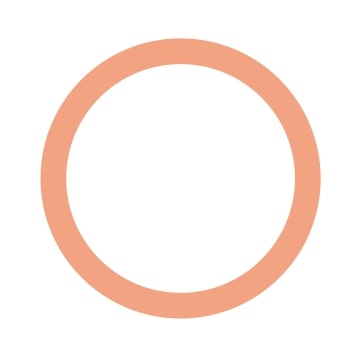}&$\mathrm{Al}$ & \cite{Wang2009} & Sputter & NA, RIE & $\mathrm{Al_{2}O_{3}}$ & $\mathrm{\lambda/4}$ & 3-10 & & 16/6.4-5/2 \\
\includegraphics[width=2.5mm]{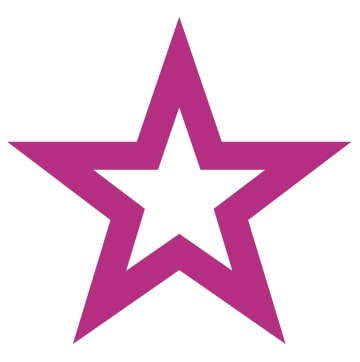}&$\mathrm{NbTiN}$ & \cite{Barends2010b} & Sputter & H-pass., RIE & $\mathrm{Si}$ & $\mathrm{\lambda/4}$ & 3 & & 3/2-6/2 \\ 
\includegraphics[width=2.5mm]{NbTiN_Barends2010.jpg}&$\mathrm{Ta}$ & \cite{Barends2010b} & Sputter & NA, RIE & $\mathrm{Si}$ & $\mathrm{\lambda/4}$ & 30 & & 5/2 \\
\includegraphics[width=2.5mm]{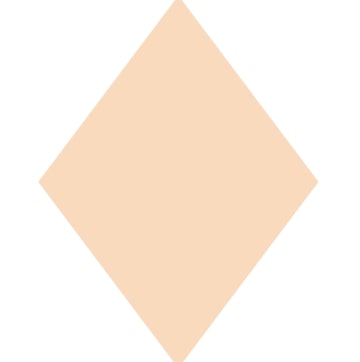}&$\mathrm{Nb}$ & \cite{Macha2010} & NA & none, dry etch & $\mathrm{Al_{2}O_{3}}$ & $\mathrm{\lambda/2}$ & & 2.4-2.6 & 50/30 \\
\includegraphics[width=2.5mm]{Nb_Macha2011.jpg}&$\mathrm{Nb}$ & \cite{Macha2010} & NA & none, dry etch & $\mathrm{Si}$ & $\mathrm{\lambda/2}$ & & 1.3, 1.6 & 50/30 \\ 
\includegraphics[width=2.5mm]{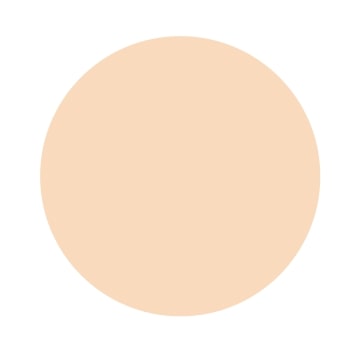}&$\mathrm{Al}$ & \cite{Macha2010} & NA & none, liftoff & $\mathrm{Al_{2}O_{3}}$ & $\mathrm{\lambda/2}$ & & 2.0 & 50/30 \\ 
\includegraphics[width=2.5mm]{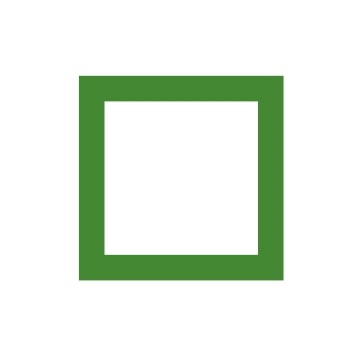}&$\mathrm{TiN}$ & \cite{Vissers2010} & Sputter & Anneal, RIE & $\mathrm{Al_{2}O_{3}}$ & $\mathrm{\lambda/4}$ & 30 & & 3/2 \\ 
\includegraphics[width=2.5mm]{TiN_Vissers2010.jpg}&$\mathrm{TiN}$ & \cite{Vissers2010} & Sputter & HF+150nm SiN, RIE & $\mathrm{Al_{2}O_{3}}$ & $\mathrm{\lambda/4}$ & 20 & & 3/2 \\ 
\includegraphics[width=2.5mm]{TiN_Vissers2010.jpg}&$\mathrm{TiN}$ & \cite{Vissers2010} & Sputter & HF+50nm SiN, RIE & $\mathrm{Al_{2}O_{3}}$ & $\mathrm{\lambda/4}$ & 10 & & 3/2 \\ 
\includegraphics[width=2.5mm]{TiN_Vissers2010.jpg}&$\mathrm{TiN}$ & \cite{Vissers2010} & Sputter & HF+nitride, RIE & Si & $\mathrm{\lambda/4}$ & 2 & & 3/2 \\ 
\includegraphics[width=2.5mm]{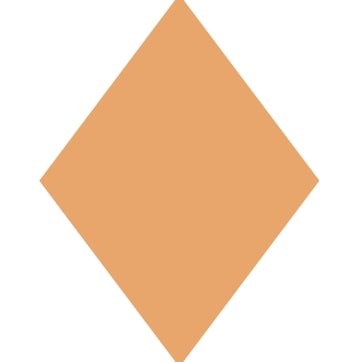}&$\mathrm{Nb}$ & \cite{Wisbey2010} & NA & RF clean, RIE & $\mathrm{Si}$ & $\mathrm{\lambda/4}$ & & 13-17 & 3/2 \\ 
\includegraphics[width=2.5mm]{Nb_Wisbey2010.jpg}&$\mathrm{Nb}$ & \cite{Wisbey2010} & NA & HF, RIE & $\mathrm{Si}$ & $\mathrm{\lambda/4}$ & & 7 & 3/2 \\ 
\includegraphics[width=2.5mm]{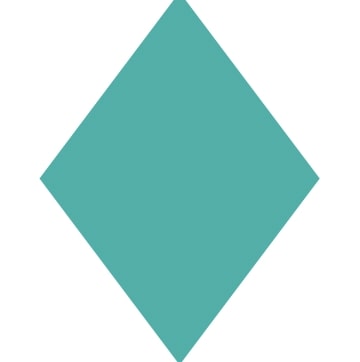}&$\mathrm{Nb}$ & \cite{Sage2011} & Sputter & H-pass., RIE & $\mathrm{Si}$ & $\mathrm{\lambda/2}$ & & 15-37  & 3/2, 10/5 \\ 
\includegraphics[width=2.5mm]{Nb_Sage2011.jpg}&$\mathrm{Nb}$ & \cite{Sage2011} & Sputter & RCA-1+Piranha, RIE & $\mathrm{Al_{2}O_{3}}$ & $\mathrm{\lambda/2}$ & & 18-23 & 3/2, 10/5 \\
\includegraphics[width=2.5mm]{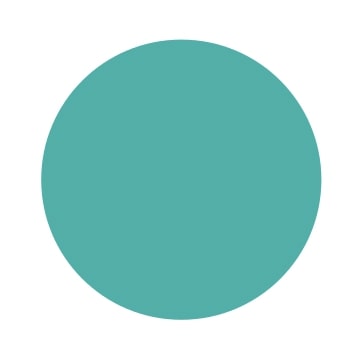}&$\mathrm{Al}$ & \cite{Sage2011} & Sputter & H-pass., RIE & $\mathrm{Si}$ & $\mathrm{\lambda/2}$ & & 1.5 & 3/2, 10/5 \\ 
\includegraphics[width=2.5mm]{Al_Sage2011.jpg}&$\mathrm{Al}$ & \cite{Sage2011} & Sputter & RCA-1/Piranha, wet etch & $\mathrm{Al_{2}O_{3}}$ & $\mathrm{\lambda/2}$ & & 1.6 & 3/2, 10/5 \\
\includegraphics[width=2.5mm]{Al_Sage2011.jpg}&$\mathrm{Al}$ & \cite{Sage2011} & MBE & none, RIE & $\mathrm{Al_{2}O_{3}}$ & $\mathrm{\lambda/2}$ & & 1.8, 6.5 & 3/2, 10/5 \\ 
\includegraphics[width=2.5mm]{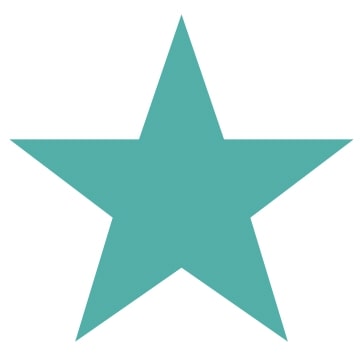}&$\mathrm{Re}$ & \cite{Sage2011} & MBE & RCA-1/Piranha, RIE & $\mathrm{Al_{2}O_{3}}$ & $\mathrm{\lambda/2}$ & & 1.8, 3.3 & 3/2, 10/5 \\ 
\includegraphics[width=2.5mm]{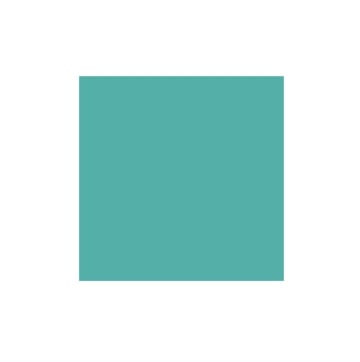}&$\mathrm{TiN}$ & \cite{Sage2011} & Sputter & H-pass., RIE & $\mathrm{Si}$ & $\mathrm{\lambda/2}$ & & 0.96, 3 & 3/2, 10/5 \\ 
\includegraphics[width=2.5mm]{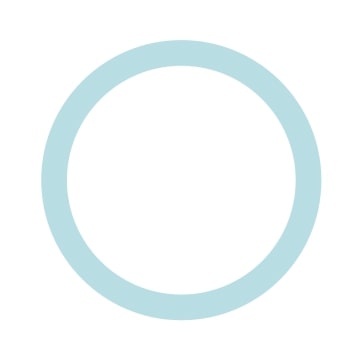}&$\mathrm{Al}$ & \cite{Megrant2012} & Sputter & Ar mill, RIE & $\mathrm{Al_{2}O_{3}}$ & $\mathrm{\lambda/4}$ & 2.5, 6.3 & & 3/2, 15/10 \\
\includegraphics[width=2.5mm]{Al_Megrant2012.jpg}&$\mathrm{Al}$ & \cite{Megrant2012} & E-beam & Ar mill, RIE & $\mathrm{Al_{2}O_{3}}$ & $\mathrm{\lambda/4}$ & 1.4, 1.5 & & 3/2, 15/10\\
\includegraphics[width=2.5mm]{Al_Megrant2012.jpg}&$\mathrm{Al}$ & \cite{Megrant2012} & MBE & Various, RIE & $\mathrm{Al_{2}O_{3}}$ & $\mathrm{\lambda/4}$ & 0.58-1.3& & 3/2, 15/10\\
&$\mathrm{Nb}$ & \cite{Burnett2013} & Sputter & NA & $\mathrm{Al_{2}O_{3}}$ & LE IDC & & 2.0 & \\
&$\mathrm{Re}$ & \cite{Cho2013} & MBE & Anneal / NA & $\mathrm{Al_{2}O_{3}}$ & LE IDC & 30-50 & & \\
\includegraphics[width=2.5mm]{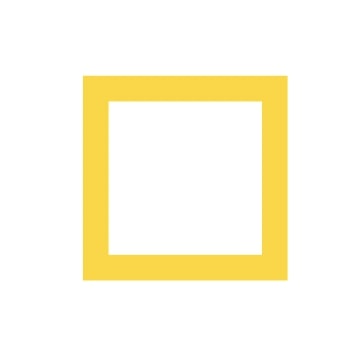}&$\mathrm{TiN}$ & \cite{Ohya2014} & Sputter & Nano-Strip+HF, RIE & $\mathrm{Si}$ & $\mathrm{\lambda/4}$ & 1 & & 15/10\\
&$\mathrm{Nb+Pt}$ & Burnett \emph{et al.} (\citeyear{Burnett2016}) & MBE & NA, RIE & $\mathrm{Al_{2}O_{3}}$ & LE IDC & & 12 & \\
&$\mathrm{Nb}$ & Burnett \emph{et al.} (\citeyear{Burnett2016}) & Sputter & NA, RIE & $\mathrm{Al_{2}O_{3}}$ & Fractal & & 1.1 & \\
\includegraphics[width=2.5mm]{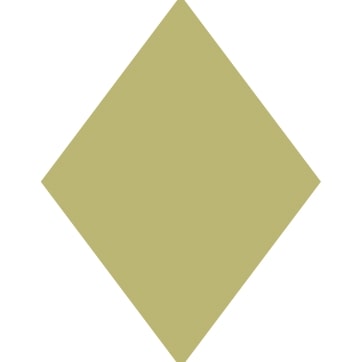}&$\mathrm{Nb}$ & \cite{Goetz2016} & Sputter & HF / RIE & $\mathrm{Si}$ & $\mathrm{\lambda/2}$ & & 0.9 & 20/12 \\
\includegraphics[width=2.5mm]{Nb_Goetz2016.jpg}&$\mathrm{Nb}$ & \cite{Goetz2016} & Sputter & Ar mill / RIE & $\mathrm{Al_{2}O_{3}}$ & $\mathrm{\lambda/2}$ & & 1.6 & 20/12\\
\includegraphics[width=2.5mm]{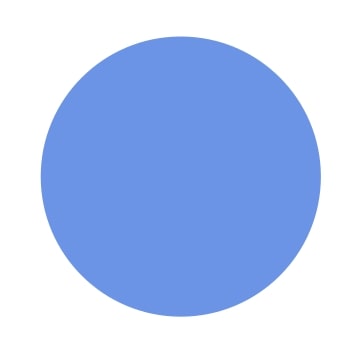}&$\mathrm{Al}$ & \cite{Richardson2016} & MBE & Various, wet etch & $\mathrm{Si}$ & $\mathrm{\lambda/4}$ & & 0.2-760 & 3/2-22/12\\
\includegraphics[width=2.5mm]{Al_Richardson2016.jpg}&$\mathrm{Al}$ & \cite{Richardson2016} & MBE & Various, RIE & $\mathrm{Si}$ & $\mathrm{\lambda/4}$ & & 0.5-4800 & 3/2-22/12 \\
\includegraphics[width=2.5mm]{Al_Richardson2016.jpg}&$\mathrm{Al}$ & \cite{Richardson2016} & MBE & Various, wet etch & $\mathrm{Al_{2}O_{3}}$ & $\mathrm{\lambda/4}$ & & 0.5-5.3 & 3/2-22/12\\
\includegraphics[width=2.5mm]{Al_Richardson2016.jpg}&$\mathrm{Al}$ & \cite{Richardson2016} & MBE & Various, RIE & $\mathrm{Al_{2}O_{3}}$ & $\mathrm{\lambda/4}$ & & 0.4-7.4 & 3/2-22/12\\
\includegraphics[width=2.5mm]{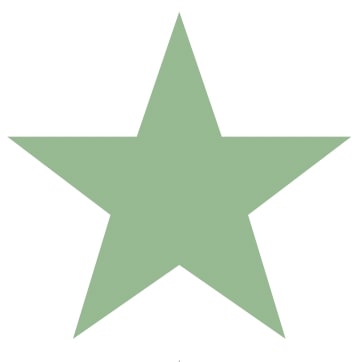}&$\mathrm{NbN}$ & \cite{DeGraaf2017} & Sputter & none, NA & $\mathrm{Al_{2}O_{3}}$ & Fractal & & 10.4-10.6 & g=2\\
\includegraphics[width=2.5mm]{NbN_deGraaf2017.jpg}&$\mathrm{NbN}$ & \cite{DeGraaf2017} & Sputter & Anneal, NA & $\mathrm{Al_{2}O_{3}}$ & Fractal & & 7.44, 7.69 & g=2\\
&$\mathrm{Al}$ & \cite{Burnett2018} & E-beam & HF, wet etch & $\mathrm{Si}$ & $\mathrm{\lambda/4}$ & 1.3 & 1.1 \\
\includegraphics[width=2.5mm]{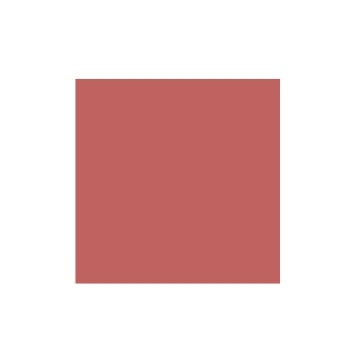}&$\mathrm{TiN}$ & \cite{Calusine2018} & Sputter & RCA, RIE & $\mathrm{Si}$ & $\mathrm{\lambda/4}$  & & 0.3 & 16/8-22/11 \\
\includegraphics[width=2.5mm]{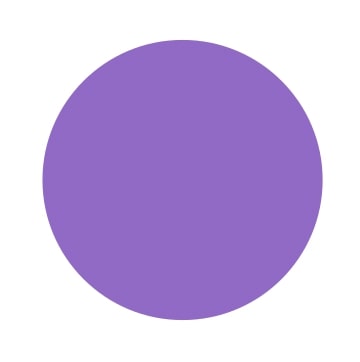}&$\mathrm{Al}$ & \cite{Earnest2018} & E-beam & none, RIE & $\mathrm{Si}$ & $\mathrm{\lambda/4}$ & 3.1 & 3.27 & 15/9 \\
\includegraphics[width=2.5mm]{Al_Earnest2018.jpg}&$\mathrm{Al}$ & \cite{Earnest2018} & E-beam & RCA-1+HF, RIE & $\mathrm{Si}$ & $\mathrm{\lambda/4}$ & 1.9 & 1.53 & 15/9 \\
\includegraphics[width=2.5mm]{Al_Earnest2018.jpg}&$\mathrm{Al}$ & \cite{Earnest2018} & E-beam & Anneal, RIE & $\mathrm{Si}$ & $\mathrm{\lambda/4}$ & 1.8 & 1.56 & 15/9 \\
\includegraphics[width=2.5mm]{Al_Earnest2018.jpg}&$\mathrm{Al}$ & \cite{Earnest2018} & E-beam & RCA-1+HF+anneal, RIE & $\mathrm{Si}$ & $\mathrm{\lambda/4}$  & 1.2 & 0.8 & 15/9 \\
\includegraphics[width=2.5mm]{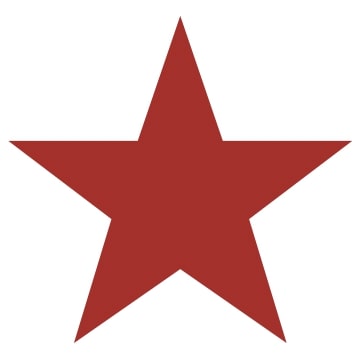}&$\mathrm{In}$ & \cite{McRae2018} & Therm. evap. & none, wet etch & $\mathrm{Si}$ & $\mathrm{\lambda/4}$ & & 40 & 12/6 \\
\includegraphics[width=2.5mm]{In_McRae2018.jpg}&$\mathrm{In}$ & \cite{McRae2018} & Therm. evap. & HF, wet etch & $\mathrm{Si}$ & $\mathrm{\lambda/4}$ & & 50 & 12/6 \\
&$\mathrm{TiN}$ & \cite{Shearrow2018} & ALD & Nano-Strip/HF, RIE & $\mathrm{Si}$ & LE IDC & 0.5-17 & \\

\includegraphics[width=2.5mm]{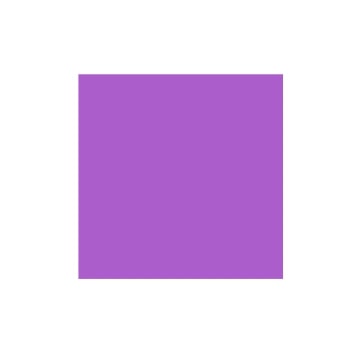}&$\mathrm{TiN}$ & \cite{Lock2019} & Sputter & HF, RIE & $\mathrm{Si}$ & $\mathrm{\lambda/4}$ & & 0.2-30 & 3/12 \\
&$\mathrm{Nb}$ & \cite{Nersisyan2019} & PVD & Various, RIE & $\mathrm{Si}$ & $\mathrm{\lambda/4}$ & 0.8-6 & \\
&$\mathrm{TiN}$ & \cite{woods2019_interface} & Sputter & RCA, RIE & $\mathrm{Si}$ & $\mathrm{\lambda/4}$ & & 0.3-1 \\
&$\mathrm{In}$ & \cite{Lei2020} & Therm. evap. & He-N-H pass. & $\mathrm{Si}$ & 3D & 5600 &  \\
&$\mathrm{Nb}$ & \cite{Romanenko2020} & Bulk & heat treatment &  & 3D & & <.00052 \\
\end{tabular}
\end{ruledtabular}
\end{table*}

\begin{table*}
\caption{\label{tab:bulkmetalstable}Bulk superconducting metal measurements. Separately from TLS loss in dielectrics, surface resistance ${R_{S}}$ causes dissipation in the region of the superconductor penetrated by microwave supercurrent. Numbers referenced here are lower bounds obtained by attributing all loss to the superconductor and then scaling the measured $Q_{i}$ by the normalized field energy stored in the surface layer of the cavity (kinetic inductance fraction, or conductor participation ratio). $\delta = 1/Q_{S} = R_{S}/\alpha\omega \mu_{0} \lambda $, using penetration depth $\lambda$ and kinetic inductance fraction $\alpha$ that is either assumed from independent measurement or extracted from cavity frequency shift versus temperature data according to Eq.~\ref{eq:surface_impedance}. }
\begin{ruledtabular}
\begin{tabular}{cccccc}
SC & Reference & Surf./Etch & Geom. & ${R_{S}}  \mathrm{ (n\Omega)}$ & $\delta$ \\ 
\hline
A1 (e-beam evap.) & \cite{Minev2013} &  & WGM ring & 250 & 2$\times 10^{-4}$  \\
Al (4N) & \cite{Reagor_10ms3D} & Al etch A & Cylindrical TE011 &  & 2.2$\times 10^{-4}$  \\
Al (4N) & \cite{Reagor_10ms3D} & Al etch A & Rectangular TE101 &  & 4.8$\times 10^{-4}$  \\
Al (4N) & \cite{Reagor2016} & Al etch A & Coaxial $\mathrm{\lambda/4}$ &  & 3$\times 10^{-4}$  \\
In & \cite{Brecht_In_cavities} & E-plated & Rectangular TE101 & 260  & 3$\times 10^{-3}$ \\
Al (6061)  & \cite{AxlineThesis} & Al etch A & Rectangular TE101 & &  8$\times 10^{-3}$ \\
Al (5N5) & \cite{AxlineThesis} & Al etch A &  Cylindrical TM010 & &   2$\times 10^{-4}$ \\
Nb & \cite{AxlineThesis} &  &  Rectangular TE101 & &  7$\times 10^{-4}$ \\
Nb & \cite{Romanenko2017} & various, EP, heat treatments & Elliptical TM010 & 1.4 to 17 & \\
In & \cite{Lei2020} & therm. evap., He-N-H pass. & Rectangular TE101 & 260  & 1.2$\times 10^{-2}$ \\
\end{tabular}
\end{ruledtabular}
\end{table*}

Tables summarizing a selection of existing measurements in literature for $\delta_{LP}$ and $F \delta^0_{TLS}$ are given for deposited dielectrics (Table \ref{tab:dielectricstable}) and interfaces associated with thin film superconducting metals (Table \ref{tab:metalstable} and Fig.~\ref{fig:review_scatter}). Conductor loss limits are listed in Table~\ref{tab:bulkmetalstable}. Variables not satisfactorily accounted for in these comparisons include the impact of resonator geometry and filling factor, resonance frequency, $Q_{i}/Q_{c}$ matching, miscellaneous fabrication differences, the presence or absence of magnetic shielding, IR filtering and isolation, and differences in measurement base temperature and sample thermalization. In particular, in order to compare results across experiments, the filling factor of the TLS material must be known. For CPWs, the filling factor can be estimated from the conductor width $w$ and gap $g$ between the conductor and ground planes, where $F \propto w,g$ roughly holds. For experiments that only report the low power loss, in order to compare to TLS loss values, the high-power loss must be orders of magnitude smaller than the low power loss, and it must be demonstrated that the low power regime is being reached.

Of special note, uncontrolled thermalization, unwanted noise, and unshielded magnetic fields can lead to systematic under- or over-reporting of loss. It is the opinion of the authors that, at minimum, base temperature and power dependent sweeps be published along with filling factors of the TLS material, $Q_c$, and resonator fitting information. Other parameters that would be highly useful include etch type and material deposition parameters.

Due to the numerous variables affecting the reported loss values, is it difficult to make definite conclusions from literature comparisons. However, some trends can be seen. They are outlined below along with noteworthy results from individual studies.

\subsection{Substrates}

High-purity crystalline substrates, float-zone intrinsic or other high resistivity Si and $\mathrm{Al_2O_3}$ are common, but other substrates have also been investigated \citep{Martinis2014}. Although their intrinsic losses may be very low, even with extreme geometries the high $F$ of substrates significantly reduces the performance of resonators and qubits \citep{Calusine2018}.

As shown in Table~\ref{tab:metalstable}, there are no significant trends in total TLS loss between devices on Si and on $\mathrm{Al_2O_3}$.

\subsection{Metal Deposition}

Deposition parameters of the metal can influence TLS loss, typically through the metal-air interface.  Stress, stoichiometry, impurities and crystal growth mode can all influence the amount of TLS loss \citep{Vissers2010,jaim2014_TiN,OConnell2008,Wang2009,Nersisyan2019}.

Epitaxial materials have generally higher performance than amorphous materials. Whether the benefit is due to crystallinity or overall cleanliness is disputed, but superconductors grown epitaxially have demonstrated lower TLS loss than other growth methods \citep{Megrant2012,Richardson2016}.

As shown in Table~\ref{tab:metalstable}, there is a wide range of performance in Al, TiN, and Nb devices. In terms of hero devices (Fig.~\ref{fig:review_scatter}), those patterned from Al and TiN have reached the lowest TLS loss values, with Nb and other superconductors such as Re and Ta trailing behind. \cite{Place2020} have shown promising results with epi-Ta electrodes in a qubit, but resonator loss measurements have not been reported.

3D cavities can have orders of magnitude smaller TLS filling factor and omit bulk dielectric substrates altogether. Therefore, they are more sensitive probes of quasiparticle loss in the bulk superconductors from which they are constructed. Upper bounds on this bulk superconductor surface loss from several recent works in 3D cavities are included in Table~\ref{tab:bulkmetalstable}. As with planar fabricated resonators, the performance is influenced by the metal purity and surface treatments.

\subsection{Interfaces}

Losses at the metal-substrate and substrate-air interfaces are thought to contribute significantly to overall loss, and much effort has been made to explore deposition and cleaning techniques to mitigate and understand these induced losses \citep{Nersisyan2019}. As the interface between the superconductor and the substrate may be the largest source of TLS loss \citep{Gao2008b,Wenner2011,Calusine2018}, surface treatments prior to deposition are especially important. Cleaning the substrate surface and removing surface oxide with RCA-1, HF, HMDS, and/or annealing has been demonstrated to improve resonator performance \citep{Wisbey2010,Bruno2015,Nersisyan2019,Earnest2018}. Additional surface treatments to passivate the Si surface have also been shown to be beneficial \citep{Earnest2018,Vissers2010,jaim2014_TiN,Lock2019}. However, ion cleaning of Si substrates in particular prior to deposition is associated with higher surface dielectric loss \citep{Wisbey2010,Nersisyan2019}.  Studies have shown that $\mathrm{Al_2O_3}$ substrates with rougher surfaces, or those with adsorbates, may contribute higher loss \citep{Megrant2012}.

The liftoff or etch process used to fabricate the resonator is also important to its performance.  Etching may affect the sidewalls of the superconductor as well as alter the surface of the underlying substrate \citep{sandberg2012etch,Richardson2016,Nersisyan2019,Lock2019}. In multilayer processes, Ar ion cleaning is used to remove metal oxides and create good electrical contact. However, this cleaning also damages the substrate and induces loss \citep{sandberg2012etch,Wisbey2010}. Liftoff processes have been optimized \citep{Nersisyan2019} to minimize the amount of substrate subject to Ar cleaning either by reducing the area of the layer, or by including a separate, small, higher loss "bandage" layer that connects the large area low loss electrodes \citep{Dunsworth2017}. Post-liftoff O$_2$ descum steps are also found to reduce loss  \citep{Quintana2014}. 

\subsection{Deposited Dielectrics}

While the highest performance substrate-based resonators are typically single-layer and do not have deposited dielectrics, more complex circuits implement multilayer processing and deposited dielectrics. However, deposited dielectrics have been found to possess widely varying losses (Table~\ref{tab:dielectricstable}). The lowest documented loss $\sim~10^{-5}$ occurs in a wide variety of materials \citep{OConnell2008,Paik2010,Cho2013,Burnett2013,Goetz2016,Brehm2017}, and materials that have demonstrated high loss ($\sim 10^{-3}$) are also varying.


\subsection{Josephson Junctions}

A crucial dielectric barrier in superconducting qubit circuits is the tunnel barrier oxide in the Josephson junction of the qubit itself. The vast majority of these Josephson junctions are made from Al$_2$O$_3$ grown at room temperature. This thin Al$_2$O$_3$ present in tunnel junctions is a source of TLS loss. Bulk loss of an amorphous Al$_2$O$_3$ junction has been measured to be on the order of $10^{-3}$ \citep{Stoutimore2012}. The fluxonium qubit, with its lower resonant frequency, can maintain long coherence times despite containing an array of large-area junctions with losses $\sim 10^{-4}$ \citep{Nuyen2019}. Crystalline Al$_2$O$_3$ barriers have shown promising losses around $6\times10^{-5}$ \citep{weides2011_epiAl2O3}.

One current design philosophy is to statistically avoid strongly coupled TLS in Josephson junctions by minimizing the junction volume \citep{Stoutimore2012}. While current qubit designs have been engineered to avoid the loss from the tunnel barrier dielectric, lower loss tunnel barriers or interfaces could enable new, less constrained designs, and continued improvements in the tunnel barrier are a fruitful topic for ongoing research.

\section{\label{sec:outlook}Conclusions and Outlook}

Superconducting microwave resonator measurements are an effective way to probe low power cryogenic losses in bulk and thin film dielectrics, interfaces between materials, and surface oxides. In this regime, TLS loss is often the dominant type of materials loss, and can be distinguished from other losses by its unique association with the capacitor and by its power and temperature dependence. Experiments probing the temperature, power, frequency, geometric and magnetic field dependence of loss can be used to determine the interplay between various loss mechanisms.

In order to perform accurate measurements, experiments must be carefully designed and executed. Common pitfalls producing inaccurate resonator measurement results include:
\begin{itemize}
    \item A lack of thermalization, which leads to artificially low reported intrinsic TLS loss;
    \item The measurement of low power loss (in the place of TLS loss) when power-independent losses are large, which produces a seemingly high intrinsic TLS loss;
    \item Measuring in reflection mode, which could report inaccurate $Q_i$ values due to impedance mismatches that cannot be corrected;
    \item Measuring in transmission mode, which has the same drawback as above plus insufficient baseline information, making separating $Q_c$ and $Q_i$ impossible; and
    \item Using a resonator design with a low filling factor for the material of interest, if other loss mechanisms are not negligible.
\end{itemize}

Many different methods have been used to determine interface, thin film, and bulk dielectric loss at low powers and temperatures in superconducting quantum circuits. A review of the current body of work in this field lacks strong conclusions due to the many discrepancies in reporting of experimental parameters. The state-of-the-art in terms of planar superconducting geometries give losses in the low $10^{-6}$ range, which has been shown for several materials sets. Deposited dielectric performance is currently limited to $\sim 10^{-5}$. 

Promising directions for future work can be divided into two subfields: loss metrology, and materials research. 

Fruitful future projects in loss metrology include investigations into: thermalization of samples (including the impact of the sample box material and design, mounting method, and thermalization time); precise and unobtrusive calibration for millikelvin $S_{21}$ measurements; time fluctuations in $Q_i$ and $Q_c$ (including the effect of sampling rates and power regime); and resonator fitting (especially regime-dependent variations between DCM and INV models). Continuing to report noise values and further develop techniques for measuring noise could also be important for obtaining a complete picture of loss mechanisms in resonators.

In terms of materials research, we anticipate continued performance improvement in both dielectrics and superconducting metals. Towards reducing TLS loss, promising avenues include interface engineering and development of novel dielectrics.  Given the diversity of substrates, metals, and passivation layers currently employed, there exist materials or conditions that have yet to be explored to their full potential. Separately, planar resonators and 3D cavities alike will benefit from development of superconducting bonding. In the former case, high quality superconducting enclosures will be helpful to mitigate radiative loss and parasitic modes. In the case of 3D cavities, low-loss superconducting bonds enable scalable geometric constructions, including multilayer microwave integrated quantum circuits.

The authors advocate that future work in this area report TLS loss rather than low power loss, and publish base temperature, associated $Q_c$ values, and filling factors of the target materials for more accurate comparison between experiments and to avoid limiting the value of published results. 

Ideally, TLS loss would be extracted using both power- and temperature-dependent sweeps to demonstrate self-consistent results, although that is not always possible due to the additional complexity in experimental set-up. The inclusion of multiple or repeated resonator measurements, periodically in the same cooldown and over repeated cooldowns, as well as the use of a traceable resonator fit, would also strongly increase experimental reliability. Finally, the simultaneous measurement of multiple resonator parameters such as loss, frequency shift, and noise can yield a more comprehensive picture of TLS behavior, as well as reveal non-TLS loss mechanisms that may be present.

\section*{Data Availability}
Data sharing is not applicable to this article as no new data were created or analyzed in this study.

\begin{acknowledgments}
The authors would like to thank David Fork for his guidance, as well as Eric Rosenthal, Johannes Hubmayr, Katarina Cicak, Sisira Kanhirathingal, Daniel Sank, and Bob Buckley for their expertise and assistance with the writing process. This research was supported by the US Department of Energy (Award No. de-sc0019199).
\end{acknowledgments}

\FloatBarrier

\bibliography{ResMaterialsReview}

\providecommand{\noopsort}[1]{}\providecommand{\singleletter}[1]{#1}
\begin{thebibliography}{147}%
\makeatletter
\providecommand \@ifxundefined [1]{%
 \@ifx{#1\undefined}
}%
\providecommand \@ifnum [1]{%
 \ifnum #1\expandafter \@firstoftwo
 \else \expandafter \@secondoftwo
 \fi
}%
\providecommand \@ifx [1]{%
 \ifx #1\expandafter \@firstoftwo
 \else \expandafter \@secondoftwo
 \fi
}%
\providecommand \natexlab [1]{#1}%
\providecommand \enquote  [1]{``#1''}%
\providecommand \bibnamefont  [1]{#1}%
\providecommand \bibfnamefont [1]{#1}%
\providecommand \citenamefont [1]{#1}%
\providecommand \href@noop [0]{\@secondoftwo}%
\providecommand \href [0]{\begingroup \@sanitize@url \@href}%
\providecommand \@href[1]{\@@startlink{#1}\@@href}%
\providecommand \@@href[1]{\endgroup#1\@@endlink}%
\providecommand \@sanitize@url [0]{\catcode `\\12\catcode `\$12\catcode
  `\&12\catcode `\#12\catcode `\^12\catcode `\_12\catcode `\%12\relax}%
\providecommand \@@startlink[1]{}%
\providecommand \@@endlink[0]{}%
\providecommand \url  [0]{\begingroup\@sanitize@url \@url }%
\providecommand \@url [1]{\endgroup\@href {#1}{\urlprefix }}%
\providecommand \urlprefix  [0]{URL }%
\providecommand \Eprint [0]{\href }%
\providecommand \doibase [0]{http://dx.doi.org/}%
\providecommand \selectlanguage [0]{\@gobble}%
\providecommand \bibinfo  [0]{\@secondoftwo}%
\providecommand \bibfield  [0]{\@secondoftwo}%
\providecommand \translation [1]{[#1]}%
\providecommand \BibitemOpen [0]{}%
\providecommand \bibitemStop [0]{}%
\providecommand \bibitemNoStop [0]{.\EOS\space}%
\providecommand \EOS [0]{\spacefactor3000\relax}%
\providecommand \BibitemShut  [1]{\csname bibitem#1\endcsname}%
\let\auto@bib@innerbib\@empty
\bibitem [{\citenamefont {Abuwasib}, \citenamefont {Krantz},\ and\
  \citenamefont {Delsing}(2013)}]{Abuwasib2013}%
  \BibitemOpen
  \bibfield  {author} {\bibinfo {author} {\bibnamefont {Abuwasib},
  \bibfnamefont {M.}}, \bibinfo {author} {\bibnamefont {Krantz}, \bibfnamefont
  {P.}}, \ and\ \bibinfo {author} {\bibnamefont {Delsing}, \bibfnamefont
  {P.}},\ }\bibfield  {title} {\enquote {\bibinfo {title} {{Fabrication of
  large dimension aluminum air-bridges for superconducting quantum
  circuits}},}\ }\href@noop {} {\bibfield  {journal} {\bibinfo  {journal} {J.
  Vac. Sci. Technol. B}\ }\textbf {\bibinfo {volume} {31}},\ \bibinfo {pages}
  {3} (\bibinfo {year} {2013})}\BibitemShut {NoStop}%
\bibitem [{\citenamefont {Anderson}, \citenamefont {Halperin},\ and\
  \citenamefont {Varma}(1972)}]{Anderson1972}%
  \BibitemOpen
  \bibfield  {author} {\bibinfo {author} {\bibnamefont {Anderson},
  \bibfnamefont {P.~W.}}, \bibinfo {author} {\bibnamefont {Halperin},
  \bibfnamefont {B.}}, \ and\ \bibinfo {author} {\bibnamefont {Varma},
  \bibfnamefont {C.~M.}},\ }\bibfield  {title} {\enquote {\bibinfo {title}
  {Anomalous low-temperature thermal properties of glasses and spin glasses},}\
  }\href@noop {} {\bibfield  {journal} {\bibinfo  {journal} {Philos. Mag.}\
  }\textbf {\bibinfo {volume} {25}},\ \bibinfo {pages} {1--9} (\bibinfo {year}
  {1972})}\BibitemShut {NoStop}%
\bibitem [{\citenamefont {Arute}\ \emph {et~al.}(2019)\citenamefont {Arute},
  \citenamefont {Arya}, \citenamefont {Babbush}, \citenamefont {Bacon},
  \citenamefont {Bardin}, \citenamefont {Barends}, \citenamefont {Biswas},
  \citenamefont {Boixo}, \citenamefont {Brandao}, \citenamefont {Buell} \emph
  {et~al.}}]{arute2019quantum}%
  \BibitemOpen
  \bibfield  {author} {\bibinfo {author} {\bibnamefont {Arute}, \bibfnamefont
  {F.}}, \bibinfo {author} {\bibnamefont {Arya}, \bibfnamefont {K.}}, \bibinfo
  {author} {\bibnamefont {Babbush}, \bibfnamefont {R.}}, \bibinfo {author}
  {\bibnamefont {Bacon}, \bibfnamefont {D.}}, \bibinfo {author} {\bibnamefont
  {Bardin}, \bibfnamefont {J.~C.}}, \bibinfo {author} {\bibnamefont {Barends},
  \bibfnamefont {R.}}, \bibinfo {author} {\bibnamefont {Biswas}, \bibfnamefont
  {R.}}, \bibinfo {author} {\bibnamefont {Boixo}, \bibfnamefont {S.}}, \bibinfo
  {author} {\bibnamefont {Brandao}, \bibfnamefont {F.~G.}}, \bibinfo {author}
  {\bibnamefont {Buell}, \bibfnamefont {D.~A.}},  \emph {et~al.},\ }\bibfield
  {title} {\enquote {\bibinfo {title} {Quantum supremacy using a programmable
  superconducting processor},}\ }\href@noop {} {\bibfield  {journal} {\bibinfo
  {journal} {Nature}\ }\textbf {\bibinfo {volume} {574}},\ \bibinfo {pages}
  {505--510} (\bibinfo {year} {2019})}\BibitemShut {NoStop}%
\bibitem [{\citenamefont {Axline}(2018)}]{AxlineThesis}%
  \BibitemOpen
  \bibfield  {author} {\bibinfo {author} {\bibnamefont {Axline}, \bibfnamefont
  {C.}},\ }\emph {\bibinfo {title} {{Building Blocks for Modular Circuit QED
  Quantum Computing}}},\ \href@noop {} {Ph.D. thesis} (\bibinfo {year}
  {2018})\BibitemShut {NoStop}%
\bibitem [{\citenamefont {Barends}\ \emph {et~al.}(2008)\citenamefont
  {Barends}, \citenamefont {Baselmans}, \citenamefont {Yates}, \citenamefont
  {Gao}, \citenamefont {Hovenier},\ and\ \citenamefont
  {Klapwijk}}]{Barends2008}%
  \BibitemOpen
  \bibfield  {author} {\bibinfo {author} {\bibnamefont {Barends}, \bibfnamefont
  {R.}}, \bibinfo {author} {\bibnamefont {Baselmans}, \bibfnamefont {J.}},
  \bibinfo {author} {\bibnamefont {Yates}, \bibfnamefont {S.}}, \bibinfo
  {author} {\bibnamefont {Gao}, \bibfnamefont {J.}}, \bibinfo {author}
  {\bibnamefont {Hovenier}, \bibfnamefont {J.}}, \ and\ \bibinfo {author}
  {\bibnamefont {Klapwijk}, \bibfnamefont {T.}},\ }\bibfield  {title} {\enquote
  {\bibinfo {title} {{Quasiparticle relaxation in optically excited high-Q
  superconducting resonators}},}\ }\href@noop {} {\bibfield  {journal}
  {\bibinfo  {journal} {Phys. Rev. Lett.}\ }\textbf {\bibinfo {volume} {100}},\
  \bibinfo {pages} {257002} (\bibinfo {year} {2008})}\BibitemShut {NoStop}%
\bibitem [{\citenamefont {Barends}\ \emph {et~al.}(2013)\citenamefont
  {Barends}, \citenamefont {Kelly}, \citenamefont {Megrant}, \citenamefont
  {Sank}, \citenamefont {Jeffrey}, \citenamefont {Chen}, \citenamefont {Yin},
  \citenamefont {Chiaro}, \citenamefont {Mutus}, \citenamefont {Neill} \emph
  {et~al.}}]{barends2013coherent}%
  \BibitemOpen
  \bibfield  {author} {\bibinfo {author} {\bibnamefont {Barends}, \bibfnamefont
  {R.}}, \bibinfo {author} {\bibnamefont {Kelly}, \bibfnamefont {J.}}, \bibinfo
  {author} {\bibnamefont {Megrant}, \bibfnamefont {A.}}, \bibinfo {author}
  {\bibnamefont {Sank}, \bibfnamefont {D.}}, \bibinfo {author} {\bibnamefont
  {Jeffrey}, \bibfnamefont {E.}}, \bibinfo {author} {\bibnamefont {Chen},
  \bibfnamefont {Y.}}, \bibinfo {author} {\bibnamefont {Yin}, \bibfnamefont
  {Y.}}, \bibinfo {author} {\bibnamefont {Chiaro}, \bibfnamefont {B.}},
  \bibinfo {author} {\bibnamefont {Mutus}, \bibfnamefont {J.}}, \bibinfo
  {author} {\bibnamefont {Neill}, \bibfnamefont {C.}},  \emph {et~al.},\
  }\bibfield  {title} {\enquote {\bibinfo {title} {Coherent josephson qubit
  suitable for scalable quantum integrated circuits},}\ }\href@noop {}
  {\bibfield  {journal} {\bibinfo  {journal} {Physical review letters}\
  }\textbf {\bibinfo {volume} {111}},\ \bibinfo {pages} {080502} (\bibinfo
  {year} {2013})}\BibitemShut {NoStop}%
\bibitem [{\citenamefont {Barends}\ \emph {et~al.}(2010)\citenamefont
  {Barends}, \citenamefont {Vercruyssen}, \citenamefont {Endo}, \citenamefont
  {De~Visser}, \citenamefont {Zijlstra}, \citenamefont {Klapwijk},
  \citenamefont {Diener}, \citenamefont {Yates},\ and\ \citenamefont
  {Baselmans}}]{Barends2010b}%
  \BibitemOpen
  \bibfield  {author} {\bibinfo {author} {\bibnamefont {Barends}, \bibfnamefont
  {R.}}, \bibinfo {author} {\bibnamefont {Vercruyssen}, \bibfnamefont {N.}},
  \bibinfo {author} {\bibnamefont {Endo}, \bibfnamefont {A.}}, \bibinfo
  {author} {\bibnamefont {De~Visser}, \bibfnamefont {P.}}, \bibinfo {author}
  {\bibnamefont {Zijlstra}, \bibfnamefont {T.}}, \bibinfo {author}
  {\bibnamefont {Klapwijk}, \bibfnamefont {T.}}, \bibinfo {author}
  {\bibnamefont {Diener}, \bibfnamefont {P.}}, \bibinfo {author} {\bibnamefont
  {Yates}, \bibfnamefont {S.}}, \ and\ \bibinfo {author} {\bibnamefont
  {Baselmans}, \bibfnamefont {J.}},\ }\bibfield  {title} {\enquote {\bibinfo
  {title} {Minimal resonator loss for circuit quantum electrodynamics},}\
  }\href@noop {} {\bibfield  {journal} {\bibinfo  {journal} {Appl. Phys.
  Lett.}\ }\textbf {\bibinfo {volume} {97}},\ \bibinfo {pages} {023508}
  (\bibinfo {year} {2010})}\BibitemShut {NoStop}%
\bibitem [{\citenamefont {Barends}\ \emph {et~al.}(2011)\citenamefont
  {Barends}, \citenamefont {Wenner}, \citenamefont {Lenander}, \citenamefont
  {Chen}, \citenamefont {Bialczak}, \citenamefont {Kelly}, \citenamefont
  {Lucero}, \citenamefont {O’Malley}, \citenamefont {Mariantoni},
  \citenamefont {Sank} \emph {et~al.}}]{Barends2011}%
  \BibitemOpen
  \bibfield  {author} {\bibinfo {author} {\bibnamefont {Barends}, \bibfnamefont
  {R.}}, \bibinfo {author} {\bibnamefont {Wenner}, \bibfnamefont {J.}},
  \bibinfo {author} {\bibnamefont {Lenander}, \bibfnamefont {M.}}, \bibinfo
  {author} {\bibnamefont {Chen}, \bibfnamefont {Y.}}, \bibinfo {author}
  {\bibnamefont {Bialczak}, \bibfnamefont {R.~C.}}, \bibinfo {author}
  {\bibnamefont {Kelly}, \bibfnamefont {J.}}, \bibinfo {author} {\bibnamefont
  {Lucero}, \bibfnamefont {E.}}, \bibinfo {author} {\bibnamefont {O’Malley},
  \bibfnamefont {P.}}, \bibinfo {author} {\bibnamefont {Mariantoni},
  \bibfnamefont {M.}}, \bibinfo {author} {\bibnamefont {Sank}, \bibfnamefont
  {D.}},  \emph {et~al.},\ }\bibfield  {title} {\enquote {\bibinfo {title}
  {Minimizing quasiparticle generation from stray infrared light in
  superconducting quantum circuits},}\ }\href@noop {} {\bibfield  {journal}
  {\bibinfo  {journal} {Appl. Phys. Lett.}\ }\textbf {\bibinfo {volume} {99}},\
  \bibinfo {pages} {113507} (\bibinfo {year} {2011})}\BibitemShut {NoStop}%
\bibitem [{\citenamefont {B{\'e}janin}\ \emph {et~al.}(2016)\citenamefont
  {B{\'e}janin}, \citenamefont {McConkey}, \citenamefont {Rinehart},
  \citenamefont {Earnest}, \citenamefont {McRae}, \citenamefont {Shiri},
  \citenamefont {Bateman}, \citenamefont {Rohanizadegan}, \citenamefont
  {Penava}, \citenamefont {Breul} \emph {et~al.}}]{Bejanin2016}%
  \BibitemOpen
  \bibfield  {author} {\bibinfo {author} {\bibnamefont {B{\'e}janin},
  \bibfnamefont {J.}}, \bibinfo {author} {\bibnamefont {McConkey},
  \bibfnamefont {T.}}, \bibinfo {author} {\bibnamefont {Rinehart},
  \bibfnamefont {J.}}, \bibinfo {author} {\bibnamefont {Earnest}, \bibfnamefont
  {C.}}, \bibinfo {author} {\bibnamefont {McRae}, \bibfnamefont {C.}}, \bibinfo
  {author} {\bibnamefont {Shiri}, \bibfnamefont {D.}}, \bibinfo {author}
  {\bibnamefont {Bateman}, \bibfnamefont {J.}}, \bibinfo {author} {\bibnamefont
  {Rohanizadegan}, \bibfnamefont {Y.}}, \bibinfo {author} {\bibnamefont
  {Penava}, \bibfnamefont {B.}}, \bibinfo {author} {\bibnamefont {Breul},
  \bibfnamefont {P.}},  \emph {et~al.},\ }\bibfield  {title} {\enquote
  {\bibinfo {title} {Three-dimensional wiring for extensible quantum computing:
  The quantum socket},}\ }\href@noop {} {\bibfield  {journal} {\bibinfo
  {journal} {Phys. Rev. Applied}\ }\textbf {\bibinfo {volume} {6}},\ \bibinfo
  {pages} {044010} (\bibinfo {year} {2016})}\BibitemShut {NoStop}%
\bibitem [{\citenamefont {Bergeal}\ \emph {et~al.}(2010)\citenamefont
  {Bergeal}, \citenamefont {Schackert}, \citenamefont {Metcalfe}, \citenamefont
  {Vijay}, \citenamefont {Manucharyan}, \citenamefont {Frunzio}, \citenamefont
  {Prober}, \citenamefont {Schoelkopf}, \citenamefont {Girvin},\ and\
  \citenamefont {Devoret}}]{bergeal2010}%
  \BibitemOpen
  \bibfield  {author} {\bibinfo {author} {\bibnamefont {Bergeal}, \bibfnamefont
  {N.}}, \bibinfo {author} {\bibnamefont {Schackert}, \bibfnamefont {F.}},
  \bibinfo {author} {\bibnamefont {Metcalfe}, \bibfnamefont {M.}}, \bibinfo
  {author} {\bibnamefont {Vijay}, \bibfnamefont {R.}}, \bibinfo {author}
  {\bibnamefont {Manucharyan}, \bibfnamefont {V.}}, \bibinfo {author}
  {\bibnamefont {Frunzio}, \bibfnamefont {L.}}, \bibinfo {author} {\bibnamefont
  {Prober}, \bibfnamefont {D.}}, \bibinfo {author} {\bibnamefont {Schoelkopf},
  \bibfnamefont {R.}}, \bibinfo {author} {\bibnamefont {Girvin}, \bibfnamefont
  {S.}}, \ and\ \bibinfo {author} {\bibnamefont {Devoret}, \bibfnamefont
  {M.}},\ }\bibfield  {title} {\enquote {\bibinfo {title} {Phase-preserving
  amplification near the quantum limit with a josephson ring modulator},}\
  }\href@noop {} {\bibfield  {journal} {\bibinfo  {journal} {Nature}\ }\textbf
  {\bibinfo {volume} {465}},\ \bibinfo {pages} {64--68} (\bibinfo {year}
  {2010})}\BibitemShut {NoStop}%
\bibitem [{\citenamefont {Blais}\ \emph {et~al.}(2004)\citenamefont {Blais},
  \citenamefont {Huang}, \citenamefont {Wallraff}, \citenamefont {Girvin},\
  and\ \citenamefont {Schoelkopf}}]{Blais2004}%
  \BibitemOpen
  \bibfield  {author} {\bibinfo {author} {\bibnamefont {Blais}, \bibfnamefont
  {A.}}, \bibinfo {author} {\bibnamefont {Huang}, \bibfnamefont {R.-S.}},
  \bibinfo {author} {\bibnamefont {Wallraff}, \bibfnamefont {A.}}, \bibinfo
  {author} {\bibnamefont {Girvin}, \bibfnamefont {S.~M.}}, \ and\ \bibinfo
  {author} {\bibnamefont {Schoelkopf}, \bibfnamefont {R.~J.}},\ }\bibfield
  {title} {\enquote {\bibinfo {title} {Cavity quantum electrodynamics for
  superconducting electrical circuits: An architecture for quantum
  computation},}\ }\href@noop {} {\bibfield  {journal} {\bibinfo  {journal}
  {Phys. Rev. A}\ }\textbf {\bibinfo {volume} {69}},\ \bibinfo {pages} {062320}
  (\bibinfo {year} {2004})}\BibitemShut {NoStop}%
\bibitem [{\citenamefont {Bothner}\ \emph {et~al.}(2011)\citenamefont
  {Bothner}, \citenamefont {Gaber}, \citenamefont {Kemmler}, \citenamefont
  {Koelle},\ and\ \citenamefont {Kleiner}}]{bothner2011improving}%
  \BibitemOpen
  \bibfield  {author} {\bibinfo {author} {\bibnamefont {Bothner}, \bibfnamefont
  {D.}}, \bibinfo {author} {\bibnamefont {Gaber}, \bibfnamefont {T.}}, \bibinfo
  {author} {\bibnamefont {Kemmler}, \bibfnamefont {M.}}, \bibinfo {author}
  {\bibnamefont {Koelle}, \bibfnamefont {D.}}, \ and\ \bibinfo {author}
  {\bibnamefont {Kleiner}, \bibfnamefont {R.}},\ }\bibfield  {title} {\enquote
  {\bibinfo {title} {Improving the performance of superconducting microwave
  resonators in magnetic fields},}\ }\href@noop {} {\bibfield  {journal}
  {\bibinfo  {journal} {Appl. Phys. Lett.}\ }\textbf {\bibinfo {volume} {98}},\
  \bibinfo {pages} {102504} (\bibinfo {year} {2011})}\BibitemShut {NoStop}%
\bibitem [{\citenamefont {Brecht}\ \emph {et~al.}(2016)\citenamefont {Brecht},
  \citenamefont {Pfaff}, \citenamefont {Wang}, \citenamefont {Chu},
  \citenamefont {Frunzio}, \citenamefont {Devoret},\ and\ \citenamefont
  {Schoelkopf}}]{brecht2016multilayer}%
  \BibitemOpen
  \bibfield  {author} {\bibinfo {author} {\bibnamefont {Brecht}, \bibfnamefont
  {T.}}, \bibinfo {author} {\bibnamefont {Pfaff}, \bibfnamefont {W.}}, \bibinfo
  {author} {\bibnamefont {Wang}, \bibfnamefont {C.}}, \bibinfo {author}
  {\bibnamefont {Chu}, \bibfnamefont {Y.}}, \bibinfo {author} {\bibnamefont
  {Frunzio}, \bibfnamefont {L.}}, \bibinfo {author} {\bibnamefont {Devoret},
  \bibfnamefont {M.~H.}}, \ and\ \bibinfo {author} {\bibnamefont {Schoelkopf},
  \bibfnamefont {R.~J.}},\ }\bibfield  {title} {\enquote {\bibinfo {title}
  {Multilayer microwave integrated quantum circuits for scalable quantum
  computing},}\ }\href@noop {} {\bibfield  {journal} {\bibinfo  {journal} {Npj
  Quantum Inf.}\ }\textbf {\bibinfo {volume} {2}},\ \bibinfo {pages} {1--4}
  (\bibinfo {year} {2016})}\BibitemShut {NoStop}%
\bibitem [{\citenamefont {Brecht}\ \emph {et~al.}(2015)\citenamefont {Brecht},
  \citenamefont {Reagor}, \citenamefont {Chu}, \citenamefont {Pfaff},
  \citenamefont {Wang}, \citenamefont {Frunzio}, \citenamefont {Devoret},\ and\
  \citenamefont {Schoelkopf}}]{Brecht_In_cavities}%
  \BibitemOpen
  \bibfield  {author} {\bibinfo {author} {\bibnamefont {Brecht}, \bibfnamefont
  {T.}}, \bibinfo {author} {\bibnamefont {Reagor}, \bibfnamefont {M.}},
  \bibinfo {author} {\bibnamefont {Chu}, \bibfnamefont {Y.}}, \bibinfo {author}
  {\bibnamefont {Pfaff}, \bibfnamefont {W.}}, \bibinfo {author} {\bibnamefont
  {Wang}, \bibfnamefont {C.}}, \bibinfo {author} {\bibnamefont {Frunzio},
  \bibfnamefont {L.}}, \bibinfo {author} {\bibnamefont {Devoret}, \bibfnamefont
  {M.~H.}}, \ and\ \bibinfo {author} {\bibnamefont {Schoelkopf}, \bibfnamefont
  {R.~J.}},\ }\bibfield  {title} {\enquote {\bibinfo {title} {Demonstration of
  superconducting micromachined cavities},}\ }\href@noop {} {\bibfield
  {journal} {\bibinfo  {journal} {Appl. Phys. Lett.}\ }\textbf {\bibinfo
  {volume} {107}},\ \bibinfo {pages} {192603} (\bibinfo {year}
  {2015})}\BibitemShut {NoStop}%
\bibitem [{\citenamefont {Brehm}\ \emph {et~al.}(2017)\citenamefont {Brehm},
  \citenamefont {Bilmes}, \citenamefont {Weiss}, \citenamefont {Ustinov},\ and\
  \citenamefont {Lisenfeld}}]{Brehm2017}%
  \BibitemOpen
  \bibfield  {author} {\bibinfo {author} {\bibnamefont {Brehm}, \bibfnamefont
  {J.~D.}}, \bibinfo {author} {\bibnamefont {Bilmes}, \bibfnamefont {A.}},
  \bibinfo {author} {\bibnamefont {Weiss}, \bibfnamefont {G.}}, \bibinfo
  {author} {\bibnamefont {Ustinov}, \bibfnamefont {A.~V.}}, \ and\ \bibinfo
  {author} {\bibnamefont {Lisenfeld}, \bibfnamefont {J.}},\ }\bibfield  {title}
  {\enquote {\bibinfo {title} {Transmission-line resonators for the study of
  individual two-level tunneling systems},}\ }\href@noop {} {\bibfield
  {journal} {\bibinfo  {journal} {Appl. Phys. Lett.}\ }\textbf {\bibinfo
  {volume} {111}},\ \bibinfo {pages} {112601} (\bibinfo {year}
  {2017})}\BibitemShut {NoStop}%
\bibitem [{\citenamefont {Bronn}\ \emph {et~al.}(2018)\citenamefont {Bronn},
  \citenamefont {Adiga}, \citenamefont {Olivadese}, \citenamefont {Wu},
  \citenamefont {Chow},\ and\ \citenamefont {Pappas}}]{Bronn2018}%
  \BibitemOpen
  \bibfield  {author} {\bibinfo {author} {\bibnamefont {Bronn}, \bibfnamefont
  {N.~T.}}, \bibinfo {author} {\bibnamefont {Adiga}, \bibfnamefont {V.~P.}},
  \bibinfo {author} {\bibnamefont {Olivadese}, \bibfnamefont {S.~B.}}, \bibinfo
  {author} {\bibnamefont {Wu}, \bibfnamefont {X.}}, \bibinfo {author}
  {\bibnamefont {Chow}, \bibfnamefont {J.~M.}}, \ and\ \bibinfo {author}
  {\bibnamefont {Pappas}, \bibfnamefont {D.~P.}},\ }\bibfield  {title}
  {\enquote {\bibinfo {title} {High coherence plane breaking packaging for
  superconducting qubits},}\ }\href@noop {} {\bibfield  {journal} {\bibinfo
  {journal} {Quantum science and technology}\ }\textbf {\bibinfo {volume}
  {3}},\ \bibinfo {pages} {024007} (\bibinfo {year} {2018})}\BibitemShut
  {NoStop}%
\bibitem [{\citenamefont {Bruno}\ \emph {et~al.}(2015)\citenamefont {Bruno},
  \citenamefont {De~Lange}, \citenamefont {Asaad}, \citenamefont {Van
  Der~Enden}, \citenamefont {Langford},\ and\ \citenamefont
  {DiCarlo}}]{Bruno2015}%
  \BibitemOpen
  \bibfield  {author} {\bibinfo {author} {\bibnamefont {Bruno}, \bibfnamefont
  {A.}}, \bibinfo {author} {\bibnamefont {De~Lange}, \bibfnamefont {G.}},
  \bibinfo {author} {\bibnamefont {Asaad}, \bibfnamefont {S.}}, \bibinfo
  {author} {\bibnamefont {Van Der~Enden}, \bibfnamefont {K.}}, \bibinfo
  {author} {\bibnamefont {Langford}, \bibfnamefont {N.}}, \ and\ \bibinfo
  {author} {\bibnamefont {DiCarlo}, \bibfnamefont {L.}},\ }\bibfield  {title}
  {\enquote {\bibinfo {title} {Reducing intrinsic loss in superconducting
  resonators by surface treatment and deep etching of silicon substrates},}\
  }\href@noop {} {\bibfield  {journal} {\bibinfo  {journal} {Appl. Phys.
  Lett.}\ }\textbf {\bibinfo {volume} {106}},\ \bibinfo {pages} {182601}
  (\bibinfo {year} {2015})}\BibitemShut {NoStop}%
\bibitem [{\citenamefont {Burnett}\ \emph {et~al.}(2018)\citenamefont
  {Burnett}, \citenamefont {Bengtsson}, \citenamefont {Niepce},\ and\
  \citenamefont {Bylander}}]{Burnett2018}%
  \BibitemOpen
  \bibfield  {author} {\bibinfo {author} {\bibnamefont {Burnett}, \bibfnamefont
  {J.}}, \bibinfo {author} {\bibnamefont {Bengtsson}, \bibfnamefont {A.}},
  \bibinfo {author} {\bibnamefont {Niepce}, \bibfnamefont {D.}}, \ and\
  \bibinfo {author} {\bibnamefont {Bylander}, \bibfnamefont {J.}},\ }\bibfield
  {title} {\enquote {\bibinfo {title} {Noise and loss of superconducting
  aluminium resonators at single photon energies},}\ }\href@noop {} {\bibfield
  {journal} {\bibinfo  {journal} {J. Phys. Conf. Ser.}\ }\textbf {\bibinfo
  {volume} {969}},\ \bibinfo {pages} {012131} (\bibinfo {year}
  {2018})}\BibitemShut {NoStop}%
\bibitem [{\citenamefont {Burnett}, \citenamefont {Faoro},\ and\ \citenamefont
  {Lindstr{\"o}m}(2016)}]{Burnett2016}%
  \BibitemOpen
  \bibfield  {author} {\bibinfo {author} {\bibnamefont {Burnett}, \bibfnamefont
  {J.}}, \bibinfo {author} {\bibnamefont {Faoro}, \bibfnamefont {L.}}, \ and\
  \bibinfo {author} {\bibnamefont {Lindstr{\"o}m}, \bibfnamefont {T.}},\
  }\bibfield  {title} {\enquote {\bibinfo {title} {Analysis of high quality
  superconducting resonators: consequences for tls properties in amorphous
  oxides},}\ }\href@noop {} {\bibfield  {journal} {\bibinfo  {journal}
  {Supercond. Sci. Technol.}\ }\textbf {\bibinfo {volume} {29}},\ \bibinfo
  {pages} {044008} (\bibinfo {year} {2016})}\BibitemShut {NoStop}%
\bibitem [{\citenamefont {Burnett}\ \emph {et~al.}(2013)\citenamefont
  {Burnett}, \citenamefont {Lindstr{\"o}m}, \citenamefont {Oxborrow},
  \citenamefont {Harada}, \citenamefont {Sekine}, \citenamefont {Meeson},\ and\
  \citenamefont {Tzalenchuk}}]{Burnett2013}%
  \BibitemOpen
  \bibfield  {author} {\bibinfo {author} {\bibnamefont {Burnett}, \bibfnamefont
  {J.}}, \bibinfo {author} {\bibnamefont {Lindstr{\"o}m}, \bibfnamefont {T.}},
  \bibinfo {author} {\bibnamefont {Oxborrow}, \bibfnamefont {M.}}, \bibinfo
  {author} {\bibnamefont {Harada}, \bibfnamefont {Y.}}, \bibinfo {author}
  {\bibnamefont {Sekine}, \bibfnamefont {Y.}}, \bibinfo {author} {\bibnamefont
  {Meeson}, \bibfnamefont {P.}}, \ and\ \bibinfo {author} {\bibnamefont
  {Tzalenchuk}, \bibfnamefont {A.~Y.}},\ }\bibfield  {title} {\enquote
  {\bibinfo {title} {Slow noise processes in superconducting resonators},}\
  }\href@noop {} {\bibfield  {journal} {\bibinfo  {journal} {Phys. Rev. B}\
  }\textbf {\bibinfo {volume} {87}},\ \bibinfo {pages} {140501} (\bibinfo
  {year} {2013})}\BibitemShut {NoStop}%
\bibitem [{\citenamefont {Calusine}\ \emph {et~al.}(2018)\citenamefont
  {Calusine}, \citenamefont {Melville}, \citenamefont {Woods}, \citenamefont
  {Das}, \citenamefont {Stull}, \citenamefont {Bolkhovsky}, \citenamefont
  {Braje}, \citenamefont {Hover}, \citenamefont {Kim}, \citenamefont {Miloshi}
  \emph {et~al.}}]{Calusine2018}%
  \BibitemOpen
  \bibfield  {author} {\bibinfo {author} {\bibnamefont {Calusine},
  \bibfnamefont {G.}}, \bibinfo {author} {\bibnamefont {Melville},
  \bibfnamefont {A.}}, \bibinfo {author} {\bibnamefont {Woods}, \bibfnamefont
  {W.}}, \bibinfo {author} {\bibnamefont {Das}, \bibfnamefont {R.}}, \bibinfo
  {author} {\bibnamefont {Stull}, \bibfnamefont {C.}}, \bibinfo {author}
  {\bibnamefont {Bolkhovsky}, \bibfnamefont {V.}}, \bibinfo {author}
  {\bibnamefont {Braje}, \bibfnamefont {D.}}, \bibinfo {author} {\bibnamefont
  {Hover}, \bibfnamefont {D.}}, \bibinfo {author} {\bibnamefont {Kim},
  \bibfnamefont {D.~K.}}, \bibinfo {author} {\bibnamefont {Miloshi},
  \bibfnamefont {X.}},  \emph {et~al.},\ }\bibfield  {title} {\enquote
  {\bibinfo {title} {Analysis and mitigation of interface losses in trenched
  superconducting coplanar waveguide resonators},}\ }\href@noop {} {\bibfield
  {journal} {\bibinfo  {journal} {Appl. Phys. Lett.}\ }\textbf {\bibinfo
  {volume} {112}},\ \bibinfo {pages} {062601} (\bibinfo {year}
  {2018})}\BibitemShut {NoStop}%
\bibitem [{\citenamefont {Cardani}\ \emph {et~al.}(2020)\citenamefont
  {Cardani}, \citenamefont {Valenti}, \citenamefont {Casali}, \citenamefont
  {Catelani}, \citenamefont {Charpentier}, \citenamefont {Clemenza},
  \citenamefont {Colantoni}, \citenamefont {Cruciani}, \citenamefont {Gironi},
  \citenamefont {Grünhaupt}, \citenamefont {Gusenkova}, \citenamefont
  {Henriques}, \citenamefont {Lagoin}, \citenamefont {Martinez}, \citenamefont
  {Pettinari}, \citenamefont {Rusconi}, \citenamefont {Sander}, \citenamefont
  {Ustinov}, \citenamefont {Weber}, \citenamefont {Wernsdorfer}, \citenamefont
  {Vignati}, \citenamefont {Pirro},\ and\ \citenamefont {Pop}}]{cardani2020}%
  \BibitemOpen
  \bibfield  {author} {\bibinfo {author} {\bibnamefont {Cardani}, \bibfnamefont
  {L.}}, \bibinfo {author} {\bibnamefont {Valenti}, \bibfnamefont {F.}},
  \bibinfo {author} {\bibnamefont {Casali}, \bibfnamefont {N.}}, \bibinfo
  {author} {\bibnamefont {Catelani}, \bibfnamefont {G.}}, \bibinfo {author}
  {\bibnamefont {Charpentier}, \bibfnamefont {T.}}, \bibinfo {author}
  {\bibnamefont {Clemenza}, \bibfnamefont {M.}}, \bibinfo {author}
  {\bibnamefont {Colantoni}, \bibfnamefont {I.}}, \bibinfo {author}
  {\bibnamefont {Cruciani}, \bibfnamefont {A.}}, \bibinfo {author}
  {\bibnamefont {Gironi}, \bibfnamefont {L.}}, \bibinfo {author} {\bibnamefont
  {Grünhaupt}, \bibfnamefont {L.}}, \bibinfo {author} {\bibnamefont
  {Gusenkova}, \bibfnamefont {D.}}, \bibinfo {author} {\bibnamefont
  {Henriques}, \bibfnamefont {F.}}, \bibinfo {author} {\bibnamefont {Lagoin},
  \bibfnamefont {M.}}, \bibinfo {author} {\bibnamefont {Martinez},
  \bibfnamefont {M.}}, \bibinfo {author} {\bibnamefont {Pettinari},
  \bibfnamefont {G.}}, \bibinfo {author} {\bibnamefont {Rusconi}, \bibfnamefont
  {C.}}, \bibinfo {author} {\bibnamefont {Sander}, \bibfnamefont {O.}},
  \bibinfo {author} {\bibnamefont {Ustinov}, \bibfnamefont {A.~V.}}, \bibinfo
  {author} {\bibnamefont {Weber}, \bibfnamefont {M.}}, \bibinfo {author}
  {\bibnamefont {Wernsdorfer}, \bibfnamefont {W.}}, \bibinfo {author}
  {\bibnamefont {Vignati}, \bibfnamefont {M.}}, \bibinfo {author} {\bibnamefont
  {Pirro}, \bibfnamefont {S.}}, \ and\ \bibinfo {author} {\bibnamefont {Pop},
  \bibfnamefont {I.~M.}},\ }\href@noop {} {\enquote {\bibinfo {title} {Reducing
  the impact of radioactivity on quantum circuits in a deep-underground
  facility},}\ } (\bibinfo {year} {2020}),\ \Eprint
  {http://arxiv.org/abs/2005.02286} {arXiv:2005.02286 [cond-mat.supr-con]}
  \BibitemShut {NoStop}%
\bibitem [{\citenamefont {Castellanos-Beltran}\ and\ \citenamefont
  {Lehnert}(2007)}]{castellanos2007widely}%
  \BibitemOpen
  \bibfield  {author} {\bibinfo {author} {\bibnamefont {Castellanos-Beltran},
  \bibfnamefont {M.}}\ and\ \bibinfo {author} {\bibnamefont {Lehnert},
  \bibfnamefont {K.}},\ }\bibfield  {title} {\enquote {\bibinfo {title} {Widely
  tunable parametric amplifier based on a superconducting quantum interference
  device array resonator},}\ }\href@noop {} {\bibfield  {journal} {\bibinfo
  {journal} {Appl. Phys. Lett.}\ }\textbf {\bibinfo {volume} {91}},\ \bibinfo
  {pages} {083509} (\bibinfo {year} {2007})}\BibitemShut {NoStop}%
\bibitem [{\citenamefont {Castellanos-Beltran}(2010)}]{manuelthesis}%
  \BibitemOpen
  \bibfield  {author} {\bibinfo {author} {\bibnamefont {Castellanos-Beltran},
  \bibfnamefont {M.~A.}},\ }\emph {\bibinfo {title} {Development of a Josephson
  parametric amplifier for the preparation and detection of nonclassical states
  of microwave fields}},\ \href@noop {} {Ph.D. thesis},\ \bibinfo  {school}
  {University of Colorado Boulder} (\bibinfo {year} {2010})\BibitemShut
  {NoStop}%
\bibitem [{\citenamefont {Catelani}\ \emph {et~al.}(2011)\citenamefont
  {Catelani}, \citenamefont {Schoelkopf}, \citenamefont {Devoret},\ and\
  \citenamefont {Glazman}}]{catelani2011relaxation}%
  \BibitemOpen
  \bibfield  {author} {\bibinfo {author} {\bibnamefont {Catelani},
  \bibfnamefont {G.}}, \bibinfo {author} {\bibnamefont {Schoelkopf},
  \bibfnamefont {R.~J.}}, \bibinfo {author} {\bibnamefont {Devoret},
  \bibfnamefont {M.~H.}}, \ and\ \bibinfo {author} {\bibnamefont {Glazman},
  \bibfnamefont {L.~I.}},\ }\bibfield  {title} {\enquote {\bibinfo {title}
  {Relaxation and frequency shifts induced by quasiparticles in superconducting
  qubits},}\ }\href@noop {} {\bibfield  {journal} {\bibinfo  {journal}
  {Physical Review B}\ }\textbf {\bibinfo {volume} {84}},\ \bibinfo {pages}
  {064517} (\bibinfo {year} {2011})}\BibitemShut {NoStop}%
\bibitem [{\citenamefont {Chang}\ \emph {et~al.}(2013)\citenamefont {Chang},
  \citenamefont {Vissers}, \citenamefont {C{\'o}rcoles}, \citenamefont
  {Sandberg}, \citenamefont {Gao}, \citenamefont {Abraham}, \citenamefont
  {Chow}, \citenamefont {Gambetta}, \citenamefont {Beth~Rothwell},
  \citenamefont {Keefe} \emph {et~al.}}]{Chang2013}%
  \BibitemOpen
  \bibfield  {author} {\bibinfo {author} {\bibnamefont {Chang}, \bibfnamefont
  {J.~B.}}, \bibinfo {author} {\bibnamefont {Vissers}, \bibfnamefont {M.~R.}},
  \bibinfo {author} {\bibnamefont {C{\'o}rcoles}, \bibfnamefont {A.~D.}},
  \bibinfo {author} {\bibnamefont {Sandberg}, \bibfnamefont {M.}}, \bibinfo
  {author} {\bibnamefont {Gao}, \bibfnamefont {J.}}, \bibinfo {author}
  {\bibnamefont {Abraham}, \bibfnamefont {D.~W.}}, \bibinfo {author}
  {\bibnamefont {Chow}, \bibfnamefont {J.~M.}}, \bibinfo {author} {\bibnamefont
  {Gambetta}, \bibfnamefont {J.~M.}}, \bibinfo {author} {\bibnamefont
  {Beth~Rothwell}, \bibfnamefont {M.}}, \bibinfo {author} {\bibnamefont
  {Keefe}, \bibfnamefont {G.~A.}},  \emph {et~al.},\ }\bibfield  {title}
  {\enquote {\bibinfo {title} {Improved superconducting qubit coherence using
  titanium nitride},}\ }\href@noop {} {\bibfield  {journal} {\bibinfo
  {journal} {Appl. Phys. Lett.}\ }\textbf {\bibinfo {volume} {103}},\ \bibinfo
  {pages} {012602} (\bibinfo {year} {2013})}\BibitemShut {NoStop}%
\bibitem [{\citenamefont {Chen}\ \emph {et~al.}(2014)\citenamefont {Chen},
  \citenamefont {Megrant}, \citenamefont {Kelly}, \citenamefont {Barends},
  \citenamefont {Bochmann}, \citenamefont {Chen}, \citenamefont {Chiaro},
  \citenamefont {Dunsworth}, \citenamefont {Jeffrey}, \citenamefont {Mutus}
  \emph {et~al.}}]{chen2014_airbridge}%
  \BibitemOpen
  \bibfield  {author} {\bibinfo {author} {\bibnamefont {Chen}, \bibfnamefont
  {Z.}}, \bibinfo {author} {\bibnamefont {Megrant}, \bibfnamefont {A.}},
  \bibinfo {author} {\bibnamefont {Kelly}, \bibfnamefont {J.}}, \bibinfo
  {author} {\bibnamefont {Barends}, \bibfnamefont {R.}}, \bibinfo {author}
  {\bibnamefont {Bochmann}, \bibfnamefont {J.}}, \bibinfo {author}
  {\bibnamefont {Chen}, \bibfnamefont {Y.}}, \bibinfo {author} {\bibnamefont
  {Chiaro}, \bibfnamefont {B.}}, \bibinfo {author} {\bibnamefont {Dunsworth},
  \bibfnamefont {A.}}, \bibinfo {author} {\bibnamefont {Jeffrey}, \bibfnamefont
  {E.}}, \bibinfo {author} {\bibnamefont {Mutus}, \bibfnamefont {J.}},  \emph
  {et~al.},\ }\bibfield  {title} {\enquote {\bibinfo {title} {Fabrication and
  characterization of aluminum airbridges for superconducting microwave
  circuits},}\ }\href@noop {} {\bibfield  {journal} {\bibinfo  {journal} {Appl.
  Phys. Lett.}\ }\textbf {\bibinfo {volume} {104}},\ \bibinfo {pages} {052602}
  (\bibinfo {year} {2014})}\BibitemShut {NoStop}%
\bibitem [{\citenamefont {Chiaro}\ \emph {et~al.}(2016)\citenamefont {Chiaro},
  \citenamefont {Megrant}, \citenamefont {Dunsworth}, \citenamefont {Chen},
  \citenamefont {Barends}, \citenamefont {Campbell}, \citenamefont {Chen},
  \citenamefont {Fowler}, \citenamefont {Hoi}, \citenamefont {Jeffrey} \emph
  {et~al.}}]{Chiaro2016}%
  \BibitemOpen
  \bibfield  {author} {\bibinfo {author} {\bibnamefont {Chiaro}, \bibfnamefont
  {B.}}, \bibinfo {author} {\bibnamefont {Megrant}, \bibfnamefont {A.}},
  \bibinfo {author} {\bibnamefont {Dunsworth}, \bibfnamefont {A.}}, \bibinfo
  {author} {\bibnamefont {Chen}, \bibfnamefont {Z.}}, \bibinfo {author}
  {\bibnamefont {Barends}, \bibfnamefont {R.}}, \bibinfo {author} {\bibnamefont
  {Campbell}, \bibfnamefont {B.}}, \bibinfo {author} {\bibnamefont {Chen},
  \bibfnamefont {Y.}}, \bibinfo {author} {\bibnamefont {Fowler}, \bibfnamefont
  {A.}}, \bibinfo {author} {\bibnamefont {Hoi}, \bibfnamefont {I.}}, \bibinfo
  {author} {\bibnamefont {Jeffrey}, \bibfnamefont {E.}},  \emph {et~al.},\
  }\bibfield  {title} {\enquote {\bibinfo {title} {Dielectric surface loss in
  superconducting resonators with flux-trapping holes},}\ }\href@noop {}
  {\bibfield  {journal} {\bibinfo  {journal} {Supercond. Sci. Technol.}\
  }\textbf {\bibinfo {volume} {29}},\ \bibinfo {pages} {104006} (\bibinfo
  {year} {2016})}\BibitemShut {NoStop}%
\bibitem [{\citenamefont {Cho}\ \emph {et~al.}(2013)\citenamefont {Cho},
  \citenamefont {Patel}, \citenamefont {Podkaminer}, \citenamefont {Gao},
  \citenamefont {Folkman}, \citenamefont {Bark}, \citenamefont {Lee},
  \citenamefont {Zhang}, \citenamefont {Pan}, \citenamefont {McDermott} \emph
  {et~al.}}]{Cho2013}%
  \BibitemOpen
  \bibfield  {author} {\bibinfo {author} {\bibnamefont {Cho}, \bibfnamefont
  {K.-H.}}, \bibinfo {author} {\bibnamefont {Patel}, \bibfnamefont {U.}},
  \bibinfo {author} {\bibnamefont {Podkaminer}, \bibfnamefont {J.}}, \bibinfo
  {author} {\bibnamefont {Gao}, \bibfnamefont {Y.}}, \bibinfo {author}
  {\bibnamefont {Folkman}, \bibfnamefont {C.}}, \bibinfo {author} {\bibnamefont
  {Bark}, \bibfnamefont {C.}}, \bibinfo {author} {\bibnamefont {Lee},
  \bibfnamefont {S.}}, \bibinfo {author} {\bibnamefont {Zhang}, \bibfnamefont
  {Y.}}, \bibinfo {author} {\bibnamefont {Pan}, \bibfnamefont {X.}}, \bibinfo
  {author} {\bibnamefont {McDermott}, \bibfnamefont {R.}},  \emph {et~al.},\
  }\bibfield  {title} {\enquote {\bibinfo {title} {Epitaxial {Al$_2$O$_3$}
  capacitors for low microwave loss superconducting quantum circuits},}\
  }\href@noop {} {\bibfield  {journal} {\bibinfo  {journal} {APL Materials}\
  }\textbf {\bibinfo {volume} {1}},\ \bibinfo {pages} {042115} (\bibinfo {year}
  {2013})}\BibitemShut {NoStop}%
\bibitem [{\citenamefont {Cicak}\ \emph {et~al.}(2010)\citenamefont {Cicak},
  \citenamefont {Li}, \citenamefont {Strong}, \citenamefont {Allman},
  \citenamefont {Altomare}, \citenamefont {Sirois}, \citenamefont {Whittaker},
  \citenamefont {Teufel},\ and\ \citenamefont {Simmonds}}]{Cicak2010}%
  \BibitemOpen
  \bibfield  {author} {\bibinfo {author} {\bibnamefont {Cicak}, \bibfnamefont
  {K.}}, \bibinfo {author} {\bibnamefont {Li}, \bibfnamefont {D.}}, \bibinfo
  {author} {\bibnamefont {Strong}, \bibfnamefont {J.~A.}}, \bibinfo {author}
  {\bibnamefont {Allman}, \bibfnamefont {M.~S.}}, \bibinfo {author}
  {\bibnamefont {Altomare}, \bibfnamefont {F.}}, \bibinfo {author}
  {\bibnamefont {Sirois}, \bibfnamefont {A.~J.}}, \bibinfo {author}
  {\bibnamefont {Whittaker}, \bibfnamefont {J.~D.}}, \bibinfo {author}
  {\bibnamefont {Teufel}, \bibfnamefont {J.~D.}}, \ and\ \bibinfo {author}
  {\bibnamefont {Simmonds}, \bibfnamefont {R.~W.}},\ }\bibfield  {title}
  {\enquote {\bibinfo {title} {Low-loss superconducting resonant circuits using
  vacuum-gap-based microwave components},}\ }\href@noop {} {\bibfield
  {journal} {\bibinfo  {journal} {Appl. Phys. Lett.}\ }\textbf {\bibinfo
  {volume} {96}},\ \bibinfo {pages} {093502} (\bibinfo {year}
  {2010})}\BibitemShut {NoStop}%
\bibitem [{\citenamefont {Córcoles}\ \emph {et~al.}(2011)\citenamefont
  {Córcoles}, \citenamefont {Chow}, \citenamefont {Gambetta}, \citenamefont
  {Rigetti}, \citenamefont {Rozen}, \citenamefont {Keefe}, \citenamefont
  {Beth~Rothwell}, \citenamefont {Ketchen},\ and\ \citenamefont
  {Steffen}}]{Corcoles2011}%
  \BibitemOpen
  \bibfield  {author} {\bibinfo {author} {\bibnamefont {Córcoles},
  \bibfnamefont {A.~D.}}, \bibinfo {author} {\bibnamefont {Chow}, \bibfnamefont
  {J.~M.}}, \bibinfo {author} {\bibnamefont {Gambetta}, \bibfnamefont {J.~M.}},
  \bibinfo {author} {\bibnamefont {Rigetti}, \bibfnamefont {C.}}, \bibinfo
  {author} {\bibnamefont {Rozen}, \bibfnamefont {J.~R.}}, \bibinfo {author}
  {\bibnamefont {Keefe}, \bibfnamefont {G.~A.}}, \bibinfo {author}
  {\bibnamefont {Beth~Rothwell}, \bibfnamefont {M.}}, \bibinfo {author}
  {\bibnamefont {Ketchen}, \bibfnamefont {M.~B.}}, \ and\ \bibinfo {author}
  {\bibnamefont {Steffen}, \bibfnamefont {M.}},\ }\bibfield  {title} {\enquote
  {\bibinfo {title} {Protecting superconducting qubits from radiation},}\
  }\href {\doibase 10.1063/1.3658630} {\bibfield  {journal} {\bibinfo
  {journal} {Applied Physics Letters}\ }\textbf {\bibinfo {volume} {99}},\
  \bibinfo {pages} {181906} (\bibinfo {year} {2011})},\ \Eprint
  {http://arxiv.org/abs/https://doi.org/10.1063/1.3658630}
  {https://doi.org/10.1063/1.3658630} \BibitemShut {NoStop}%
\bibitem [{\citenamefont {De~Graaf}\ \emph {et~al.}(2018)\citenamefont
  {De~Graaf}, \citenamefont {Faoro}, \citenamefont {Burnett}, \citenamefont
  {Adamyan}, \citenamefont {Tzalenchuk}, \citenamefont {Kubatkin},
  \citenamefont {Lindstr{\"o}m},\ and\ \citenamefont {Danilov}}]{DeGraaf2017}%
  \BibitemOpen
  \bibfield  {author} {\bibinfo {author} {\bibnamefont {De~Graaf},
  \bibfnamefont {S.}}, \bibinfo {author} {\bibnamefont {Faoro}, \bibfnamefont
  {L.}}, \bibinfo {author} {\bibnamefont {Burnett}, \bibfnamefont {J.}},
  \bibinfo {author} {\bibnamefont {Adamyan}, \bibfnamefont {A.}}, \bibinfo
  {author} {\bibnamefont {Tzalenchuk}, \bibfnamefont {A.~Y.}}, \bibinfo
  {author} {\bibnamefont {Kubatkin}, \bibfnamefont {S.}}, \bibinfo {author}
  {\bibnamefont {Lindstr{\"o}m}, \bibfnamefont {T.}}, \ and\ \bibinfo {author}
  {\bibnamefont {Danilov}, \bibfnamefont {A.}},\ }\bibfield  {title} {\enquote
  {\bibinfo {title} {Suppression of low-frequency charge noise in
  superconducting resonators by surface spin desorption},}\ }\href@noop {}
  {\bibfield  {journal} {\bibinfo  {journal} {Nat. Commun.}\ }\textbf {\bibinfo
  {volume} {9}},\ \bibinfo {pages} {1--6} (\bibinfo {year} {2018})}\BibitemShut
  {NoStop}%
\bibitem [{\citenamefont {De~Visser}\ \emph {et~al.}(2014)\citenamefont
  {De~Visser}, \citenamefont {Goldie}, \citenamefont {Diener}, \citenamefont
  {Withington}, \citenamefont {Baselmans},\ and\ \citenamefont
  {Klapwijk}}]{DeVisser2014}%
  \BibitemOpen
  \bibfield  {author} {\bibinfo {author} {\bibnamefont {De~Visser},
  \bibfnamefont {P.}}, \bibinfo {author} {\bibnamefont {Goldie}, \bibfnamefont
  {D.}}, \bibinfo {author} {\bibnamefont {Diener}, \bibfnamefont {P.}},
  \bibinfo {author} {\bibnamefont {Withington}, \bibfnamefont {S.}}, \bibinfo
  {author} {\bibnamefont {Baselmans}, \bibfnamefont {J.}}, \ and\ \bibinfo
  {author} {\bibnamefont {Klapwijk}, \bibfnamefont {T.}},\ }\bibfield  {title}
  {\enquote {\bibinfo {title} {Evidence of a nonequilibrium distribution of
  quasiparticles in the microwave response of a superconducting aluminum
  resonator},}\ }\href@noop {} {\bibfield  {journal} {\bibinfo  {journal}
  {Phys. Rev. Lett.}\ }\textbf {\bibinfo {volume} {112}},\ \bibinfo {pages}
  {047004} (\bibinfo {year} {2014})}\BibitemShut {NoStop}%
\bibitem [{\citenamefont {Deng}, \citenamefont {Otto},\ and\ \citenamefont
  {Lupascu}(2013)}]{Deng2013}%
  \BibitemOpen
  \bibfield  {author} {\bibinfo {author} {\bibnamefont {Deng}, \bibfnamefont
  {C.}}, \bibinfo {author} {\bibnamefont {Otto}, \bibfnamefont {M.}}, \ and\
  \bibinfo {author} {\bibnamefont {Lupascu}, \bibfnamefont {A.}},\ }\bibfield
  {title} {\enquote {\bibinfo {title} {An analysis method for transmission
  measurements of superconducting resonators with applications to
  quantum-regime dielectric-loss measurements},}\ }\href@noop {} {\bibfield
  {journal} {\bibinfo  {journal} {J. Appl. Phys.}\ }\textbf {\bibinfo {volume}
  {114}},\ \bibinfo {pages} {054504} (\bibinfo {year} {2013})}\BibitemShut
  {NoStop}%
\bibitem [{\citenamefont {Deng}, \citenamefont {Otto},\ and\ \citenamefont
  {Lupascu}(2014)}]{Deng2014}%
  \BibitemOpen
  \bibfield  {author} {\bibinfo {author} {\bibnamefont {Deng}, \bibfnamefont
  {C.}}, \bibinfo {author} {\bibnamefont {Otto}, \bibfnamefont {M.}}, \ and\
  \bibinfo {author} {\bibnamefont {Lupascu}, \bibfnamefont {A.}},\ }\bibfield
  {title} {\enquote {\bibinfo {title} {Characterization of low-temperature
  microwave loss of thin aluminum oxide formed by plasma oxidation},}\
  }\href@noop {} {\bibfield  {journal} {\bibinfo  {journal} {Appl. Phys.
  Lett.}\ }\textbf {\bibinfo {volume} {104}},\ \bibinfo {pages} {043506}
  (\bibinfo {year} {2014})}\BibitemShut {NoStop}%
\bibitem [{\citenamefont {Dial}\ \emph {et~al.}(2016)\citenamefont {Dial},
  \citenamefont {McClure}, \citenamefont {Poletto}, \citenamefont {Keefe},
  \citenamefont {Rothwell}, \citenamefont {Gambetta}, \citenamefont {Abraham},
  \citenamefont {Chow},\ and\ \citenamefont {Steffen}}]{dial2016bulk_surf}%
  \BibitemOpen
  \bibfield  {author} {\bibinfo {author} {\bibnamefont {Dial}, \bibfnamefont
  {O.}}, \bibinfo {author} {\bibnamefont {McClure}, \bibfnamefont {D.~T.}},
  \bibinfo {author} {\bibnamefont {Poletto}, \bibfnamefont {S.}}, \bibinfo
  {author} {\bibnamefont {Keefe}, \bibfnamefont {G.}}, \bibinfo {author}
  {\bibnamefont {Rothwell}, \bibfnamefont {M.~B.}}, \bibinfo {author}
  {\bibnamefont {Gambetta}, \bibfnamefont {J.~M.}}, \bibinfo {author}
  {\bibnamefont {Abraham}, \bibfnamefont {D.~W.}}, \bibinfo {author}
  {\bibnamefont {Chow}, \bibfnamefont {J.~M.}}, \ and\ \bibinfo {author}
  {\bibnamefont {Steffen}, \bibfnamefont {M.}},\ }\bibfield  {title} {\enquote
  {\bibinfo {title} {Bulk and surface loss in superconducting transmon
  qubits},}\ }\href@noop {} {\bibfield  {journal} {\bibinfo  {journal}
  {Supercond. Sci. Technol.}\ }\textbf {\bibinfo {volume} {29}},\ \bibinfo
  {pages} {044001} (\bibinfo {year} {2016})}\BibitemShut {NoStop}%
\bibitem [{\citenamefont {Duff}\ \emph {et~al.}(2016)\citenamefont {Duff},
  \citenamefont {Austermann}, \citenamefont {Beall}, \citenamefont {Becker},
  \citenamefont {Datta}, \citenamefont {Gallardo}, \citenamefont {Henderson},
  \citenamefont {Hilton}, \citenamefont {Ho}, \citenamefont {Hubmayr} \emph
  {et~al.}}]{duff2016_SiN}%
  \BibitemOpen
  \bibfield  {author} {\bibinfo {author} {\bibnamefont {Duff}, \bibfnamefont
  {S.~M.}}, \bibinfo {author} {\bibnamefont {Austermann}, \bibfnamefont {J.}},
  \bibinfo {author} {\bibnamefont {Beall}, \bibfnamefont {J.}}, \bibinfo
  {author} {\bibnamefont {Becker}, \bibfnamefont {D.}}, \bibinfo {author}
  {\bibnamefont {Datta}, \bibfnamefont {R.}}, \bibinfo {author} {\bibnamefont
  {Gallardo}, \bibfnamefont {P.}}, \bibinfo {author} {\bibnamefont {Henderson},
  \bibfnamefont {S.}}, \bibinfo {author} {\bibnamefont {Hilton}, \bibfnamefont
  {G.}}, \bibinfo {author} {\bibnamefont {Ho}, \bibfnamefont {S.}}, \bibinfo
  {author} {\bibnamefont {Hubmayr}, \bibfnamefont {J.}},  \emph {et~al.},\
  }\bibfield  {title} {\enquote {\bibinfo {title} {Advanced actpol multichroic
  polarimeter array fabrication process for 150 mm wafers},}\ }\href@noop {}
  {\bibfield  {journal} {\bibinfo  {journal} {J. Low Temp. Phys.}\ }\textbf
  {\bibinfo {volume} {184}},\ \bibinfo {pages} {634--641} (\bibinfo {year}
  {2016})}\BibitemShut {NoStop}%
\bibitem [{\citenamefont {Dunsworth}(2018)}]{dunsworth2018thesis}%
  \BibitemOpen
  \bibfield  {author} {\bibinfo {author} {\bibnamefont {Dunsworth},
  \bibfnamefont {A.}},\ }\emph {\bibinfo {title} {High Fidelity Entangling
  Gates in Superconducting Qubits}},\ \href@noop {} {Ph.D. thesis},\ \bibinfo
  {school} {University of California, Santa Barbara} (\bibinfo {year} {2018}),\
  \bibinfo {note} {embargoed until August 2020}\BibitemShut {NoStop}%
\bibitem [{\citenamefont {Dunsworth}\ \emph {et~al.}(2018)\citenamefont
  {Dunsworth}, \citenamefont {Barends}, \citenamefont {Chen}, \citenamefont
  {Chen}, \citenamefont {Chiaro}, \citenamefont {Fowler}, \citenamefont
  {Foxen}, \citenamefont {Jeffrey}, \citenamefont {Kelly}, \citenamefont
  {Klimov} \emph {et~al.}}]{dunsworth2018_airbridge}%
  \BibitemOpen
  \bibfield  {author} {\bibinfo {author} {\bibnamefont {Dunsworth},
  \bibfnamefont {A.}}, \bibinfo {author} {\bibnamefont {Barends}, \bibfnamefont
  {R.}}, \bibinfo {author} {\bibnamefont {Chen}, \bibfnamefont {Y.}}, \bibinfo
  {author} {\bibnamefont {Chen}, \bibfnamefont {Z.}}, \bibinfo {author}
  {\bibnamefont {Chiaro}, \bibfnamefont {B.}}, \bibinfo {author} {\bibnamefont
  {Fowler}, \bibfnamefont {A.}}, \bibinfo {author} {\bibnamefont {Foxen},
  \bibfnamefont {B.}}, \bibinfo {author} {\bibnamefont {Jeffrey}, \bibfnamefont
  {E.}}, \bibinfo {author} {\bibnamefont {Kelly}, \bibfnamefont {J.}}, \bibinfo
  {author} {\bibnamefont {Klimov}, \bibfnamefont {P.}},  \emph {et~al.},\
  }\bibfield  {title} {\enquote {\bibinfo {title} {A method for building low
  loss multi-layer wiring for superconducting microwave devices},}\ }\href@noop
  {} {\bibfield  {journal} {\bibinfo  {journal} {Appl. Phys. Lett.}\ }\textbf
  {\bibinfo {volume} {112}},\ \bibinfo {pages} {063502} (\bibinfo {year}
  {2018})}\BibitemShut {NoStop}%
\bibitem [{\citenamefont {Dunsworth}\ \emph {et~al.}(2017)\citenamefont
  {Dunsworth}, \citenamefont {Megrant}, \citenamefont {Quintana}, \citenamefont
  {Chen}, \citenamefont {Barends}, \citenamefont {Burkett}, \citenamefont
  {Foxen}, \citenamefont {Chen}, \citenamefont {Chiaro}, \citenamefont
  {Fowler}, \citenamefont {Graff}, \citenamefont {Jeffrey}, \citenamefont
  {Kelly}, \citenamefont {Lucero}, \citenamefont {Mutus}, \citenamefont
  {Neeley}, \citenamefont {Neill}, \citenamefont {Roushan}, \citenamefont
  {Sank}, \citenamefont {Vainsencher}, \citenamefont {Wenner}, \citenamefont
  {White},\ and\ \citenamefont {Martinis}}]{Dunsworth2017}%
  \BibitemOpen
  \bibfield  {author} {\bibinfo {author} {\bibnamefont {Dunsworth},
  \bibfnamefont {A.}}, \bibinfo {author} {\bibnamefont {Megrant}, \bibfnamefont
  {A.}}, \bibinfo {author} {\bibnamefont {Quintana}, \bibfnamefont {C.}},
  \bibinfo {author} {\bibnamefont {Chen}, \bibfnamefont {Z.}}, \bibinfo
  {author} {\bibnamefont {Barends}, \bibfnamefont {R.}}, \bibinfo {author}
  {\bibnamefont {Burkett}, \bibfnamefont {B.}}, \bibinfo {author} {\bibnamefont
  {Foxen}, \bibfnamefont {B.}}, \bibinfo {author} {\bibnamefont {Chen},
  \bibfnamefont {Y.}}, \bibinfo {author} {\bibnamefont {Chiaro}, \bibfnamefont
  {B.}}, \bibinfo {author} {\bibnamefont {Fowler}, \bibfnamefont {A.}},
  \bibinfo {author} {\bibnamefont {Graff}, \bibfnamefont {R.}}, \bibinfo
  {author} {\bibnamefont {Jeffrey}, \bibfnamefont {E.}}, \bibinfo {author}
  {\bibnamefont {Kelly}, \bibfnamefont {J.}}, \bibinfo {author} {\bibnamefont
  {Lucero}, \bibfnamefont {E.}}, \bibinfo {author} {\bibnamefont {Mutus},
  \bibfnamefont {J.~Y.}}, \bibinfo {author} {\bibnamefont {Neeley},
  \bibfnamefont {M.}}, \bibinfo {author} {\bibnamefont {Neill}, \bibfnamefont
  {C.}}, \bibinfo {author} {\bibnamefont {Roushan}, \bibfnamefont {P.}},
  \bibinfo {author} {\bibnamefont {Sank}, \bibfnamefont {D.}}, \bibinfo
  {author} {\bibnamefont {Vainsencher}, \bibfnamefont {A.}}, \bibinfo {author}
  {\bibnamefont {Wenner}, \bibfnamefont {J.}}, \bibinfo {author} {\bibnamefont
  {White}, \bibfnamefont {T.~C.}}, \ and\ \bibinfo {author} {\bibnamefont
  {Martinis}, \bibfnamefont {J.~M.}},\ }\bibfield  {title} {\enquote {\bibinfo
  {title} {{Characterization and reduction of capacitive loss induced by
  sub-micron Josephson junction fabrication in superconducting qubits}},}\
  }\href {\doibase 10.1063/1.4993577} {\bibfield  {journal} {\bibinfo
  {journal} {Appl. Phys. Lett.}\ }\textbf {\bibinfo {volume} {111}} (\bibinfo
  {year} {2017}),\ 10.1063/1.4993577},\ \Eprint
  {http://arxiv.org/abs/1706.00879} {arXiv:1706.00879} \BibitemShut {NoStop}%
\bibitem [{\citenamefont {Earnest}\ \emph {et~al.}(2018)\citenamefont
  {Earnest}, \citenamefont {B{\'e}janin}, \citenamefont {McConkey},
  \citenamefont {Peters}, \citenamefont {Korinek}, \citenamefont {Yuan},\ and\
  \citenamefont {Mariantoni}}]{Earnest2018}%
  \BibitemOpen
  \bibfield  {author} {\bibinfo {author} {\bibnamefont {Earnest}, \bibfnamefont
  {C.~T.}}, \bibinfo {author} {\bibnamefont {B{\'e}janin}, \bibfnamefont
  {J.~H.}}, \bibinfo {author} {\bibnamefont {McConkey}, \bibfnamefont {T.~G.}},
  \bibinfo {author} {\bibnamefont {Peters}, \bibfnamefont {E.~A.}}, \bibinfo
  {author} {\bibnamefont {Korinek}, \bibfnamefont {A.}}, \bibinfo {author}
  {\bibnamefont {Yuan}, \bibfnamefont {H.}}, \ and\ \bibinfo {author}
  {\bibnamefont {Mariantoni}, \bibfnamefont {M.}},\ }\bibfield  {title}
  {\enquote {\bibinfo {title} {Substrate surface engineering for high-quality
  silicon/aluminum superconducting resonators},}\ }\href@noop {} {\bibfield
  {journal} {\bibinfo  {journal} {Supercond. Sci. Technol.}\ }\textbf {\bibinfo
  {volume} {31}},\ \bibinfo {pages} {125013} (\bibinfo {year}
  {2018})}\BibitemShut {NoStop}%
\bibitem [{\citenamefont {Foxen}\ \emph {et~al.}(2018)\citenamefont {Foxen},
  \citenamefont {Mutus}, \citenamefont {Lucero}, \citenamefont {Graff},
  \citenamefont {Megrant}, \citenamefont {Chen}, \citenamefont {Quintana},
  \citenamefont {Burkett}, \citenamefont {Kelly}, \citenamefont {Jeffrey},
  \citenamefont {Yang}, \citenamefont {Yu}, \citenamefont {Arya}, \citenamefont
  {Barends}, \citenamefont {Chen}, \citenamefont {Chiaro}, \citenamefont
  {Dunsworth}, \citenamefont {Fowler}, \citenamefont {Gidney}, \citenamefont
  {Giustina}, \citenamefont {Huang}, \citenamefont {Klimov}, \citenamefont
  {Neeley}, \citenamefont {Neill}, \citenamefont {Roushan}, \citenamefont
  {Sank}, \citenamefont {Vainsencher}, \citenamefont {Wenner}, \citenamefont
  {White},\ and\ \citenamefont {Martinis}}]{Foxen2018}%
  \BibitemOpen
  \bibfield  {author} {\bibinfo {author} {\bibnamefont {Foxen}, \bibfnamefont
  {B.}}, \bibinfo {author} {\bibnamefont {Mutus}, \bibfnamefont {J.~Y.}},
  \bibinfo {author} {\bibnamefont {Lucero}, \bibfnamefont {E.}}, \bibinfo
  {author} {\bibnamefont {Graff}, \bibfnamefont {R.}}, \bibinfo {author}
  {\bibnamefont {Megrant}, \bibfnamefont {A.}}, \bibinfo {author} {\bibnamefont
  {Chen}, \bibfnamefont {Y.}}, \bibinfo {author} {\bibnamefont {Quintana},
  \bibfnamefont {C.}}, \bibinfo {author} {\bibnamefont {Burkett}, \bibfnamefont
  {B.}}, \bibinfo {author} {\bibnamefont {Kelly}, \bibfnamefont {J.}}, \bibinfo
  {author} {\bibnamefont {Jeffrey}, \bibfnamefont {E.}}, \bibinfo {author}
  {\bibnamefont {Yang}, \bibfnamefont {Y.}}, \bibinfo {author} {\bibnamefont
  {Yu}, \bibfnamefont {A.}}, \bibinfo {author} {\bibnamefont {Arya},
  \bibfnamefont {K.}}, \bibinfo {author} {\bibnamefont {Barends}, \bibfnamefont
  {R.}}, \bibinfo {author} {\bibnamefont {Chen}, \bibfnamefont {Z.}}, \bibinfo
  {author} {\bibnamefont {Chiaro}, \bibfnamefont {B.}}, \bibinfo {author}
  {\bibnamefont {Dunsworth}, \bibfnamefont {A.}}, \bibinfo {author}
  {\bibnamefont {Fowler}, \bibfnamefont {A.}}, \bibinfo {author} {\bibnamefont
  {Gidney}, \bibfnamefont {C.}}, \bibinfo {author} {\bibnamefont {Giustina},
  \bibfnamefont {M.}}, \bibinfo {author} {\bibnamefont {Huang}, \bibfnamefont
  {T.}}, \bibinfo {author} {\bibnamefont {Klimov}, \bibfnamefont {P.}},
  \bibinfo {author} {\bibnamefont {Neeley}, \bibfnamefont {M.}}, \bibinfo
  {author} {\bibnamefont {Neill}, \bibfnamefont {C.}}, \bibinfo {author}
  {\bibnamefont {Roushan}, \bibfnamefont {P.}}, \bibinfo {author} {\bibnamefont
  {Sank}, \bibfnamefont {D.}}, \bibinfo {author} {\bibnamefont {Vainsencher},
  \bibfnamefont {A.}}, \bibinfo {author} {\bibnamefont {Wenner}, \bibfnamefont
  {J.}}, \bibinfo {author} {\bibnamefont {White}, \bibfnamefont {T.~C.}}, \
  and\ \bibinfo {author} {\bibnamefont {Martinis}, \bibfnamefont {J.~M.}},\
  }\bibfield  {title} {\enquote {\bibinfo {title} {{Qubit compatible
  superconducting interconnects}},}\ }\href@noop {} {\bibfield  {journal}
  {\bibinfo  {journal} {Quant. Sci. Tech.}\ }\textbf {\bibinfo {volume} {3}},\
  \bibinfo {pages} {014005} (\bibinfo {year} {2018})}\BibitemShut {NoStop}%
\bibitem [{\citenamefont {Gao}(2008)}]{GaoThesis}%
  \BibitemOpen
  \bibfield  {author} {\bibinfo {author} {\bibnamefont {Gao}, \bibfnamefont
  {J.}},\ }\emph {\bibinfo {title} {The physics of superconducting microwave
  resonators}},\ \href@noop {} {Ph.D. thesis},\ \bibinfo  {school} {California
  Institute of Technology} (\bibinfo {year} {2008})\BibitemShut {NoStop}%
\bibitem [{\citenamefont {Gao}\ \emph {et~al.}(2008{\natexlab{a}})\citenamefont
  {Gao}, \citenamefont {Daal}, \citenamefont {Martinis}, \citenamefont
  {Vayonakis}, \citenamefont {Zmuidzinas}, \citenamefont {Sadoulet},
  \citenamefont {Mazin}, \citenamefont {Day},\ and\ \citenamefont
  {Leduc}}]{Gao_2Dnoise}%
  \BibitemOpen
  \bibfield  {author} {\bibinfo {author} {\bibnamefont {Gao}, \bibfnamefont
  {J.}}, \bibinfo {author} {\bibnamefont {Daal}, \bibfnamefont {M.}}, \bibinfo
  {author} {\bibnamefont {Martinis}, \bibfnamefont {J.~M.}}, \bibinfo {author}
  {\bibnamefont {Vayonakis}, \bibfnamefont {A.}}, \bibinfo {author}
  {\bibnamefont {Zmuidzinas}, \bibfnamefont {J.}}, \bibinfo {author}
  {\bibnamefont {Sadoulet}, \bibfnamefont {B.}}, \bibinfo {author}
  {\bibnamefont {Mazin}, \bibfnamefont {B.~A.}}, \bibinfo {author}
  {\bibnamefont {Day}, \bibfnamefont {P.~K.}}, \ and\ \bibinfo {author}
  {\bibnamefont {Leduc}, \bibfnamefont {H.~G.}},\ }\bibfield  {title} {\enquote
  {\bibinfo {title} {A semiempirical model for two-level system noise in
  superconducting microresonators},}\ }\href@noop {} {\bibfield  {journal}
  {\bibinfo  {journal} {Appl. Phys. Lett.}\ }\textbf {\bibinfo {volume} {92}},\
  \bibinfo {pages} {212504} (\bibinfo {year} {2008}{\natexlab{a}})}\BibitemShut
  {NoStop}%
\bibitem [{\citenamefont {Gao}\ \emph {et~al.}(2008{\natexlab{b}})\citenamefont
  {Gao}, \citenamefont {Daal}, \citenamefont {Martinis}, \citenamefont
  {Vayonakis}, \citenamefont {Zmuidzinas}, \citenamefont {Sadoulet},
  \citenamefont {Mazin}, \citenamefont {Day},\ and\ \citenamefont
  {Leduc}}]{Gao2008a}%
  \BibitemOpen
  \bibfield  {author} {\bibinfo {author} {\bibnamefont {Gao}, \bibfnamefont
  {J.}}, \bibinfo {author} {\bibnamefont {Daal}, \bibfnamefont {M.}}, \bibinfo
  {author} {\bibnamefont {Martinis}, \bibfnamefont {J.~M.}}, \bibinfo {author}
  {\bibnamefont {Vayonakis}, \bibfnamefont {A.}}, \bibinfo {author}
  {\bibnamefont {Zmuidzinas}, \bibfnamefont {J.}}, \bibinfo {author}
  {\bibnamefont {Sadoulet}, \bibfnamefont {B.}}, \bibinfo {author}
  {\bibnamefont {Mazin}, \bibfnamefont {B.~A.}}, \bibinfo {author}
  {\bibnamefont {Day}, \bibfnamefont {P.~K.}}, \ and\ \bibinfo {author}
  {\bibnamefont {Leduc}, \bibfnamefont {H.~G.}},\ }\bibfield  {title} {\enquote
  {\bibinfo {title} {A semiempirical model for two-level system noise in
  superconducting microresonators},}\ }\href@noop {} {\bibfield  {journal}
  {\bibinfo  {journal} {Appl. Phys. Lett.}\ }\textbf {\bibinfo {volume} {92}},\
  \bibinfo {pages} {212504} (\bibinfo {year} {2008}{\natexlab{b}})}\BibitemShut
  {NoStop}%
\bibitem [{\citenamefont {Gao}\ \emph {et~al.}(2008{\natexlab{c}})\citenamefont
  {Gao}, \citenamefont {Daal}, \citenamefont {Vayonakis}, \citenamefont
  {Kumar}, \citenamefont {Zmuidzinas}, \citenamefont {Sadoulet}, \citenamefont
  {Mazin}, \citenamefont {Day},\ and\ \citenamefont {Leduc}}]{Gao2008b}%
  \BibitemOpen
  \bibfield  {author} {\bibinfo {author} {\bibnamefont {Gao}, \bibfnamefont
  {J.}}, \bibinfo {author} {\bibnamefont {Daal}, \bibfnamefont {M.}}, \bibinfo
  {author} {\bibnamefont {Vayonakis}, \bibfnamefont {A.}}, \bibinfo {author}
  {\bibnamefont {Kumar}, \bibfnamefont {S.}}, \bibinfo {author} {\bibnamefont
  {Zmuidzinas}, \bibfnamefont {J.}}, \bibinfo {author} {\bibnamefont
  {Sadoulet}, \bibfnamefont {B.}}, \bibinfo {author} {\bibnamefont {Mazin},
  \bibfnamefont {B.~A.}}, \bibinfo {author} {\bibnamefont {Day}, \bibfnamefont
  {P.~K.}}, \ and\ \bibinfo {author} {\bibnamefont {Leduc}, \bibfnamefont
  {H.~G.}},\ }\bibfield  {title} {\enquote {\bibinfo {title} {Experimental
  evidence for a surface distribution of two-level systems in superconducting
  lithographed microwave resonators},}\ }\href@noop {} {\bibfield  {journal}
  {\bibinfo  {journal} {Appl. Phys. Lett.}\ }\textbf {\bibinfo {volume} {92}},\
  \bibinfo {pages} {152505} (\bibinfo {year} {2008}{\natexlab{c}})}\BibitemShut
  {NoStop}%
\bibitem [{\citenamefont {Gao}\ \emph {et~al.}(2011)\citenamefont {Gao},
  \citenamefont {Vale}, \citenamefont {Mates}, \citenamefont {Schmidt},
  \citenamefont {Hilton}, \citenamefont {Irwin}, \citenamefont {Mallet},
  \citenamefont {Castellanos-Beltran}, \citenamefont {Lehnert}, \citenamefont
  {Zmuidzinas} \emph {et~al.}}]{gao2011strongly}%
  \BibitemOpen
  \bibfield  {author} {\bibinfo {author} {\bibnamefont {Gao}, \bibfnamefont
  {J.}}, \bibinfo {author} {\bibnamefont {Vale}, \bibfnamefont {L.}}, \bibinfo
  {author} {\bibnamefont {Mates}, \bibfnamefont {J.}}, \bibinfo {author}
  {\bibnamefont {Schmidt}, \bibfnamefont {D.}}, \bibinfo {author} {\bibnamefont
  {Hilton}, \bibfnamefont {G.}}, \bibinfo {author} {\bibnamefont {Irwin},
  \bibfnamefont {K.}}, \bibinfo {author} {\bibnamefont {Mallet}, \bibfnamefont
  {F.}}, \bibinfo {author} {\bibnamefont {Castellanos-Beltran}, \bibfnamefont
  {M.}}, \bibinfo {author} {\bibnamefont {Lehnert}, \bibfnamefont {K.}},
  \bibinfo {author} {\bibnamefont {Zmuidzinas}, \bibfnamefont {J.}},  \emph
  {et~al.},\ }\bibfield  {title} {\enquote {\bibinfo {title} {Strongly
  quadrature-dependent noise in superconducting microresonators measured at the
  vacuum-noise limit},}\ }\href@noop {} {\bibfield  {journal} {\bibinfo
  {journal} {Appl. Phys. Lett.}\ }\textbf {\bibinfo {volume} {98}},\ \bibinfo
  {pages} {232508} (\bibinfo {year} {2011})}\BibitemShut {NoStop}%
\bibitem [{\citenamefont {Gao}\ \emph {et~al.}(2009)\citenamefont {Gao},
  \citenamefont {Vayonakis}, \citenamefont {Noroozian}, \citenamefont
  {Zmuidzinas}, \citenamefont {Day},\ and\ \citenamefont
  {Leduc}}]{gao2009_150GHz}%
  \BibitemOpen
  \bibfield  {author} {\bibinfo {author} {\bibnamefont {Gao}, \bibfnamefont
  {J.}}, \bibinfo {author} {\bibnamefont {Vayonakis}, \bibfnamefont {A.}},
  \bibinfo {author} {\bibnamefont {Noroozian}, \bibfnamefont {O.}}, \bibinfo
  {author} {\bibnamefont {Zmuidzinas}, \bibfnamefont {J.}}, \bibinfo {author}
  {\bibnamefont {Day}, \bibfnamefont {P.~K.}}, \ and\ \bibinfo {author}
  {\bibnamefont {Leduc}, \bibfnamefont {H.~G.}},\ }\bibfield  {title} {\enquote
  {\bibinfo {title} {Measurement of loss in superconducting microstrip at
  millimeter-wave frequencies},}\ }in\ \href@noop {} {\emph {\bibinfo
  {booktitle} {AIP Conference Proceedings}}},\ Vol.\ \bibinfo {volume} {1185}\
  (\bibinfo {organization} {American Institute of Physics},\ \bibinfo {year}
  {2009})\ pp.\ \bibinfo {pages} {164--167}\BibitemShut {NoStop}%
\bibitem [{\citenamefont {Gao}\ \emph {et~al.}(2007)\citenamefont {Gao},
  \citenamefont {Zmuidzinas}, \citenamefont {Mazin}, \citenamefont {LeDuc},\
  and\ \citenamefont {Day}}]{Gao2007}%
  \BibitemOpen
  \bibfield  {author} {\bibinfo {author} {\bibnamefont {Gao}, \bibfnamefont
  {J.}}, \bibinfo {author} {\bibnamefont {Zmuidzinas}, \bibfnamefont {J.}},
  \bibinfo {author} {\bibnamefont {Mazin}, \bibfnamefont {B.~A.}}, \bibinfo
  {author} {\bibnamefont {LeDuc}, \bibfnamefont {H.~G.}}, \ and\ \bibinfo
  {author} {\bibnamefont {Day}, \bibfnamefont {P.~K.}},\ }\bibfield  {title}
  {\enquote {\bibinfo {title} {Noise properties of superconducting coplanar
  waveguide microwave resonators},}\ }\href@noop {} {\bibfield  {journal}
  {\bibinfo  {journal} {Appl. Phys. Lett.}\ }\textbf {\bibinfo {volume} {90}},\
  \bibinfo {pages} {102507} (\bibinfo {year} {2007})}\BibitemShut {NoStop}%
\bibitem [{\citenamefont {Geerlings}\ \emph {et~al.}(2012)\citenamefont
  {Geerlings}, \citenamefont {Shankar}, \citenamefont {Edwards}, \citenamefont
  {Frunzio}, \citenamefont {Schoelkopf},\ and\ \citenamefont
  {Devoret}}]{geerlings2012}%
  \BibitemOpen
  \bibfield  {author} {\bibinfo {author} {\bibnamefont {Geerlings},
  \bibfnamefont {K.}}, \bibinfo {author} {\bibnamefont {Shankar}, \bibfnamefont
  {S.}}, \bibinfo {author} {\bibnamefont {Edwards}, \bibfnamefont {E.}},
  \bibinfo {author} {\bibnamefont {Frunzio}, \bibfnamefont {L.}}, \bibinfo
  {author} {\bibnamefont {Schoelkopf}, \bibfnamefont {R.}}, \ and\ \bibinfo
  {author} {\bibnamefont {Devoret}, \bibfnamefont {M.}},\ }\bibfield  {title}
  {\enquote {\bibinfo {title} {Improving the quality factor of microwave
  compact resonators by optimizing their geometrical parameters},}\ }\href@noop
  {} {\bibfield  {journal} {\bibinfo  {journal} {Appl. Phys. Lett.}\ }\textbf
  {\bibinfo {volume} {100}},\ \bibinfo {pages} {192601} (\bibinfo {year}
  {2012})}\BibitemShut {NoStop}%
\bibitem [{\citenamefont {Goetz}\ \emph {et~al.}(2016)\citenamefont {Goetz},
  \citenamefont {Deppe}, \citenamefont {Haeberlein}, \citenamefont {Wulschner},
  \citenamefont {Zollitsch}, \citenamefont {Meier}, \citenamefont {Fischer},
  \citenamefont {Eder}, \citenamefont {Xie}, \citenamefont {Fedorov} \emph
  {et~al.}}]{Goetz2016}%
  \BibitemOpen
  \bibfield  {author} {\bibinfo {author} {\bibnamefont {Goetz}, \bibfnamefont
  {J.}}, \bibinfo {author} {\bibnamefont {Deppe}, \bibfnamefont {F.}}, \bibinfo
  {author} {\bibnamefont {Haeberlein}, \bibfnamefont {M.}}, \bibinfo {author}
  {\bibnamefont {Wulschner}, \bibfnamefont {F.}}, \bibinfo {author}
  {\bibnamefont {Zollitsch}, \bibfnamefont {C.~W.}}, \bibinfo {author}
  {\bibnamefont {Meier}, \bibfnamefont {S.}}, \bibinfo {author} {\bibnamefont
  {Fischer}, \bibfnamefont {M.}}, \bibinfo {author} {\bibnamefont {Eder},
  \bibfnamefont {P.}}, \bibinfo {author} {\bibnamefont {Xie}, \bibfnamefont
  {E.}}, \bibinfo {author} {\bibnamefont {Fedorov}, \bibfnamefont {K.~G.}},
  \emph {et~al.},\ }\bibfield  {title} {\enquote {\bibinfo {title} {Loss
  mechanisms in superconducting thin film microwave resonators},}\ }\href@noop
  {} {\bibfield  {journal} {\bibinfo  {journal} {J. Appl. Phys.}\ }\textbf
  {\bibinfo {volume} {119}},\ \bibinfo {pages} {015304} (\bibinfo {year}
  {2016})}\BibitemShut {NoStop}%
\bibitem [{\citenamefont {Goldie}\ and\ \citenamefont
  {Withington}(2012)}]{Goldie2012}%
  \BibitemOpen
  \bibfield  {author} {\bibinfo {author} {\bibnamefont {Goldie}, \bibfnamefont
  {D.}}\ and\ \bibinfo {author} {\bibnamefont {Withington}, \bibfnamefont
  {S.}},\ }\bibfield  {title} {\enquote {\bibinfo {title} {Non-equilibrium
  superconductivity in quantum-sensing superconducting resonators},}\
  }\href@noop {} {\bibfield  {journal} {\bibinfo  {journal} {Supercond. Sci.
  Technol.}\ }\textbf {\bibinfo {volume} {26}},\ \bibinfo {pages} {015004}
  (\bibinfo {year} {2012})}\BibitemShut {NoStop}%
\bibitem [{\citenamefont {G{\"o}ppl}\ \emph {et~al.}(2008)\citenamefont
  {G{\"o}ppl}, \citenamefont {Fragner}, \citenamefont {Baur}, \citenamefont
  {Bianchetti}, \citenamefont {Filipp}, \citenamefont {Fink}, \citenamefont
  {Leek}, \citenamefont {Puebla}, \citenamefont {Steffen},\ and\ \citenamefont
  {Wallraff}}]{goppl2008coplanar}%
  \BibitemOpen
  \bibfield  {author} {\bibinfo {author} {\bibnamefont {G{\"o}ppl},
  \bibfnamefont {M.}}, \bibinfo {author} {\bibnamefont {Fragner}, \bibfnamefont
  {A.}}, \bibinfo {author} {\bibnamefont {Baur}, \bibfnamefont {M.}}, \bibinfo
  {author} {\bibnamefont {Bianchetti}, \bibfnamefont {R.}}, \bibinfo {author}
  {\bibnamefont {Filipp}, \bibfnamefont {S.}}, \bibinfo {author} {\bibnamefont
  {Fink}, \bibfnamefont {J.}}, \bibinfo {author} {\bibnamefont {Leek},
  \bibfnamefont {P.}}, \bibinfo {author} {\bibnamefont {Puebla}, \bibfnamefont
  {G.}}, \bibinfo {author} {\bibnamefont {Steffen}, \bibfnamefont {L.}}, \ and\
  \bibinfo {author} {\bibnamefont {Wallraff}, \bibfnamefont {A.}},\ }\bibfield
  {title} {\enquote {\bibinfo {title} {Coplanar waveguide resonators for
  circuit quantum electrodynamics},}\ }\href@noop {} {\bibfield  {journal}
  {\bibinfo  {journal} {J. Appl. Phys.}\ }\textbf {\bibinfo {volume} {104}},\
  \bibinfo {pages} {113904} (\bibinfo {year} {2008})}\BibitemShut {NoStop}%
\bibitem [{\citenamefont {Hofheinz}\ \emph {et~al.}(2009)\citenamefont
  {Hofheinz}, \citenamefont {Wang}, \citenamefont {Ansmann}, \citenamefont
  {Bialczak}, \citenamefont {Lucero}, \citenamefont {Neeley}, \citenamefont
  {O'connell}, \citenamefont {Sank}, \citenamefont {Wenner}, \citenamefont
  {Martinis} \emph {et~al.}}]{Hofheinz2009}%
  \BibitemOpen
  \bibfield  {author} {\bibinfo {author} {\bibnamefont {Hofheinz},
  \bibfnamefont {M.}}, \bibinfo {author} {\bibnamefont {Wang}, \bibfnamefont
  {H.}}, \bibinfo {author} {\bibnamefont {Ansmann}, \bibfnamefont {M.}},
  \bibinfo {author} {\bibnamefont {Bialczak}, \bibfnamefont {R.~C.}}, \bibinfo
  {author} {\bibnamefont {Lucero}, \bibfnamefont {E.}}, \bibinfo {author}
  {\bibnamefont {Neeley}, \bibfnamefont {M.}}, \bibinfo {author} {\bibnamefont
  {O'connell}, \bibfnamefont {A.}}, \bibinfo {author} {\bibnamefont {Sank},
  \bibfnamefont {D.}}, \bibinfo {author} {\bibnamefont {Wenner}, \bibfnamefont
  {J.}}, \bibinfo {author} {\bibnamefont {Martinis}, \bibfnamefont {J.~M.}},
  \emph {et~al.},\ }\bibfield  {title} {\enquote {\bibinfo {title}
  {Synthesizing arbitrary quantum states in a superconducting resonator},}\
  }\href@noop {} {\bibfield  {journal} {\bibinfo  {journal} {Nature}\ }\textbf
  {\bibinfo {volume} {459}},\ \bibinfo {pages} {546--549} (\bibinfo {year}
  {2009})}\BibitemShut {NoStop}%
\bibitem [{\citenamefont {Houck}\ \emph {et~al.}(2008)\citenamefont {Houck},
  \citenamefont {Schreier}, \citenamefont {Johnson}, \citenamefont {Chow},
  \citenamefont {Koch}, \citenamefont {Gambetta}, \citenamefont {Schuster},
  \citenamefont {Frunzio}, \citenamefont {Devoret}, \citenamefont {Girvin},\
  and\ \citenamefont {Schoelkopf}}]{Houck2008}%
  \BibitemOpen
  \bibfield  {author} {\bibinfo {author} {\bibnamefont {Houck}, \bibfnamefont
  {A.~A.}}, \bibinfo {author} {\bibnamefont {Schreier}, \bibfnamefont {J.~A.}},
  \bibinfo {author} {\bibnamefont {Johnson}, \bibfnamefont {B.~R.}}, \bibinfo
  {author} {\bibnamefont {Chow}, \bibfnamefont {J.~M.}}, \bibinfo {author}
  {\bibnamefont {Koch}, \bibfnamefont {J.}}, \bibinfo {author} {\bibnamefont
  {Gambetta}, \bibfnamefont {J.~M.}}, \bibinfo {author} {\bibnamefont
  {Schuster}, \bibfnamefont {D.~I.}}, \bibinfo {author} {\bibnamefont
  {Frunzio}, \bibfnamefont {L.}}, \bibinfo {author} {\bibnamefont {Devoret},
  \bibfnamefont {M.~H.}}, \bibinfo {author} {\bibnamefont {Girvin},
  \bibfnamefont {S.~M.}}, \ and\ \bibinfo {author} {\bibnamefont {Schoelkopf},
  \bibfnamefont {R.~J.}},\ }\bibfield  {title} {\enquote {\bibinfo {title}
  {{Controlling the spontaneous emission of a superconducting transmon
  qubit}},}\ }\href@noop {} {\bibfield  {journal} {\bibinfo  {journal} {Phys.
  Rev. Lett.}\ }\textbf {\bibinfo {volume} {101}},\ \bibinfo {pages} {080502}
  (\bibinfo {year} {2008})}\BibitemShut {NoStop}%
\bibitem [{\citenamefont {Jaim}\ \emph {et~al.}(2014)\citenamefont {Jaim},
  \citenamefont {Aguilar}, \citenamefont {Sarabi}, \citenamefont {Rosen},
  \citenamefont {Ramanayaka}, \citenamefont {Lock}, \citenamefont
  {Richardson},\ and\ \citenamefont {Osborn}}]{jaim2014_TiN}%
  \BibitemOpen
  \bibfield  {author} {\bibinfo {author} {\bibnamefont {Jaim}, \bibfnamefont
  {H.~I.}}, \bibinfo {author} {\bibnamefont {Aguilar}, \bibfnamefont {J.}},
  \bibinfo {author} {\bibnamefont {Sarabi}, \bibfnamefont {B.}}, \bibinfo
  {author} {\bibnamefont {Rosen}, \bibfnamefont {Y.}}, \bibinfo {author}
  {\bibnamefont {Ramanayaka}, \bibfnamefont {A.}}, \bibinfo {author}
  {\bibnamefont {Lock}, \bibfnamefont {E.}}, \bibinfo {author} {\bibnamefont
  {Richardson}, \bibfnamefont {C.}}, \ and\ \bibinfo {author} {\bibnamefont
  {Osborn}, \bibfnamefont {K.}},\ }\bibfield  {title} {\enquote {\bibinfo
  {title} {Superconducting tin films sputtered over a large range of substrate
  dc bias},}\ }\href@noop {} {\bibfield  {journal} {\bibinfo  {journal} {IEEE
  Trans. Appl. Supercond.}\ }\textbf {\bibinfo {volume} {25}},\ \bibinfo
  {pages} {1--5} (\bibinfo {year} {2014})}\BibitemShut {NoStop}%
\bibitem [{\citenamefont {Kaiser}\ \emph {et~al.}(2010)\citenamefont {Kaiser},
  \citenamefont {Skacel}, \citenamefont {W{\"u}nsch}, \citenamefont {Dolata},
  \citenamefont {Mackrodt}, \citenamefont {Zorin},\ and\ \citenamefont
  {Siegel}}]{kaiser2010_nb2O5}%
  \BibitemOpen
  \bibfield  {author} {\bibinfo {author} {\bibnamefont {Kaiser}, \bibfnamefont
  {C.}}, \bibinfo {author} {\bibnamefont {Skacel}, \bibfnamefont {S.}},
  \bibinfo {author} {\bibnamefont {W{\"u}nsch}, \bibfnamefont {S.}}, \bibinfo
  {author} {\bibnamefont {Dolata}, \bibfnamefont {R.}}, \bibinfo {author}
  {\bibnamefont {Mackrodt}, \bibfnamefont {B.}}, \bibinfo {author}
  {\bibnamefont {Zorin}, \bibfnamefont {A.}}, \ and\ \bibinfo {author}
  {\bibnamefont {Siegel}, \bibfnamefont {M.}},\ }\bibfield  {title} {\enquote
  {\bibinfo {title} {Measurement of dielectric losses in amorphous thin films
  at gigahertz frequencies using superconducting resonators},}\ }\href@noop {}
  {\bibfield  {journal} {\bibinfo  {journal} {Supercond. Sci. Technol.}\
  }\textbf {\bibinfo {volume} {23}},\ \bibinfo {pages} {075008} (\bibinfo
  {year} {2010})}\BibitemShut {NoStop}%
\bibitem [{\citenamefont {Karatsu}\ \emph {et~al.}(2019)\citenamefont
  {Karatsu}, \citenamefont {Endo}, \citenamefont {Bueno}, \citenamefont
  {de~Visser}, \citenamefont {Barends}, \citenamefont {Thoen}, \citenamefont
  {Murugesan}, \citenamefont {Tomita},\ and\ \citenamefont
  {Baselmans}}]{Karatsu2019}%
  \BibitemOpen
  \bibfield  {author} {\bibinfo {author} {\bibnamefont {Karatsu}, \bibfnamefont
  {K.}}, \bibinfo {author} {\bibnamefont {Endo}, \bibfnamefont {A.}}, \bibinfo
  {author} {\bibnamefont {Bueno}, \bibfnamefont {J.}}, \bibinfo {author}
  {\bibnamefont {de~Visser}, \bibfnamefont {P.~J.}}, \bibinfo {author}
  {\bibnamefont {Barends}, \bibfnamefont {R.}}, \bibinfo {author} {\bibnamefont
  {Thoen}, \bibfnamefont {D.~J.}}, \bibinfo {author} {\bibnamefont {Murugesan},
  \bibfnamefont {V.}}, \bibinfo {author} {\bibnamefont {Tomita}, \bibfnamefont
  {N.}}, \ and\ \bibinfo {author} {\bibnamefont {Baselmans}, \bibfnamefont
  {J.~J.~A.}},\ }\bibfield  {title} {\enquote {\bibinfo {title} {Mitigation of
  cosmic ray effect on microwave kinetic inductance detector arrays},}\ }\href
  {\doibase 10.1063/1.5052419} {\bibfield  {journal} {\bibinfo  {journal}
  {Applied Physics Letters}\ }\textbf {\bibinfo {volume} {114}},\ \bibinfo
  {pages} {032601} (\bibinfo {year} {2019})},\ \Eprint
  {http://arxiv.org/abs/https://doi.org/10.1063/1.5052419}
  {https://doi.org/10.1063/1.5052419} \BibitemShut {NoStop}%
\bibitem [{\citenamefont {Khalil}\ \emph {et~al.}(2012)\citenamefont {Khalil},
  \citenamefont {Stoutimore}, \citenamefont {Wellstood},\ and\ \citenamefont
  {Osborn}}]{Khalil2012}%
  \BibitemOpen
  \bibfield  {author} {\bibinfo {author} {\bibnamefont {Khalil}, \bibfnamefont
  {M.}}, \bibinfo {author} {\bibnamefont {Stoutimore}, \bibfnamefont {M.}},
  \bibinfo {author} {\bibnamefont {Wellstood}, \bibfnamefont {F.}}, \ and\
  \bibinfo {author} {\bibnamefont {Osborn}, \bibfnamefont {K.}},\ }\bibfield
  {title} {\enquote {\bibinfo {title} {An analysis method for asymmetric
  resonator transmission applied to superconducting devices},}\ }\href@noop {}
  {\bibfield  {journal} {\bibinfo  {journal} {J. Appl. Phys.}\ }\textbf
  {\bibinfo {volume} {111}},\ \bibinfo {pages} {054510} (\bibinfo {year}
  {2012})}\BibitemShut {NoStop}%
\bibitem [{\citenamefont {Kraus}\ \emph {et~al.}(1986)\citenamefont {Kraus},
  \citenamefont {Tiuri}, \citenamefont {R{\"a}is{\"a}nen},\ and\ \citenamefont
  {Carr}}]{kraus1986}%
  \BibitemOpen
  \bibfield  {author} {\bibinfo {author} {\bibnamefont {Kraus}, \bibfnamefont
  {J.~D.}}, \bibinfo {author} {\bibnamefont {Tiuri}, \bibfnamefont {M.}},
  \bibinfo {author} {\bibnamefont {R{\"a}is{\"a}nen}, \bibfnamefont {A.~V.}}, \
  and\ \bibinfo {author} {\bibnamefont {Carr}, \bibfnamefont {T.~D.}},\
  }\href@noop {} {\emph {\bibinfo {title} {Radio astronomy}}},\ Vol.~\bibinfo
  {volume} {69}\ (\bibinfo  {publisher} {Cygnus-Quasar Books Powell, Ohio},\
  \bibinfo {year} {1986})\BibitemShut {NoStop}%
\bibitem [{\citenamefont {Kreikebaum}\ \emph {et~al.}(2016)\citenamefont
  {Kreikebaum}, \citenamefont {Dove}, \citenamefont {Livingston}, \citenamefont
  {Kim},\ and\ \citenamefont {Siddiqi}}]{kreikebaum2016optimization}%
  \BibitemOpen
  \bibfield  {author} {\bibinfo {author} {\bibnamefont {Kreikebaum},
  \bibfnamefont {J.~M.}}, \bibinfo {author} {\bibnamefont {Dove}, \bibfnamefont
  {A.}}, \bibinfo {author} {\bibnamefont {Livingston}, \bibfnamefont {W.}},
  \bibinfo {author} {\bibnamefont {Kim}, \bibfnamefont {E.}}, \ and\ \bibinfo
  {author} {\bibnamefont {Siddiqi}, \bibfnamefont {I.}},\ }\bibfield  {title}
  {\enquote {\bibinfo {title} {Optimization of infrared and magnetic shielding
  of superconducting tin and al coplanar microwave resonators},}\ }\href@noop
  {} {\bibfield  {journal} {\bibinfo  {journal} {Supercond. Sci. Technol.}\
  }\textbf {\bibinfo {volume} {29}},\ \bibinfo {pages} {104002} (\bibinfo
  {year} {2016})}\BibitemShut {NoStop}%
\bibitem [{\citenamefont {Krinner}\ \emph {et~al.}(2019)\citenamefont
  {Krinner}, \citenamefont {Storz}, \citenamefont {Kurpiers}, \citenamefont
  {Magnard}, \citenamefont {Heinsoo}, \citenamefont {Keller}, \citenamefont
  {Luetolf}, \citenamefont {Eichler},\ and\ \citenamefont
  {Wallraff}}]{krinner2019engineering}%
  \BibitemOpen
  \bibfield  {author} {\bibinfo {author} {\bibnamefont {Krinner}, \bibfnamefont
  {S.}}, \bibinfo {author} {\bibnamefont {Storz}, \bibfnamefont {S.}}, \bibinfo
  {author} {\bibnamefont {Kurpiers}, \bibfnamefont {P.}}, \bibinfo {author}
  {\bibnamefont {Magnard}, \bibfnamefont {P.}}, \bibinfo {author} {\bibnamefont
  {Heinsoo}, \bibfnamefont {J.}}, \bibinfo {author} {\bibnamefont {Keller},
  \bibfnamefont {R.}}, \bibinfo {author} {\bibnamefont {Luetolf}, \bibfnamefont
  {J.}}, \bibinfo {author} {\bibnamefont {Eichler}, \bibfnamefont {C.}}, \ and\
  \bibinfo {author} {\bibnamefont {Wallraff}, \bibfnamefont {A.}},\ }\bibfield
  {title} {\enquote {\bibinfo {title} {Engineering cryogenic setups for
  100-qubit scale superconducting circuit systems},}\ }\href@noop {} {\bibfield
   {journal} {\bibinfo  {journal} {EPJ Quantum Technol.}\ }\textbf {\bibinfo
  {volume} {6}},\ \bibinfo {pages} {2} (\bibinfo {year} {2019})}\BibitemShut
  {NoStop}%
\bibitem [{\citenamefont {Krupka}\ \emph {et~al.}(1999)\citenamefont {Krupka},
  \citenamefont {Derzakowski}, \citenamefont {Abramowicz}, \citenamefont
  {Tobar},\ and\ \citenamefont {Geyer}}]{Krupka1999}%
  \BibitemOpen
  \bibfield  {author} {\bibinfo {author} {\bibnamefont {Krupka}, \bibfnamefont
  {J.}}, \bibinfo {author} {\bibnamefont {Derzakowski}, \bibfnamefont {K.}},
  \bibinfo {author} {\bibnamefont {Abramowicz}, \bibfnamefont {A.}}, \bibinfo
  {author} {\bibnamefont {Tobar}, \bibfnamefont {M.~E.}}, \ and\ \bibinfo
  {author} {\bibnamefont {Geyer}, \bibfnamefont {R.~G.}},\ }\bibfield  {title}
  {\enquote {\bibinfo {title} {{Use of whispering-gallery modes for complex
  permittivity determinations of ultra-low-loss dielectric materials}},}\
  }\href {\doibase 10.1109/22.769347} {\bibfield  {journal} {\bibinfo
  {journal} {Trans. Microw. Theory Tech.}\ }\textbf {\bibinfo {volume} {47}},\
  \bibinfo {pages} {752--759} (\bibinfo {year} {1999})}\BibitemShut {NoStop}%
\bibitem [{\citenamefont {Kumar}\ \emph {et~al.}(2008)\citenamefont {Kumar},
  \citenamefont {Gao}, \citenamefont {Zmuidzinas}, \citenamefont {Mazin},
  \citenamefont {LeDuc},\ and\ \citenamefont {Day}}]{Kumar2008}%
  \BibitemOpen
  \bibfield  {author} {\bibinfo {author} {\bibnamefont {Kumar}, \bibfnamefont
  {S.}}, \bibinfo {author} {\bibnamefont {Gao}, \bibfnamefont {J.}}, \bibinfo
  {author} {\bibnamefont {Zmuidzinas}, \bibfnamefont {J.}}, \bibinfo {author}
  {\bibnamefont {Mazin}, \bibfnamefont {B.~A.}}, \bibinfo {author}
  {\bibnamefont {LeDuc}, \bibfnamefont {H.~G.}}, \ and\ \bibinfo {author}
  {\bibnamefont {Day}, \bibfnamefont {P.~K.}},\ }\bibfield  {title} {\enquote
  {\bibinfo {title} {Temperature dependence of the frequency and noise of
  superconducting coplanar waveguide resonators},}\ }\href@noop {} {\bibfield
  {journal} {\bibinfo  {journal} {Appl. Phys. Lett.}\ }\textbf {\bibinfo
  {volume} {92}},\ \bibinfo {pages} {123503} (\bibinfo {year}
  {2008})}\BibitemShut {NoStop}%
\bibitem [{\citenamefont {Le~Floch}\ \emph {et~al.}(2014)\citenamefont
  {Le~Floch}, \citenamefont {Fan}, \citenamefont {Humbert}, \citenamefont
  {Shan}, \citenamefont {F{\'e}rachou}, \citenamefont {Bara-Maillet},
  \citenamefont {Aubourg}, \citenamefont {Hartnett}, \citenamefont
  {Madrangeas}, \citenamefont {Cros} \emph {et~al.}}]{LeFloch2014}%
  \BibitemOpen
  \bibfield  {author} {\bibinfo {author} {\bibnamefont {Le~Floch},
  \bibfnamefont {J.-M.}}, \bibinfo {author} {\bibnamefont {Fan}, \bibfnamefont
  {Y.}}, \bibinfo {author} {\bibnamefont {Humbert}, \bibfnamefont {G.}},
  \bibinfo {author} {\bibnamefont {Shan}, \bibfnamefont {Q.}}, \bibinfo
  {author} {\bibnamefont {F{\'e}rachou}, \bibfnamefont {D.}}, \bibinfo {author}
  {\bibnamefont {Bara-Maillet}, \bibfnamefont {R.}}, \bibinfo {author}
  {\bibnamefont {Aubourg}, \bibfnamefont {M.}}, \bibinfo {author} {\bibnamefont
  {Hartnett}, \bibfnamefont {J.~G.}}, \bibinfo {author} {\bibnamefont
  {Madrangeas}, \bibfnamefont {V.}}, \bibinfo {author} {\bibnamefont {Cros},
  \bibfnamefont {D.}},  \emph {et~al.},\ }\bibfield  {title} {\enquote
  {\bibinfo {title} {Invited article: Dielectric material characterization
  techniques and designs of high-q resonators for applications from micro to
  millimeter-waves frequencies applicable at room and cryogenic
  temperatures},}\ }\href@noop {} {\bibfield  {journal} {\bibinfo  {journal}
  {Rev. Sci. Instrum.}\ }\textbf {\bibinfo {volume} {85}},\ \bibinfo {pages}
  {031301} (\bibinfo {year} {2014})}\BibitemShut {NoStop}%
\bibitem [{\citenamefont {Lecocq}\ \emph {et~al.}(2017)\citenamefont {Lecocq},
  \citenamefont {Ranzani}, \citenamefont {Peterson}, \citenamefont {Cicak},
  \citenamefont {Simmonds}, \citenamefont {Teufel},\ and\ \citenamefont
  {Aumentado}}]{Lecocq2017}%
  \BibitemOpen
  \bibfield  {author} {\bibinfo {author} {\bibnamefont {Lecocq}, \bibfnamefont
  {F.}}, \bibinfo {author} {\bibnamefont {Ranzani}, \bibfnamefont {L.}},
  \bibinfo {author} {\bibnamefont {Peterson}, \bibfnamefont {G.}}, \bibinfo
  {author} {\bibnamefont {Cicak}, \bibfnamefont {K.}}, \bibinfo {author}
  {\bibnamefont {Simmonds}, \bibfnamefont {R.}}, \bibinfo {author}
  {\bibnamefont {Teufel}, \bibfnamefont {J.}}, \ and\ \bibinfo {author}
  {\bibnamefont {Aumentado}, \bibfnamefont {J.}},\ }\bibfield  {title}
  {\enquote {\bibinfo {title} {Nonreciprocal microwave signal processing with a
  field-programmable josephson amplifier},}\ }\href@noop {} {\bibfield
  {journal} {\bibinfo  {journal} {Phys. Rev. Applied}\ }\textbf {\bibinfo
  {volume} {7}},\ \bibinfo {pages} {024028} (\bibinfo {year}
  {2017})}\BibitemShut {NoStop}%
\bibitem [{\citenamefont {Lei}\ \emph {et~al.}(2020)\citenamefont {Lei},
  \citenamefont {Krayzman}, \citenamefont {Ganjam}, \citenamefont {Frunzio},\
  and\ \citenamefont {Schoelkopf}}]{Lei2020}%
  \BibitemOpen
  \bibfield  {author} {\bibinfo {author} {\bibnamefont {Lei}, \bibfnamefont
  {C.~U.}}, \bibinfo {author} {\bibnamefont {Krayzman}, \bibfnamefont {L.}},
  \bibinfo {author} {\bibnamefont {Ganjam}, \bibfnamefont {S.}}, \bibinfo
  {author} {\bibnamefont {Frunzio}, \bibfnamefont {L.}}, \ and\ \bibinfo
  {author} {\bibnamefont {Schoelkopf}, \bibfnamefont {R.~J.}},\ }\bibfield
  {title} {\enquote {\bibinfo {title} {High coherence superconducting microwave
  cavities with indium bump bonding},}\ }\href@noop {} {\bibfield  {journal}
  {\bibinfo  {journal} {Appl. Phys. Lett.}\ }\textbf {\bibinfo {volume}
  {116}},\ \bibinfo {pages} {154002} (\bibinfo {year} {2020})}\BibitemShut
  {NoStop}%
\bibitem [{\citenamefont {Lewis}, \citenamefont {Henry},\ and\ \citenamefont
  {Schroeder}(2017)}]{Lewis2017}%
  \BibitemOpen
  \bibfield  {author} {\bibinfo {author} {\bibnamefont {Lewis}, \bibfnamefont
  {R.~M.}}, \bibinfo {author} {\bibnamefont {Henry}, \bibfnamefont {M.~D.}}, \
  and\ \bibinfo {author} {\bibnamefont {Schroeder}, \bibfnamefont {K.}},\
  }\bibfield  {title} {\enquote {\bibinfo {title} {Vacuum gap microstrip
  microwave resonators for 2.5-d integration in quantum computing},}\
  }\href@noop {} {\bibfield  {journal} {\bibinfo  {journal} {IEEE Trans. Appl.
  Supercond.}\ }\textbf {\bibinfo {volume} {27}},\ \bibinfo {pages} {1--4}
  (\bibinfo {year} {2017})}\BibitemShut {NoStop}%
\bibitem [{\citenamefont {Li}\ \emph {et~al.}(2013)\citenamefont {Li},
  \citenamefont {Gao}, \citenamefont {Austermann}, \citenamefont {Beall},
  \citenamefont {Becker}, \citenamefont {Cho}, \citenamefont {Fox},
  \citenamefont {Halverson}, \citenamefont {Henning}, \citenamefont {Hilton}
  \emph {et~al.}}]{li2013_siox}%
  \BibitemOpen
  \bibfield  {author} {\bibinfo {author} {\bibnamefont {Li}, \bibfnamefont
  {D.}}, \bibinfo {author} {\bibnamefont {Gao}, \bibfnamefont {J.}}, \bibinfo
  {author} {\bibnamefont {Austermann}, \bibfnamefont {J.}}, \bibinfo {author}
  {\bibnamefont {Beall}, \bibfnamefont {J.}}, \bibinfo {author} {\bibnamefont
  {Becker}, \bibfnamefont {D.}}, \bibinfo {author} {\bibnamefont {Cho},
  \bibfnamefont {H.-M.}}, \bibinfo {author} {\bibnamefont {Fox}, \bibfnamefont
  {A.~E.}}, \bibinfo {author} {\bibnamefont {Halverson}, \bibfnamefont {N.}},
  \bibinfo {author} {\bibnamefont {Henning}, \bibfnamefont {J.}}, \bibinfo
  {author} {\bibnamefont {Hilton}, \bibfnamefont {G.}},  \emph {et~al.},\
  }\bibfield  {title} {\enquote {\bibinfo {title} {Improvements in silicon
  oxide dielectric loss for superconducting microwave detector circuits},}\
  }\href@noop {} {\bibfield  {journal} {\bibinfo  {journal} {IEEE Trans. Appl.
  Supercond.}\ }\textbf {\bibinfo {volume} {23}},\ \bibinfo {pages}
  {1501204--1501204} (\bibinfo {year} {2013})}\BibitemShut {NoStop}%
\bibitem [{\citenamefont {Lienhard}\ \emph {et~al.}(2019)\citenamefont
  {Lienhard}, \citenamefont {Braum{\"u}ller}, \citenamefont {Woods},
  \citenamefont {Rosenberg}, \citenamefont {Calusine}, \citenamefont {Weber},
  \citenamefont {Veps{\"a}l{\"a}inen}, \citenamefont {O'Brien}, \citenamefont
  {Orlando}, \citenamefont {Gustavsson} \emph {et~al.}}]{Lienhard2019}%
  \BibitemOpen
  \bibfield  {author} {\bibinfo {author} {\bibnamefont {Lienhard},
  \bibfnamefont {B.}}, \bibinfo {author} {\bibnamefont {Braum{\"u}ller},
  \bibfnamefont {J.}}, \bibinfo {author} {\bibnamefont {Woods}, \bibfnamefont
  {W.}}, \bibinfo {author} {\bibnamefont {Rosenberg}, \bibfnamefont {D.}},
  \bibinfo {author} {\bibnamefont {Calusine}, \bibfnamefont {G.}}, \bibinfo
  {author} {\bibnamefont {Weber}, \bibfnamefont {S.}}, \bibinfo {author}
  {\bibnamefont {Veps{\"a}l{\"a}inen}, \bibfnamefont {A.}}, \bibinfo {author}
  {\bibnamefont {O'Brien}, \bibfnamefont {K.}}, \bibinfo {author} {\bibnamefont
  {Orlando}, \bibfnamefont {T.~P.}}, \bibinfo {author} {\bibnamefont
  {Gustavsson}, \bibfnamefont {S.}},  \emph {et~al.},\ }\bibfield  {title}
  {\enquote {\bibinfo {title} {Microwave packaging for superconducting
  qubits},}\ }\href@noop {} {\bibfield  {journal} {\bibinfo  {journal} {arXiv
  preprint arXiv:1906.05425}\ } (\bibinfo {year} {2019})}\BibitemShut {NoStop}%
\bibitem [{\citenamefont {Lindstr{\"o}m}\ \emph {et~al.}(2009)\citenamefont
  {Lindstr{\"o}m}, \citenamefont {Healey}, \citenamefont {Colclough},
  \citenamefont {Muirhead},\ and\ \citenamefont
  {Tzalenchuk}}]{Lindstrom_2009_TLS}%
  \BibitemOpen
  \bibfield  {author} {\bibinfo {author} {\bibnamefont {Lindstr{\"o}m},
  \bibfnamefont {T.}}, \bibinfo {author} {\bibnamefont {Healey}, \bibfnamefont
  {J.}}, \bibinfo {author} {\bibnamefont {Colclough}, \bibfnamefont {M.}},
  \bibinfo {author} {\bibnamefont {Muirhead}, \bibfnamefont {C.}}, \ and\
  \bibinfo {author} {\bibnamefont {Tzalenchuk}, \bibfnamefont {A.~Y.}},\
  }\bibfield  {title} {\enquote {\bibinfo {title} {Properties of
  superconducting planar resonators at millikelvin temperatures},}\ }\href@noop
  {} {\bibfield  {journal} {\bibinfo  {journal} {Phys. Rev. B}\ }\textbf
  {\bibinfo {volume} {80}},\ \bibinfo {pages} {132501} (\bibinfo {year}
  {2009})}\BibitemShut {NoStop}%
\bibitem [{\citenamefont {Lock}\ \emph {et~al.}(2019)\citenamefont {Lock},
  \citenamefont {Xu}, \citenamefont {Kohler}, \citenamefont {Camacho},
  \citenamefont {Prestigiacomo}, \citenamefont {Rosen},\ and\ \citenamefont
  {Osborn}}]{Lock2019}%
  \BibitemOpen
  \bibfield  {author} {\bibinfo {author} {\bibnamefont {Lock}, \bibfnamefont
  {E.~H.}}, \bibinfo {author} {\bibnamefont {Xu}, \bibfnamefont {P.}}, \bibinfo
  {author} {\bibnamefont {Kohler}, \bibfnamefont {T.}}, \bibinfo {author}
  {\bibnamefont {Camacho}, \bibfnamefont {L.}}, \bibinfo {author} {\bibnamefont
  {Prestigiacomo}, \bibfnamefont {J.}}, \bibinfo {author} {\bibnamefont
  {Rosen}, \bibfnamefont {Y.~J.}}, \ and\ \bibinfo {author} {\bibnamefont
  {Osborn}, \bibfnamefont {K.~D.}},\ }\bibfield  {title} {\enquote {\bibinfo
  {title} {Using surface engineering to modulate superconducting coplanar
  microwave resonator performance},}\ }\href@noop {} {\bibfield  {journal}
  {\bibinfo  {journal} {IEEE Trans. Appl. Supercond.}\ }\textbf {\bibinfo
  {volume} {29}},\ \bibinfo {pages} {1--8} (\bibinfo {year}
  {2019})}\BibitemShut {NoStop}%
\bibitem [{\citenamefont {Macha}\ \emph {et~al.}(2010)\citenamefont {Macha},
  \citenamefont {van Der~Ploeg}, \citenamefont {Oelsner}, \citenamefont
  {Il’ichev}, \citenamefont {Meyer}, \citenamefont {W{\"u}nsch},\ and\
  \citenamefont {Siegel}}]{Macha2010}%
  \BibitemOpen
  \bibfield  {author} {\bibinfo {author} {\bibnamefont {Macha}, \bibfnamefont
  {P.}}, \bibinfo {author} {\bibnamefont {van Der~Ploeg}, \bibfnamefont {S.}},
  \bibinfo {author} {\bibnamefont {Oelsner}, \bibfnamefont {G.}}, \bibinfo
  {author} {\bibnamefont {Il’ichev}, \bibfnamefont {E.}}, \bibinfo {author}
  {\bibnamefont {Meyer}, \bibfnamefont {H.-G.}}, \bibinfo {author}
  {\bibnamefont {W{\"u}nsch}, \bibfnamefont {S.}}, \ and\ \bibinfo {author}
  {\bibnamefont {Siegel}, \bibfnamefont {M.}},\ }\bibfield  {title} {\enquote
  {\bibinfo {title} {Losses in coplanar waveguide resonators at millikelvin
  temperatures},}\ }\href@noop {} {\bibfield  {journal} {\bibinfo  {journal}
  {Appl. Phys. Lett.}\ }\textbf {\bibinfo {volume} {96}},\ \bibinfo {pages}
  {062503} (\bibinfo {year} {2010})}\BibitemShut {NoStop}%
\bibitem [{\citenamefont {Mariantoni}\ \emph {et~al.}(2011)\citenamefont
  {Mariantoni}, \citenamefont {Wang}, \citenamefont {Yamamoto}, \citenamefont
  {Neeley}, \citenamefont {Bialczak}, \citenamefont {Chen}, \citenamefont
  {Lenander}, \citenamefont {Lucero}, \citenamefont {O’Connell},
  \citenamefont {Sank} \emph {et~al.}}]{Mariantoni2011}%
  \BibitemOpen
  \bibfield  {author} {\bibinfo {author} {\bibnamefont {Mariantoni},
  \bibfnamefont {M.}}, \bibinfo {author} {\bibnamefont {Wang}, \bibfnamefont
  {H.}}, \bibinfo {author} {\bibnamefont {Yamamoto}, \bibfnamefont {T.}},
  \bibinfo {author} {\bibnamefont {Neeley}, \bibfnamefont {M.}}, \bibinfo
  {author} {\bibnamefont {Bialczak}, \bibfnamefont {R.~C.}}, \bibinfo {author}
  {\bibnamefont {Chen}, \bibfnamefont {Y.}}, \bibinfo {author} {\bibnamefont
  {Lenander}, \bibfnamefont {M.}}, \bibinfo {author} {\bibnamefont {Lucero},
  \bibfnamefont {E.}}, \bibinfo {author} {\bibnamefont {O’Connell},
  \bibfnamefont {A.~D.}}, \bibinfo {author} {\bibnamefont {Sank}, \bibfnamefont
  {D.}},  \emph {et~al.},\ }\bibfield  {title} {\enquote {\bibinfo {title}
  {Implementing the quantum von neumann architecture with superconducting
  circuits},}\ }\href@noop {} {\bibfield  {journal} {\bibinfo  {journal}
  {Science}\ }\textbf {\bibinfo {volume} {334}},\ \bibinfo {pages} {61--65}
  (\bibinfo {year} {2011})}\BibitemShut {NoStop}%
\bibitem [{\citenamefont {Martinis}\ \emph {et~al.}(2005)\citenamefont
  {Martinis}, \citenamefont {Cooper}, \citenamefont {McDermott}, \citenamefont
  {Steffen}, \citenamefont {Ansmann}, \citenamefont {Osborn}, \citenamefont
  {Cicak}, \citenamefont {Oh}, \citenamefont {Pappas}, \citenamefont {Simmonds}
  \emph {et~al.}}]{Martinis2005}%
  \BibitemOpen
  \bibfield  {author} {\bibinfo {author} {\bibnamefont {Martinis},
  \bibfnamefont {J.~M.}}, \bibinfo {author} {\bibnamefont {Cooper},
  \bibfnamefont {K.~B.}}, \bibinfo {author} {\bibnamefont {McDermott},
  \bibfnamefont {R.}}, \bibinfo {author} {\bibnamefont {Steffen}, \bibfnamefont
  {M.}}, \bibinfo {author} {\bibnamefont {Ansmann}, \bibfnamefont {M.}},
  \bibinfo {author} {\bibnamefont {Osborn}, \bibfnamefont {K.}}, \bibinfo
  {author} {\bibnamefont {Cicak}, \bibfnamefont {K.}}, \bibinfo {author}
  {\bibnamefont {Oh}, \bibfnamefont {S.}}, \bibinfo {author} {\bibnamefont
  {Pappas}, \bibfnamefont {D.~P.}}, \bibinfo {author} {\bibnamefont {Simmonds},
  \bibfnamefont {R.~W.}},  \emph {et~al.},\ }\bibfield  {title} {\enquote
  {\bibinfo {title} {Decoherence in josephson qubits from dielectric loss},}\
  }\href@noop {} {\bibfield  {journal} {\bibinfo  {journal} {Phys. Rev. Lett.}\
  }\textbf {\bibinfo {volume} {95}},\ \bibinfo {pages} {210503} (\bibinfo
  {year} {2005})}\BibitemShut {NoStop}%
\bibitem [{\citenamefont {Martinis}\ and\ \citenamefont
  {Megrant}(2014)}]{Martinis2014}%
  \BibitemOpen
  \bibfield  {author} {\bibinfo {author} {\bibnamefont {Martinis},
  \bibfnamefont {J.~M.}}\ and\ \bibinfo {author} {\bibnamefont {Megrant},
  \bibfnamefont {A.}},\ }\href@noop {} {\enquote {\bibinfo {title} {{UCSB final
  report for the CSQ program : Review of decoherence and materials physics for
  superconducting qubits}},}\ }\bibinfo {type} {Tech. Rep.}\ (\bibinfo {year}
  {2014})\BibitemShut {NoStop}%
\bibitem [{\citenamefont {Mattis}\ and\ \citenamefont
  {Bardeen}(1958)}]{Mattis1958}%
  \BibitemOpen
  \bibfield  {author} {\bibinfo {author} {\bibnamefont {Mattis}, \bibfnamefont
  {D.}}\ and\ \bibinfo {author} {\bibnamefont {Bardeen}, \bibfnamefont {J.}},\
  }\bibfield  {title} {\enquote {\bibinfo {title} {Theory of the anomalous skin
  effect in normal and superconducting metals},}\ }\href@noop {} {\bibfield
  {journal} {\bibinfo  {journal} {Physical Review}\ }\textbf {\bibinfo {volume}
  {111}},\ \bibinfo {pages} {412} (\bibinfo {year} {1958})}\BibitemShut
  {NoStop}%
\bibitem [{\citenamefont {Mazin}\ \emph {et~al.}(2012)\citenamefont {Mazin},
  \citenamefont {Bumble}, \citenamefont {Meeker}, \citenamefont {O’Brien},
  \citenamefont {McHugh},\ and\ \citenamefont {Langman}}]{Mazin_LE_MKIDArray}%
  \BibitemOpen
  \bibfield  {author} {\bibinfo {author} {\bibnamefont {Mazin}, \bibfnamefont
  {B.~A.}}, \bibinfo {author} {\bibnamefont {Bumble}, \bibfnamefont {B.}},
  \bibinfo {author} {\bibnamefont {Meeker}, \bibfnamefont {S.~R.}}, \bibinfo
  {author} {\bibnamefont {O’Brien}, \bibfnamefont {K.}}, \bibinfo {author}
  {\bibnamefont {McHugh}, \bibfnamefont {S.}}, \ and\ \bibinfo {author}
  {\bibnamefont {Langman}, \bibfnamefont {E.}},\ }\bibfield  {title} {\enquote
  {\bibinfo {title} {A superconducting focal plane array for ultraviolet,
  optical, and near-infrared astrophysics},}\ }\href@noop {} {\bibfield
  {journal} {\bibinfo  {journal} {Opt. Express}\ }\textbf {\bibinfo {volume}
  {20}},\ \bibinfo {pages} {1503--1511} (\bibinfo {year} {2012})}\BibitemShut
  {NoStop}%
\bibitem [{\citenamefont {McConkey}\ \emph {et~al.}(2018)\citenamefont
  {McConkey}, \citenamefont {B{\'e}janin}, \citenamefont {Earnest},
  \citenamefont {McRae}, \citenamefont {Pagel}, \citenamefont {Rinehart},\ and\
  \citenamefont {Mariantoni}}]{Mcconkey2017}%
  \BibitemOpen
  \bibfield  {author} {\bibinfo {author} {\bibnamefont {McConkey},
  \bibfnamefont {T.}}, \bibinfo {author} {\bibnamefont {B{\'e}janin},
  \bibfnamefont {J.}}, \bibinfo {author} {\bibnamefont {Earnest}, \bibfnamefont
  {C.}}, \bibinfo {author} {\bibnamefont {McRae}, \bibfnamefont {C.}}, \bibinfo
  {author} {\bibnamefont {Pagel}, \bibfnamefont {Z.}}, \bibinfo {author}
  {\bibnamefont {Rinehart}, \bibfnamefont {J.}}, \ and\ \bibinfo {author}
  {\bibnamefont {Mariantoni}, \bibfnamefont {M.}},\ }\bibfield  {title}
  {\enquote {\bibinfo {title} {Mitigating leakage errors due to cavity modes in
  a superconducting quantum computer},}\ }\href@noop {} {\bibfield  {journal}
  {\bibinfo  {journal} {Quantum Science and Technology}\ }\textbf {\bibinfo
  {volume} {3}},\ \bibinfo {pages} {034004} (\bibinfo {year}
  {2018})}\BibitemShut {NoStop}%
\bibitem [{\citenamefont {McRae}\ \emph {et~al.}(2018)\citenamefont {McRae},
  \citenamefont {B{\'e}janin}, \citenamefont {Earnest}, \citenamefont
  {McConkey}, \citenamefont {Rinehart}, \citenamefont {Deimert}, \citenamefont
  {Thomas}, \citenamefont {Wasilewski},\ and\ \citenamefont
  {Mariantoni}}]{McRae2018}%
  \BibitemOpen
  \bibfield  {author} {\bibinfo {author} {\bibnamefont {McRae}, \bibfnamefont
  {C.}}, \bibinfo {author} {\bibnamefont {B{\'e}janin}, \bibfnamefont {J.}},
  \bibinfo {author} {\bibnamefont {Earnest}, \bibfnamefont {C.}}, \bibinfo
  {author} {\bibnamefont {McConkey}, \bibfnamefont {T.}}, \bibinfo {author}
  {\bibnamefont {Rinehart}, \bibfnamefont {J.}}, \bibinfo {author}
  {\bibnamefont {Deimert}, \bibfnamefont {C.}}, \bibinfo {author} {\bibnamefont
  {Thomas}, \bibfnamefont {J.}}, \bibinfo {author} {\bibnamefont {Wasilewski},
  \bibfnamefont {Z.}}, \ and\ \bibinfo {author} {\bibnamefont {Mariantoni},
  \bibfnamefont {M.}},\ }\bibfield  {title} {\enquote {\bibinfo {title} {Thin
  film metrology and microwave loss characterization of indium and
  aluminum/indium superconducting planar resonators},}\ }\href@noop {}
  {\bibfield  {journal} {\bibinfo  {journal} {J. Appl. Phys.}\ }\textbf
  {\bibinfo {volume} {123}},\ \bibinfo {pages} {205304} (\bibinfo {year}
  {2018})}\BibitemShut {NoStop}%
\bibitem [{\citenamefont {McRae}\ \emph {et~al.}(2020)\citenamefont {McRae},
  \citenamefont {Lake}, \citenamefont {Long}, \citenamefont {Bal},
  \citenamefont {Wu}, \citenamefont {Jugdersuren}, \citenamefont {Metcalf},
  \citenamefont {Liu},\ and\ \citenamefont {Pappas}}]{McRae2020}%
  \BibitemOpen
  \bibfield  {author} {\bibinfo {author} {\bibnamefont {McRae}, \bibfnamefont
  {C.~R.~H.}}, \bibinfo {author} {\bibnamefont {Lake}, \bibfnamefont {R.~E.}},
  \bibinfo {author} {\bibnamefont {Long}, \bibfnamefont {J.~L.}}, \bibinfo
  {author} {\bibnamefont {Bal}, \bibfnamefont {M.}}, \bibinfo {author}
  {\bibnamefont {Wu}, \bibfnamefont {X.}}, \bibinfo {author} {\bibnamefont
  {Jugdersuren}, \bibfnamefont {B.}}, \bibinfo {author} {\bibnamefont
  {Metcalf}, \bibfnamefont {T.~H.}}, \bibinfo {author} {\bibnamefont {Liu},
  \bibfnamefont {X.}}, \ and\ \bibinfo {author} {\bibnamefont {Pappas},
  \bibfnamefont {D.~P.}},\ }\bibfield  {title} {\enquote {\bibinfo {title}
  {{Dielectric loss extraction for superconducting microwave resonators}},}\
  }\href {http://arxiv.org/abs/1909.07428} {\bibfield  {journal} {\bibinfo
  {journal} {Appl. Phys. Lett.}\ }\textbf {\bibinfo {volume} {116}},\ \bibinfo
  {pages} {194003} (\bibinfo {year} {2020})}\BibitemShut {NoStop}%
\bibitem [{\citenamefont {Megrant}\ \emph {et~al.}(2012)\citenamefont
  {Megrant}, \citenamefont {Neill}, \citenamefont {Barends}, \citenamefont
  {Chiaro}, \citenamefont {Chen}, \citenamefont {Feigl}, \citenamefont {Kelly},
  \citenamefont {Lucero}, \citenamefont {Mariantoni}, \citenamefont
  {O’Malley} \emph {et~al.}}]{Megrant2012}%
  \BibitemOpen
  \bibfield  {author} {\bibinfo {author} {\bibnamefont {Megrant}, \bibfnamefont
  {A.}}, \bibinfo {author} {\bibnamefont {Neill}, \bibfnamefont {C.}}, \bibinfo
  {author} {\bibnamefont {Barends}, \bibfnamefont {R.}}, \bibinfo {author}
  {\bibnamefont {Chiaro}, \bibfnamefont {B.}}, \bibinfo {author} {\bibnamefont
  {Chen}, \bibfnamefont {Y.}}, \bibinfo {author} {\bibnamefont {Feigl},
  \bibfnamefont {L.}}, \bibinfo {author} {\bibnamefont {Kelly}, \bibfnamefont
  {J.}}, \bibinfo {author} {\bibnamefont {Lucero}, \bibfnamefont {E.}},
  \bibinfo {author} {\bibnamefont {Mariantoni}, \bibfnamefont {M.}}, \bibinfo
  {author} {\bibnamefont {O’Malley}, \bibfnamefont {P.~J.}},  \emph
  {et~al.},\ }\bibfield  {title} {\enquote {\bibinfo {title} {Planar
  superconducting resonators with internal quality factors above one
  million},}\ }\href@noop {} {\bibfield  {journal} {\bibinfo  {journal} {Appl.
  Phys. Lett.}\ }\textbf {\bibinfo {volume} {100}},\ \bibinfo {pages} {113510}
  (\bibinfo {year} {2012})}\BibitemShut {NoStop}%
\bibitem [{\citenamefont {Minev}, \citenamefont {Pop},\ and\ \citenamefont
  {Devoret}(2013)}]{Minev2013}%
  \BibitemOpen
  \bibfield  {author} {\bibinfo {author} {\bibnamefont {Minev}, \bibfnamefont
  {Z.}}, \bibinfo {author} {\bibnamefont {Pop}, \bibfnamefont {I.}}, \ and\
  \bibinfo {author} {\bibnamefont {Devoret}, \bibfnamefont {M.}},\ }\bibfield
  {title} {\enquote {\bibinfo {title} {Planar superconducting whispering
  gallery mode resonators},}\ }\href@noop {} {\bibfield  {journal} {\bibinfo
  {journal} {Appl. Phys. Lett.}\ }\textbf {\bibinfo {volume} {103}},\ \bibinfo
  {pages} {142604} (\bibinfo {year} {2013})}\BibitemShut {NoStop}%
\bibitem [{\citenamefont {M{\"u}ller}, \citenamefont {Cole},\ and\
  \citenamefont {Lisenfeld}(2019)}]{Muller2019}%
  \BibitemOpen
  \bibfield  {author} {\bibinfo {author} {\bibnamefont {M{\"u}ller},
  \bibfnamefont {C.}}, \bibinfo {author} {\bibnamefont {Cole}, \bibfnamefont
  {J.~H.}}, \ and\ \bibinfo {author} {\bibnamefont {Lisenfeld}, \bibfnamefont
  {J.}},\ }\bibfield  {title} {\enquote {\bibinfo {title} {Towards
  understanding two-level-systems in amorphous solids: insights from quantum
  circuits},}\ }\href@noop {} {\bibfield  {journal} {\bibinfo  {journal} {Rep.
  Prog. Phys.}\ }\textbf {\bibinfo {volume} {82}},\ \bibinfo {pages} {124501}
  (\bibinfo {year} {2019})}\BibitemShut {NoStop}%
\bibitem [{\citenamefont {Neill}\ \emph {et~al.}(2013)\citenamefont {Neill},
  \citenamefont {Megrant}, \citenamefont {Barends}, \citenamefont {Chen},
  \citenamefont {Chiaro}, \citenamefont {Kelly}, \citenamefont {Mutus},
  \citenamefont {O'Malley}, \citenamefont {Sank}, \citenamefont {Wenner} \emph
  {et~al.}}]{neill2013fluctuations}%
  \BibitemOpen
  \bibfield  {author} {\bibinfo {author} {\bibnamefont {Neill}, \bibfnamefont
  {C.}}, \bibinfo {author} {\bibnamefont {Megrant}, \bibfnamefont {A.}},
  \bibinfo {author} {\bibnamefont {Barends}, \bibfnamefont {R.}}, \bibinfo
  {author} {\bibnamefont {Chen}, \bibfnamefont {Y.}}, \bibinfo {author}
  {\bibnamefont {Chiaro}, \bibfnamefont {B.}}, \bibinfo {author} {\bibnamefont
  {Kelly}, \bibfnamefont {J.}}, \bibinfo {author} {\bibnamefont {Mutus},
  \bibfnamefont {J.}}, \bibinfo {author} {\bibnamefont {O'Malley},
  \bibfnamefont {P.}}, \bibinfo {author} {\bibnamefont {Sank}, \bibfnamefont
  {D.}}, \bibinfo {author} {\bibnamefont {Wenner}, \bibfnamefont {J.}},  \emph
  {et~al.},\ }\bibfield  {title} {\enquote {\bibinfo {title} {Fluctuations from
  edge defects in superconducting resonators},}\ }\href@noop {} {\bibfield
  {journal} {\bibinfo  {journal} {Applied Physics Letters}\ }\textbf {\bibinfo
  {volume} {103}},\ \bibinfo {pages} {072601} (\bibinfo {year}
  {2013})}\BibitemShut {NoStop}%
\bibitem [{\citenamefont {Nersisyan}\ \emph {et~al.}(2019)\citenamefont
  {Nersisyan}, \citenamefont {Poletto}, \citenamefont {Alidoust}, \citenamefont
  {Manenti}, \citenamefont {Renzas}, \citenamefont {Bui}, \citenamefont {Vu},
  \citenamefont {Whyland}, \citenamefont {Mohan}, \citenamefont {Sete} \emph
  {et~al.}}]{Nersisyan2019}%
  \BibitemOpen
  \bibfield  {author} {\bibinfo {author} {\bibnamefont {Nersisyan},
  \bibfnamefont {A.}}, \bibinfo {author} {\bibnamefont {Poletto}, \bibfnamefont
  {S.}}, \bibinfo {author} {\bibnamefont {Alidoust}, \bibfnamefont {N.}},
  \bibinfo {author} {\bibnamefont {Manenti}, \bibfnamefont {R.}}, \bibinfo
  {author} {\bibnamefont {Renzas}, \bibfnamefont {R.}}, \bibinfo {author}
  {\bibnamefont {Bui}, \bibfnamefont {C.-V.}}, \bibinfo {author} {\bibnamefont
  {Vu}, \bibfnamefont {K.}}, \bibinfo {author} {\bibnamefont {Whyland},
  \bibfnamefont {T.}}, \bibinfo {author} {\bibnamefont {Mohan}, \bibfnamefont
  {Y.}}, \bibinfo {author} {\bibnamefont {Sete}, \bibfnamefont {E.~A.}},  \emph
  {et~al.},\ }\bibfield  {title} {\enquote {\bibinfo {title} {Manufacturing low
  dissipation superconducting quantum processors},}\ }\href@noop {} {\bibfield
  {journal} {\bibinfo  {journal} {arXiv preprint arXiv:1901.08042}\ } (\bibinfo
  {year} {2019})}\BibitemShut {NoStop}%
\bibitem [{\citenamefont {Nguyen}\ \emph {et~al.}(2019)\citenamefont {Nguyen},
  \citenamefont {Lin}, \citenamefont {Somoroff}, \citenamefont {Mencia},
  \citenamefont {Grabon},\ and\ \citenamefont {Manucharyan}}]{Nuyen2019}%
  \BibitemOpen
  \bibfield  {author} {\bibinfo {author} {\bibnamefont {Nguyen}, \bibfnamefont
  {L.~B.}}, \bibinfo {author} {\bibnamefont {Lin}, \bibfnamefont {Y.-H.}},
  \bibinfo {author} {\bibnamefont {Somoroff}, \bibfnamefont {A.}}, \bibinfo
  {author} {\bibnamefont {Mencia}, \bibfnamefont {R.}}, \bibinfo {author}
  {\bibnamefont {Grabon}, \bibfnamefont {N.}}, \ and\ \bibinfo {author}
  {\bibnamefont {Manucharyan}, \bibfnamefont {V.~E.}},\ }\bibfield  {title}
  {\enquote {\bibinfo {title} {High-coherence fluxonium qubit},}\ }\href
  {\doibase 10.1103/PhysRevX.9.041041} {\bibfield  {journal} {\bibinfo
  {journal} {Phys. Rev. X}\ }\textbf {\bibinfo {volume} {9}},\ \bibinfo {pages}
  {041041} (\bibinfo {year} {2019})}\BibitemShut {NoStop}%
\bibitem [{\citenamefont {Niepce}\ \emph {et~al.}(2020)\citenamefont {Niepce},
  \citenamefont {Burnett}, \citenamefont {Latorre},\ and\ \citenamefont
  {Bylander}}]{Niepce_2020}%
  \BibitemOpen
  \bibfield  {author} {\bibinfo {author} {\bibnamefont {Niepce}, \bibfnamefont
  {D.}}, \bibinfo {author} {\bibnamefont {Burnett}, \bibfnamefont {J.~J.}},
  \bibinfo {author} {\bibnamefont {Latorre}, \bibfnamefont {M.~G.}}, \ and\
  \bibinfo {author} {\bibnamefont {Bylander}, \bibfnamefont {J.}},\ }\bibfield
  {title} {\enquote {\bibinfo {title} {Geometric scaling of two-level-system
  loss in superconducting resonators},}\ }\href@noop {} {\bibfield  {journal}
  {\bibinfo  {journal} {Supercond. Sci. Technol.}\ }\textbf {\bibinfo {volume}
  {33}},\ \bibinfo {pages} {025013} (\bibinfo {year} {2020})}\BibitemShut
  {NoStop}%
\bibitem [{\citenamefont {Nsanzineza}(2016)}]{nsanzineza2016}%
  \BibitemOpen
  \bibfield  {author} {\bibinfo {author} {\bibnamefont {Nsanzineza},
  \bibfnamefont {I.}},\ }\emph {\bibinfo {title} {{Vortices and Quasiparticles
  in Superconducting Microwave Resonators}}},\ \href@noop {} {Ph.D. thesis},\
  \bibinfo  {school} {Syracuse University} (\bibinfo {year} {2016})\BibitemShut
  {NoStop}%
\bibitem [{\citenamefont {Nsanzineza}\ and\ \citenamefont
  {Plourde}(2014)}]{nsanzineza2014trapping}%
  \BibitemOpen
  \bibfield  {author} {\bibinfo {author} {\bibnamefont {Nsanzineza},
  \bibfnamefont {I.}}\ and\ \bibinfo {author} {\bibnamefont {Plourde},
  \bibfnamefont {B.}},\ }\bibfield  {title} {\enquote {\bibinfo {title}
  {Trapping a single vortex and reducing quasiparticles in a superconducting
  resonator},}\ }\href@noop {} {\bibfield  {journal} {\bibinfo  {journal}
  {Phys. Rev. Lett.}\ }\textbf {\bibinfo {volume} {113}},\ \bibinfo {pages}
  {117002} (\bibinfo {year} {2014})}\BibitemShut {NoStop}%
\bibitem [{\citenamefont {Ohya}\ \emph {et~al.}(2013)\citenamefont {Ohya},
  \citenamefont {Chiaro}, \citenamefont {Megrant}, \citenamefont {Neill},
  \citenamefont {Barends}, \citenamefont {Chen}, \citenamefont {Kelly},
  \citenamefont {Low}, \citenamefont {Mutus}, \citenamefont {O’Malley} \emph
  {et~al.}}]{Ohya2014}%
  \BibitemOpen
  \bibfield  {author} {\bibinfo {author} {\bibnamefont {Ohya}, \bibfnamefont
  {S.}}, \bibinfo {author} {\bibnamefont {Chiaro}, \bibfnamefont {B.}},
  \bibinfo {author} {\bibnamefont {Megrant}, \bibfnamefont {A.}}, \bibinfo
  {author} {\bibnamefont {Neill}, \bibfnamefont {C.}}, \bibinfo {author}
  {\bibnamefont {Barends}, \bibfnamefont {R.}}, \bibinfo {author} {\bibnamefont
  {Chen}, \bibfnamefont {Y.}}, \bibinfo {author} {\bibnamefont {Kelly},
  \bibfnamefont {J.}}, \bibinfo {author} {\bibnamefont {Low}, \bibfnamefont
  {D.}}, \bibinfo {author} {\bibnamefont {Mutus}, \bibfnamefont {J.}}, \bibinfo
  {author} {\bibnamefont {O’Malley}, \bibfnamefont {P.}},  \emph {et~al.},\
  }\bibfield  {title} {\enquote {\bibinfo {title} {Room temperature deposition
  of sputtered tin films for superconducting coplanar waveguide resonators},}\
  }\href@noop {} {\bibfield  {journal} {\bibinfo  {journal} {Supercond. Sci.
  Technol.}\ }\textbf {\bibinfo {volume} {27}},\ \bibinfo {pages} {015009}
  (\bibinfo {year} {2013})}\BibitemShut {NoStop}%
\bibitem [{\citenamefont {O’Connell}\ \emph {et~al.}(2008)\citenamefont
  {O’Connell}, \citenamefont {Ansmann}, \citenamefont {Bialczak},
  \citenamefont {Hofheinz}, \citenamefont {Katz}, \citenamefont {Lucero},
  \citenamefont {McKenney}, \citenamefont {Neeley}, \citenamefont {Wang},
  \citenamefont {Weig} \emph {et~al.}}]{OConnell2008}%
  \BibitemOpen
  \bibfield  {author} {\bibinfo {author} {\bibnamefont {O’Connell},
  \bibfnamefont {A.~D.}}, \bibinfo {author} {\bibnamefont {Ansmann},
  \bibfnamefont {M.}}, \bibinfo {author} {\bibnamefont {Bialczak},
  \bibfnamefont {R.~C.}}, \bibinfo {author} {\bibnamefont {Hofheinz},
  \bibfnamefont {M.}}, \bibinfo {author} {\bibnamefont {Katz}, \bibfnamefont
  {N.}}, \bibinfo {author} {\bibnamefont {Lucero}, \bibfnamefont {E.}},
  \bibinfo {author} {\bibnamefont {McKenney}, \bibfnamefont {C.}}, \bibinfo
  {author} {\bibnamefont {Neeley}, \bibfnamefont {M.}}, \bibinfo {author}
  {\bibnamefont {Wang}, \bibfnamefont {H.}}, \bibinfo {author} {\bibnamefont
  {Weig}, \bibfnamefont {E.~M.}},  \emph {et~al.},\ }\bibfield  {title}
  {\enquote {\bibinfo {title} {Microwave dielectric loss at single photon
  energies and millikelvin temperatures},}\ }\href@noop {} {\bibfield
  {journal} {\bibinfo  {journal} {Appl. Phys. Lett.}\ }\textbf {\bibinfo
  {volume} {92}},\ \bibinfo {pages} {112903} (\bibinfo {year}
  {2008})}\BibitemShut {NoStop}%
\bibitem [{\citenamefont {Paik}\ and\ \citenamefont {Osborn}(2010)}]{Paik2010}%
  \BibitemOpen
  \bibfield  {author} {\bibinfo {author} {\bibnamefont {Paik}, \bibfnamefont
  {H.}}\ and\ \bibinfo {author} {\bibnamefont {Osborn}, \bibfnamefont
  {K.~D.}},\ }\bibfield  {title} {\enquote {\bibinfo {title} {Reducing
  quantum-regime dielectric loss of silicon nitride for superconducting quantum
  circuits},}\ }\href@noop {} {\bibfield  {journal} {\bibinfo  {journal} {Appl.
  Phys. Lett.}\ }\textbf {\bibinfo {volume} {96}},\ \bibinfo {pages} {072505}
  (\bibinfo {year} {2010})}\BibitemShut {NoStop}%
\bibitem [{\citenamefont {Paik}\ \emph {et~al.}(2011)\citenamefont {Paik},
  \citenamefont {Schuster}, \citenamefont {Bishop}, \citenamefont {Kirchmair},
  \citenamefont {Catelani}, \citenamefont {Sears}, \citenamefont {Johnson},
  \citenamefont {Reagor}, \citenamefont {Frunzio}, \citenamefont {Glazman}
  \emph {et~al.}}]{Paik_PhysRevLett.107.240501}%
  \BibitemOpen
  \bibfield  {author} {\bibinfo {author} {\bibnamefont {Paik}, \bibfnamefont
  {H.}}, \bibinfo {author} {\bibnamefont {Schuster}, \bibfnamefont {D.}},
  \bibinfo {author} {\bibnamefont {Bishop}, \bibfnamefont {L.~S.}}, \bibinfo
  {author} {\bibnamefont {Kirchmair}, \bibfnamefont {G.}}, \bibinfo {author}
  {\bibnamefont {Catelani}, \bibfnamefont {G.}}, \bibinfo {author}
  {\bibnamefont {Sears}, \bibfnamefont {A.}}, \bibinfo {author} {\bibnamefont
  {Johnson}, \bibfnamefont {B.}}, \bibinfo {author} {\bibnamefont {Reagor},
  \bibfnamefont {M.}}, \bibinfo {author} {\bibnamefont {Frunzio}, \bibfnamefont
  {L.}}, \bibinfo {author} {\bibnamefont {Glazman}, \bibfnamefont {L.}},  \emph
  {et~al.},\ }\bibfield  {title} {\enquote {\bibinfo {title} {Observation of
  high coherence in josephson junction qubits measured in a three-dimensional
  circuit qed architecture},}\ }\href@noop {} {\bibfield  {journal} {\bibinfo
  {journal} {Phys. Rev. Lett.}\ }\textbf {\bibinfo {volume} {107}},\ \bibinfo
  {pages} {240501} (\bibinfo {year} {2011})}\BibitemShut {NoStop}%
\bibitem [{\citenamefont {Pappas}\ \emph {et~al.}(2011)\citenamefont {Pappas},
  \citenamefont {Vissers}, \citenamefont {Wisbey}, \citenamefont {Kline},\ and\
  \citenamefont {Gao}}]{Pappas2011}%
  \BibitemOpen
  \bibfield  {author} {\bibinfo {author} {\bibnamefont {Pappas}, \bibfnamefont
  {D.~P.}}, \bibinfo {author} {\bibnamefont {Vissers}, \bibfnamefont {M.~R.}},
  \bibinfo {author} {\bibnamefont {Wisbey}, \bibfnamefont {D.~S.}}, \bibinfo
  {author} {\bibnamefont {Kline}, \bibfnamefont {J.~S.}}, \ and\ \bibinfo
  {author} {\bibnamefont {Gao}, \bibfnamefont {J.}},\ }\bibfield  {title}
  {\enquote {\bibinfo {title} {Two level system loss in superconducting
  microwave resonators},}\ }\href@noop {} {\bibfield  {journal} {\bibinfo
  {journal} {IEEE Trans. Appl. Supercond.}\ }\textbf {\bibinfo {volume} {21}},\
  \bibinfo {pages} {871--874} (\bibinfo {year} {2011})}\BibitemShut {NoStop}%
\bibitem [{\citenamefont {Partanen}\ \emph {et~al.}(2016)\citenamefont
  {Partanen}, \citenamefont {Tan}, \citenamefont {Govenius}, \citenamefont
  {Lake}, \citenamefont {M{\"a}kel{\"a}}, \citenamefont {Tanttu},\ and\
  \citenamefont {M{\"o}tt{\"o}nen}}]{Partanen2016}%
  \BibitemOpen
  \bibfield  {author} {\bibinfo {author} {\bibnamefont {Partanen},
  \bibfnamefont {M.}}, \bibinfo {author} {\bibnamefont {Tan}, \bibfnamefont
  {K.~Y.}}, \bibinfo {author} {\bibnamefont {Govenius}, \bibfnamefont {J.}},
  \bibinfo {author} {\bibnamefont {Lake}, \bibfnamefont {R.~E.}}, \bibinfo
  {author} {\bibnamefont {M{\"a}kel{\"a}}, \bibfnamefont {M.~K.}}, \bibinfo
  {author} {\bibnamefont {Tanttu}, \bibfnamefont {T.}}, \ and\ \bibinfo
  {author} {\bibnamefont {M{\"o}tt{\"o}nen}, \bibfnamefont {M.}},\ }\bibfield
  {title} {\enquote {\bibinfo {title} {Quantum-limited heat conduction over
  macroscopic distances},}\ }\href@noop {} {\bibfield  {journal} {\bibinfo
  {journal} {Nat. Phys.}\ }\textbf {\bibinfo {volume} {12}},\ \bibinfo {pages}
  {460--464} (\bibinfo {year} {2016})}\BibitemShut {NoStop}%
\bibitem [{\citenamefont {Petersan}\ and\ \citenamefont
  {Anlage}(1998)}]{petersan1998measurement}%
  \BibitemOpen
  \bibfield  {author} {\bibinfo {author} {\bibnamefont {Petersan},
  \bibfnamefont {P.~J.}}\ and\ \bibinfo {author} {\bibnamefont {Anlage},
  \bibfnamefont {S.~M.}},\ }\bibfield  {title} {\enquote {\bibinfo {title}
  {Measurement of resonant frequency and quality factor of microwave
  resonators: Comparison of methods},}\ }\href@noop {} {\bibfield  {journal}
  {\bibinfo  {journal} {J. Appl. Phys.}\ }\textbf {\bibinfo {volume} {84}},\
  \bibinfo {pages} {3392--3402} (\bibinfo {year} {1998})}\BibitemShut {NoStop}%
\bibitem [{\citenamefont {Phillips}(1987)}]{Phillips}%
  \BibitemOpen
  \bibfield  {author} {\bibinfo {author} {\bibnamefont {Phillips},
  \bibfnamefont {W.}},\ }\bibfield  {title} {\enquote {\bibinfo {title}
  {Two-level states in glasses},}\ }\href@noop {} {\bibfield  {journal}
  {\bibinfo  {journal} {Rep. Prog. Phys.}\ }\textbf {\bibinfo {volume} {50}},\
  \bibinfo {pages} {1657} (\bibinfo {year} {1987})}\BibitemShut {NoStop}%
\bibitem [{\citenamefont {Place}\ \emph {et~al.}(2020)\citenamefont {Place},
  \citenamefont {Rodgers}, \citenamefont {Mundada}, \citenamefont {Smitham},
  \citenamefont {Fitzpatrick}, \citenamefont {Leng}, \citenamefont {Premkumar},
  \citenamefont {Bryon}, \citenamefont {Sussman}, \citenamefont {Cheng},
  \citenamefont {Madhavan}, \citenamefont {Babla}, \citenamefont {Jaeck},
  \citenamefont {Gyenis}, \citenamefont {Yao}, \citenamefont {Cava},
  \citenamefont {de~Leon},\ and\ \citenamefont {Houck}}]{Place2020}%
  \BibitemOpen
  \bibfield  {author} {\bibinfo {author} {\bibnamefont {Place}, \bibfnamefont
  {A.~P.~M.}}, \bibinfo {author} {\bibnamefont {Rodgers}, \bibfnamefont
  {L.~V.~H.}}, \bibinfo {author} {\bibnamefont {Mundada}, \bibfnamefont {P.}},
  \bibinfo {author} {\bibnamefont {Smitham}, \bibfnamefont {B.~M.}}, \bibinfo
  {author} {\bibnamefont {Fitzpatrick}, \bibfnamefont {M.}}, \bibinfo {author}
  {\bibnamefont {Leng}, \bibfnamefont {Z.}}, \bibinfo {author} {\bibnamefont
  {Premkumar}, \bibfnamefont {A.}}, \bibinfo {author} {\bibnamefont {Bryon},
  \bibfnamefont {J.}}, \bibinfo {author} {\bibnamefont {Sussman}, \bibfnamefont
  {S.}}, \bibinfo {author} {\bibnamefont {Cheng}, \bibfnamefont {G.}}, \bibinfo
  {author} {\bibnamefont {Madhavan}, \bibfnamefont {T.}}, \bibinfo {author}
  {\bibnamefont {Babla}, \bibfnamefont {H.~K.}}, \bibinfo {author}
  {\bibnamefont {Jaeck}, \bibfnamefont {B.}}, \bibinfo {author} {\bibnamefont
  {Gyenis}, \bibfnamefont {A.}}, \bibinfo {author} {\bibnamefont {Yao},
  \bibfnamefont {N.}}, \bibinfo {author} {\bibnamefont {Cava}, \bibfnamefont
  {R.~J.}}, \bibinfo {author} {\bibnamefont {de~Leon}, \bibfnamefont {N.~P.}},
  \ and\ \bibinfo {author} {\bibnamefont {Houck}, \bibfnamefont {A.~A.}},\
  }\bibfield  {title} {\enquote {\bibinfo {title} {{New material platform for
  superconducting transmon qubits with coherence times exceeding 0.3
  milliseconds}},}\ }\href@noop {} {\  (\bibinfo {year} {2020})},\ \Eprint
  {http://arxiv.org/abs/2003.00024} {arXiv:2003.00024} \BibitemShut {NoStop}%
\bibitem [{\citenamefont {Pogue}\ \emph {et~al.}(2012)\citenamefont {Pogue},
  \citenamefont {McIntyre}, \citenamefont {Sattarov},\ and\ \citenamefont
  {Reece}}]{Pogue2012}%
  \BibitemOpen
  \bibfield  {author} {\bibinfo {author} {\bibnamefont {Pogue}, \bibfnamefont
  {N.}}, \bibinfo {author} {\bibnamefont {McIntyre}, \bibfnamefont {P.}},
  \bibinfo {author} {\bibnamefont {Sattarov}, \bibfnamefont {A.}}, \ and\
  \bibinfo {author} {\bibnamefont {Reece}, \bibfnamefont {C.}},\ }\bibfield
  {title} {\enquote {\bibinfo {title} {Measurement of the dielectric properties
  of high-purity sapphire at 1.865 ghz from 2-10 kelvin},}\ }\href {\doibase
  10.1063/1.4707011} {\bibfield  {journal} {\bibinfo  {journal} {AIP Conf
  Proc.}\ }\textbf {\bibinfo {volume} {1434}},\ \bibinfo {pages} {945--952}
  (\bibinfo {year} {2012})},\ \Eprint
  {http://arxiv.org/abs/https://aip.scitation.org/doi/pdf/10.1063/1.4707011}
  {https://aip.scitation.org/doi/pdf/10.1063/1.4707011} \BibitemShut {NoStop}%
\bibitem [{\citenamefont {Pozar}(2012)}]{pozar}%
  \BibitemOpen
  \bibfield  {author} {\bibinfo {author} {\bibnamefont {Pozar}, \bibfnamefont
  {D.~M.}},\ }\href@noop {} {\emph {\bibinfo {title} {{Microwave
  Engineering}}}},\ \bibinfo {edition} {4th}\ ed.\ (\bibinfo  {publisher} {John
  Wiley {\&} Sons},\ \bibinfo {year} {2012})\BibitemShut {NoStop}%
\bibitem [{\citenamefont {Probst}\ \emph {et~al.}(2015)\citenamefont {Probst},
  \citenamefont {Song}, \citenamefont {Bushev}, \citenamefont {Ustinov},\ and\
  \citenamefont {Weides}}]{Probst2015}%
  \BibitemOpen
  \bibfield  {author} {\bibinfo {author} {\bibnamefont {Probst}, \bibfnamefont
  {S.}}, \bibinfo {author} {\bibnamefont {Song}, \bibfnamefont {F.}}, \bibinfo
  {author} {\bibnamefont {Bushev}, \bibfnamefont {P.}}, \bibinfo {author}
  {\bibnamefont {Ustinov}, \bibfnamefont {A.}}, \ and\ \bibinfo {author}
  {\bibnamefont {Weides}, \bibfnamefont {M.}},\ }\bibfield  {title} {\enquote
  {\bibinfo {title} {Efficient and robust analysis of complex scattering data
  under noise in microwave resonators},}\ }\href@noop {} {\bibfield  {journal}
  {\bibinfo  {journal} {Rev. Sci. Instrum.}\ }\textbf {\bibinfo {volume}
  {86}},\ \bibinfo {pages} {024706} (\bibinfo {year} {2015})}\BibitemShut
  {NoStop}%
\bibitem [{\citenamefont {Quintana}\ \emph {et~al.}(2014)\citenamefont
  {Quintana}, \citenamefont {Megrant}, \citenamefont {Chen}, \citenamefont
  {Dunsworth}, \citenamefont {Chiaro}, \citenamefont {Barends}, \citenamefont
  {Campbell}, \citenamefont {Chen}, \citenamefont {Hoi}, \citenamefont
  {Jeffrey} \emph {et~al.}}]{Quintana2014}%
  \BibitemOpen
  \bibfield  {author} {\bibinfo {author} {\bibnamefont {Quintana},
  \bibfnamefont {C.}}, \bibinfo {author} {\bibnamefont {Megrant}, \bibfnamefont
  {A.}}, \bibinfo {author} {\bibnamefont {Chen}, \bibfnamefont {Z.}}, \bibinfo
  {author} {\bibnamefont {Dunsworth}, \bibfnamefont {A.}}, \bibinfo {author}
  {\bibnamefont {Chiaro}, \bibfnamefont {B.}}, \bibinfo {author} {\bibnamefont
  {Barends}, \bibfnamefont {R.}}, \bibinfo {author} {\bibnamefont {Campbell},
  \bibfnamefont {B.}}, \bibinfo {author} {\bibnamefont {Chen}, \bibfnamefont
  {Y.}}, \bibinfo {author} {\bibnamefont {Hoi}, \bibfnamefont {I.-C.}},
  \bibinfo {author} {\bibnamefont {Jeffrey}, \bibfnamefont {E.}},  \emph
  {et~al.},\ }\bibfield  {title} {\enquote {\bibinfo {title} {Characterization
  and reduction of microfabrication-induced decoherence in superconducting
  quantum circuits},}\ }\href@noop {} {\bibfield  {journal} {\bibinfo
  {journal} {Appl. Phys. Lett.}\ }\textbf {\bibinfo {volume} {105}},\ \bibinfo
  {pages} {062601} (\bibinfo {year} {2014})}\BibitemShut {NoStop}%
\bibitem [{\citenamefont {Ranzani}\ \emph {et~al.}(2013)\citenamefont
  {Ranzani}, \citenamefont {Spietz}, \citenamefont {Popovic},\ and\
  \citenamefont {Aumentado}}]{ranzani2013two}%
  \BibitemOpen
  \bibfield  {author} {\bibinfo {author} {\bibnamefont {Ranzani}, \bibfnamefont
  {L.}}, \bibinfo {author} {\bibnamefont {Spietz}, \bibfnamefont {L.}},
  \bibinfo {author} {\bibnamefont {Popovic}, \bibfnamefont {Z.}}, \ and\
  \bibinfo {author} {\bibnamefont {Aumentado}, \bibfnamefont {J.}},\ }\bibfield
   {title} {\enquote {\bibinfo {title} {Two-port microwave calibration at
  millikelvin temperatures},}\ }\href@noop {} {\bibfield  {journal} {\bibinfo
  {journal} {Rev. Sci. Instrum.}\ }\textbf {\bibinfo {volume} {84}},\ \bibinfo
  {pages} {034704} (\bibinfo {year} {2013})}\BibitemShut {NoStop}%
\bibitem [{\citenamefont {Reagor}(2015)}]{ReagorThesis}%
  \BibitemOpen
  \bibfield  {author} {\bibinfo {author} {\bibnamefont {Reagor}, \bibfnamefont
  {M.}},\ }\emph {\bibinfo {title} {{Superconducting Cavities for Circuit
  Quantum Electrodynamics}}},\ \href@noop {} {Ph.D. thesis},\ \bibinfo
  {school} {Yale University} (\bibinfo {year} {2015})\BibitemShut {NoStop}%
\bibitem [{\citenamefont {Reagor}\ \emph {et~al.}(2013)\citenamefont {Reagor},
  \citenamefont {Paik}, \citenamefont {Catelani}, \citenamefont {Sun},
  \citenamefont {Axline}, \citenamefont {Holland}, \citenamefont {Pop},
  \citenamefont {Masluk}, \citenamefont {Brecht}, \citenamefont {Frunzio} \emph
  {et~al.}}]{Reagor_10ms3D}%
  \BibitemOpen
  \bibfield  {author} {\bibinfo {author} {\bibnamefont {Reagor}, \bibfnamefont
  {M.}}, \bibinfo {author} {\bibnamefont {Paik}, \bibfnamefont {H.}}, \bibinfo
  {author} {\bibnamefont {Catelani}, \bibfnamefont {G.}}, \bibinfo {author}
  {\bibnamefont {Sun}, \bibfnamefont {L.}}, \bibinfo {author} {\bibnamefont
  {Axline}, \bibfnamefont {C.}}, \bibinfo {author} {\bibnamefont {Holland},
  \bibfnamefont {E.}}, \bibinfo {author} {\bibnamefont {Pop}, \bibfnamefont
  {I.~M.}}, \bibinfo {author} {\bibnamefont {Masluk}, \bibfnamefont {N.~A.}},
  \bibinfo {author} {\bibnamefont {Brecht}, \bibfnamefont {T.}}, \bibinfo
  {author} {\bibnamefont {Frunzio}, \bibfnamefont {L.}},  \emph {et~al.},\
  }\bibfield  {title} {\enquote {\bibinfo {title} {Reaching 10 ms single photon
  lifetimes for superconducting aluminum cavities},}\ }\href@noop {} {\bibfield
   {journal} {\bibinfo  {journal} {Appl. Phys. Lett.}\ }\textbf {\bibinfo
  {volume} {102}},\ \bibinfo {pages} {192604} (\bibinfo {year}
  {2013})}\BibitemShut {NoStop}%
\bibitem [{\citenamefont {Reagor}\ \emph {et~al.}(2016)\citenamefont {Reagor},
  \citenamefont {Pfaff}, \citenamefont {Axline}, \citenamefont {Heeres},
  \citenamefont {Ofek}, \citenamefont {Sliwa}, \citenamefont {Holland},
  \citenamefont {Wang}, \citenamefont {Blumoff}, \citenamefont {Chou} \emph
  {et~al.}}]{Reagor2016}%
  \BibitemOpen
  \bibfield  {author} {\bibinfo {author} {\bibnamefont {Reagor}, \bibfnamefont
  {M.}}, \bibinfo {author} {\bibnamefont {Pfaff}, \bibfnamefont {W.}}, \bibinfo
  {author} {\bibnamefont {Axline}, \bibfnamefont {C.}}, \bibinfo {author}
  {\bibnamefont {Heeres}, \bibfnamefont {R.~W.}}, \bibinfo {author}
  {\bibnamefont {Ofek}, \bibfnamefont {N.}}, \bibinfo {author} {\bibnamefont
  {Sliwa}, \bibfnamefont {K.}}, \bibinfo {author} {\bibnamefont {Holland},
  \bibfnamefont {E.}}, \bibinfo {author} {\bibnamefont {Wang}, \bibfnamefont
  {C.}}, \bibinfo {author} {\bibnamefont {Blumoff}, \bibfnamefont {J.}},
  \bibinfo {author} {\bibnamefont {Chou}, \bibfnamefont {K.}},  \emph
  {et~al.},\ }\bibfield  {title} {\enquote {\bibinfo {title} {Quantum memory
  with millisecond coherence in circuit qed},}\ }\href@noop {} {\bibfield
  {journal} {\bibinfo  {journal} {Phys. Rev. B}\ }\textbf {\bibinfo {volume}
  {94}},\ \bibinfo {pages} {014506} (\bibinfo {year} {2016})}\BibitemShut
  {NoStop}%
\bibitem [{\citenamefont {Richardson}\ \emph {et~al.}(2016)\citenamefont
  {Richardson}, \citenamefont {Siwak}, \citenamefont {Hackley}, \citenamefont
  {Keane}, \citenamefont {Robinson}, \citenamefont {Arey}, \citenamefont
  {Arslan},\ and\ \citenamefont {Palmer}}]{Richardson2016}%
  \BibitemOpen
  \bibfield  {author} {\bibinfo {author} {\bibnamefont {Richardson},
  \bibfnamefont {C.}}, \bibinfo {author} {\bibnamefont {Siwak}, \bibfnamefont
  {N.}}, \bibinfo {author} {\bibnamefont {Hackley}, \bibfnamefont {J.}},
  \bibinfo {author} {\bibnamefont {Keane}, \bibfnamefont {Z.}}, \bibinfo
  {author} {\bibnamefont {Robinson}, \bibfnamefont {J.}}, \bibinfo {author}
  {\bibnamefont {Arey}, \bibfnamefont {B.}}, \bibinfo {author} {\bibnamefont
  {Arslan}, \bibfnamefont {I.}}, \ and\ \bibinfo {author} {\bibnamefont
  {Palmer}, \bibfnamefont {B.}},\ }\bibfield  {title} {\enquote {\bibinfo
  {title} {Fabrication artifacts and parallel loss channels in metamorphic
  epitaxial aluminum superconducting resonators},}\ }\href@noop {} {\bibfield
  {journal} {\bibinfo  {journal} {Supercond. Sci. Technol.}\ }\textbf {\bibinfo
  {volume} {29}},\ \bibinfo {pages} {064003} (\bibinfo {year}
  {2016})}\BibitemShut {NoStop}%
\bibitem [{\citenamefont {Romanenko}\ \emph {et~al.}(2020)\citenamefont
  {Romanenko}, \citenamefont {Pilipenko}, \citenamefont {Zorzetti},
  \citenamefont {Frolov}, \citenamefont {Awida}, \citenamefont {Belomestnykh},
  \citenamefont {Posen},\ and\ \citenamefont {Grassellino}}]{Romanenko2020}%
  \BibitemOpen
  \bibfield  {author} {\bibinfo {author} {\bibnamefont {Romanenko},
  \bibfnamefont {A.}}, \bibinfo {author} {\bibnamefont {Pilipenko},
  \bibfnamefont {R.}}, \bibinfo {author} {\bibnamefont {Zorzetti},
  \bibfnamefont {S.}}, \bibinfo {author} {\bibnamefont {Frolov}, \bibfnamefont
  {D.}}, \bibinfo {author} {\bibnamefont {Awida}, \bibfnamefont {M.}}, \bibinfo
  {author} {\bibnamefont {Belomestnykh}, \bibfnamefont {S.}}, \bibinfo {author}
  {\bibnamefont {Posen}, \bibfnamefont {S.}}, \ and\ \bibinfo {author}
  {\bibnamefont {Grassellino}, \bibfnamefont {A.}},\ }\bibfield  {title}
  {\enquote {\bibinfo {title} {{Three-dimensional superconducting resonators at
  $T< 20$ mK with the photon lifetime up to $\tau= 2$ seconds}},}\ }\href@noop
  {} {\bibfield  {journal} {\bibinfo  {journal} {Phys. Rev. Applied}\ }\textbf
  {\bibinfo {volume} {13}},\ \bibinfo {pages} {034032} (\bibinfo {year}
  {2020})}\BibitemShut {NoStop}%
\bibitem [{\citenamefont {Romanenko}\ and\ \citenamefont
  {Schuster}(2017)}]{Romanenko2017}%
  \BibitemOpen
  \bibfield  {author} {\bibinfo {author} {\bibnamefont {Romanenko},
  \bibfnamefont {A.}}\ and\ \bibinfo {author} {\bibnamefont {Schuster},
  \bibfnamefont {D.}},\ }\bibfield  {title} {\enquote {\bibinfo {title}
  {Understanding quality factor degradation in superconducting niobium cavities
  at low microwave field amplitudes},}\ }\href@noop {} {\bibfield  {journal}
  {\bibinfo  {journal} {Phys. Rev. Lett.}\ }\textbf {\bibinfo {volume} {119}},\
  \bibinfo {pages} {264801} (\bibinfo {year} {2017})}\BibitemShut {NoStop}%
\bibitem [{\citenamefont {Rosenberg}\ \emph {et~al.}(2017)\citenamefont
  {Rosenberg}, \citenamefont {Kim}, \citenamefont {Das}, \citenamefont {Yost},
  \citenamefont {Gustavsson}, \citenamefont {Hover}, \citenamefont {Krantz},
  \citenamefont {Melville}, \citenamefont {Racz}, \citenamefont {Samach},
  \citenamefont {Weber}, \citenamefont {Yan}, \citenamefont {Yoder},
  \citenamefont {Kerman},\ and\ \citenamefont {Oliver}}]{Rosenberg2017}%
  \BibitemOpen
  \bibfield  {author} {\bibinfo {author} {\bibnamefont {Rosenberg},
  \bibfnamefont {D.}}, \bibinfo {author} {\bibnamefont {Kim}, \bibfnamefont
  {D.}}, \bibinfo {author} {\bibnamefont {Das}, \bibfnamefont {R.}}, \bibinfo
  {author} {\bibnamefont {Yost}, \bibfnamefont {D.}}, \bibinfo {author}
  {\bibnamefont {Gustavsson}, \bibfnamefont {S.}}, \bibinfo {author}
  {\bibnamefont {Hover}, \bibfnamefont {D.}}, \bibinfo {author} {\bibnamefont
  {Krantz}, \bibfnamefont {P.}}, \bibinfo {author} {\bibnamefont {Melville},
  \bibfnamefont {A.}}, \bibinfo {author} {\bibnamefont {Racz}, \bibfnamefont
  {L.}}, \bibinfo {author} {\bibnamefont {Samach}, \bibfnamefont {G.~O.}},
  \bibinfo {author} {\bibnamefont {Weber}, \bibfnamefont {S.~J.}}, \bibinfo
  {author} {\bibnamefont {Yan}, \bibfnamefont {F.}}, \bibinfo {author}
  {\bibnamefont {Yoder}, \bibfnamefont {J.}}, \bibinfo {author} {\bibnamefont
  {Kerman}, \bibfnamefont {A.~J.}}, \ and\ \bibinfo {author} {\bibnamefont
  {Oliver}, \bibfnamefont {W.~D.}},\ }\bibfield  {title} {\enquote {\bibinfo
  {title} {{3D integrated superconducting qubits}},}\ }\href@noop {} {\bibfield
   {journal} {\bibinfo  {journal} {npj Quantum Inf.}\ }\textbf {\bibinfo
  {volume} {3}} (\bibinfo {year} {2017})}\BibitemShut {NoStop}%
\bibitem [{\citenamefont {Rosenberg}\ \emph {et~al.}(2019)\citenamefont
  {Rosenberg}, \citenamefont {Weber}, \citenamefont {Conway}, \citenamefont
  {Yost}, \citenamefont {Mallek}, \citenamefont {Calusine}, \citenamefont
  {Das}, \citenamefont {Kim}, \citenamefont {Schwartz}, \citenamefont {Woods}
  \emph {et~al.}}]{Rosenberg2019}%
  \BibitemOpen
  \bibfield  {author} {\bibinfo {author} {\bibnamefont {Rosenberg},
  \bibfnamefont {D.}}, \bibinfo {author} {\bibnamefont {Weber}, \bibfnamefont
  {S.}}, \bibinfo {author} {\bibnamefont {Conway}, \bibfnamefont {D.}},
  \bibinfo {author} {\bibnamefont {Yost}, \bibfnamefont {D.}}, \bibinfo
  {author} {\bibnamefont {Mallek}, \bibfnamefont {J.}}, \bibinfo {author}
  {\bibnamefont {Calusine}, \bibfnamefont {G.}}, \bibinfo {author}
  {\bibnamefont {Das}, \bibfnamefont {R.}}, \bibinfo {author} {\bibnamefont
  {Kim}, \bibfnamefont {D.}}, \bibinfo {author} {\bibnamefont {Schwartz},
  \bibfnamefont {M.}}, \bibinfo {author} {\bibnamefont {Woods}, \bibfnamefont
  {W.}},  \emph {et~al.},\ }\bibfield  {title} {\enquote {\bibinfo {title} {3d
  integration and packaging for solid-state qubits},}\ }\href@noop {}
  {\bibfield  {journal} {\bibinfo  {journal} {arXiv preprint arXiv:1906.11146}\
  } (\bibinfo {year} {2019})}\BibitemShut {NoStop}%
\bibitem [{\citenamefont {Sage}\ \emph {et~al.}(2011)\citenamefont {Sage},
  \citenamefont {Bolkhovsky}, \citenamefont {Oliver}, \citenamefont {Turek},\
  and\ \citenamefont {Welander}}]{Sage2011}%
  \BibitemOpen
  \bibfield  {author} {\bibinfo {author} {\bibnamefont {Sage}, \bibfnamefont
  {J.~M.}}, \bibinfo {author} {\bibnamefont {Bolkhovsky}, \bibfnamefont {V.}},
  \bibinfo {author} {\bibnamefont {Oliver}, \bibfnamefont {W.~D.}}, \bibinfo
  {author} {\bibnamefont {Turek}, \bibfnamefont {B.}}, \ and\ \bibinfo {author}
  {\bibnamefont {Welander}, \bibfnamefont {P.~B.}},\ }\bibfield  {title}
  {\enquote {\bibinfo {title} {Study of loss in superconducting coplanar
  waveguide resonators},}\ }\href@noop {} {\bibfield  {journal} {\bibinfo
  {journal} {J. Appl. Phys.}\ }\textbf {\bibinfo {volume} {109}},\ \bibinfo
  {pages} {063915} (\bibinfo {year} {2011})}\BibitemShut {NoStop}%
\bibitem [{\citenamefont {Sandberg}\ \emph {et~al.}(2012)\citenamefont
  {Sandberg}, \citenamefont {Vissers}, \citenamefont {Kline}, \citenamefont
  {Weides}, \citenamefont {Gao}, \citenamefont {Wisbey},\ and\ \citenamefont
  {Pappas}}]{sandberg2012etch}%
  \BibitemOpen
  \bibfield  {author} {\bibinfo {author} {\bibnamefont {Sandberg},
  \bibfnamefont {M.}}, \bibinfo {author} {\bibnamefont {Vissers}, \bibfnamefont
  {M.~R.}}, \bibinfo {author} {\bibnamefont {Kline}, \bibfnamefont {J.~S.}},
  \bibinfo {author} {\bibnamefont {Weides}, \bibfnamefont {M.}}, \bibinfo
  {author} {\bibnamefont {Gao}, \bibfnamefont {J.}}, \bibinfo {author}
  {\bibnamefont {Wisbey}, \bibfnamefont {D.~S.}}, \ and\ \bibinfo {author}
  {\bibnamefont {Pappas}, \bibfnamefont {D.~P.}},\ }\bibfield  {title}
  {\enquote {\bibinfo {title} {Etch induced microwave losses in titanium
  nitride superconducting resonators},}\ }\href@noop {} {\bibfield  {journal}
  {\bibinfo  {journal} {Appl. Phys. Lett.}\ }\textbf {\bibinfo {volume}
  {100}},\ \bibinfo {pages} {262605} (\bibinfo {year} {2012})}\BibitemShut
  {NoStop}%
\bibitem [{\citenamefont {Sandberg}\ \emph {et~al.}(2013)\citenamefont
  {Sandberg}, \citenamefont {Vissers}, \citenamefont {Ohki}, \citenamefont
  {Gao}, \citenamefont {Aumentado}, \citenamefont {Weides},\ and\ \citenamefont
  {Pappas}}]{Sandberg2013}%
  \BibitemOpen
  \bibfield  {author} {\bibinfo {author} {\bibnamefont {Sandberg},
  \bibfnamefont {M.}}, \bibinfo {author} {\bibnamefont {Vissers}, \bibfnamefont
  {M.~R.}}, \bibinfo {author} {\bibnamefont {Ohki}, \bibfnamefont {T.~A.}},
  \bibinfo {author} {\bibnamefont {Gao}, \bibfnamefont {J.}}, \bibinfo {author}
  {\bibnamefont {Aumentado}, \bibfnamefont {J.}}, \bibinfo {author}
  {\bibnamefont {Weides}, \bibfnamefont {M.}}, \ and\ \bibinfo {author}
  {\bibnamefont {Pappas}, \bibfnamefont {D.~P.}},\ }\bibfield  {title}
  {\enquote {\bibinfo {title} {Radiation-suppressed superconducting quantum bit
  in a planar geometry},}\ }\href@noop {} {\bibfield  {journal} {\bibinfo
  {journal} {Appl. Phys. Lett.}\ }\textbf {\bibinfo {volume} {102}},\ \bibinfo
  {pages} {072601} (\bibinfo {year} {2013})}\BibitemShut {NoStop}%
\bibitem [{\citenamefont {Sarabi}\ \emph {et~al.}(2016)\citenamefont {Sarabi},
  \citenamefont {Ramanayaka}, \citenamefont {Burin}, \citenamefont
  {Wellstood},\ and\ \citenamefont {Osborn}}]{Sarabi2016}%
  \BibitemOpen
  \bibfield  {author} {\bibinfo {author} {\bibnamefont {Sarabi}, \bibfnamefont
  {B.}}, \bibinfo {author} {\bibnamefont {Ramanayaka}, \bibfnamefont {A.~N.}},
  \bibinfo {author} {\bibnamefont {Burin}, \bibfnamefont {A.~L.}}, \bibinfo
  {author} {\bibnamefont {Wellstood}, \bibfnamefont {F.~C.}}, \ and\ \bibinfo
  {author} {\bibnamefont {Osborn}, \bibfnamefont {K.~D.}},\ }\bibfield  {title}
  {\enquote {\bibinfo {title} {Projected dipole moments of individual two-level
  defects extracted using circuit quantum electrodynamics},}\ }\href@noop {}
  {\bibfield  {journal} {\bibinfo  {journal} {Phys. Rev. Lett.}\ }\textbf
  {\bibinfo {volume} {116}},\ \bibinfo {pages} {167002} (\bibinfo {year}
  {2016})}\BibitemShut {NoStop}%
\bibitem [{\citenamefont {Schuster}(2007)}]{schuster2007}%
  \BibitemOpen
  \bibfield  {author} {\bibinfo {author} {\bibnamefont {Schuster},
  \bibfnamefont {D.~I.}},\ }\emph {\bibinfo {title} {Circuit quantum
  electrodynamics}},\ \href@noop {} {Ph.D. thesis},\ \bibinfo  {school} {Yale
  University} (\bibinfo {year} {2007})\BibitemShut {NoStop}%
\bibitem [{\citenamefont {Scigliuzzo}\ \emph {et~al.}(2020)\citenamefont
  {Scigliuzzo}, \citenamefont {Bruhat}, \citenamefont {Bengtsson},
  \citenamefont {Burnett}, \citenamefont {Roudsari},\ and\ \citenamefont
  {Delsing}}]{Scigliuzzo2020}%
  \BibitemOpen
  \bibfield  {author} {\bibinfo {author} {\bibnamefont {Scigliuzzo},
  \bibfnamefont {M.}}, \bibinfo {author} {\bibnamefont {Bruhat}, \bibfnamefont
  {L.~E.}}, \bibinfo {author} {\bibnamefont {Bengtsson}, \bibfnamefont {A.}},
  \bibinfo {author} {\bibnamefont {Burnett}, \bibfnamefont {J.}}, \bibinfo
  {author} {\bibnamefont {Roudsari}, \bibfnamefont {A.~F.}}, \ and\ \bibinfo
  {author} {\bibnamefont {Delsing}, \bibfnamefont {P.}},\ }\bibfield  {title}
  {\enquote {\bibinfo {title} {Phononic loss in superconducting resonators on
  piezoelectric substrates},}\ }\href@noop {} {\bibfield  {journal} {\bibinfo
  {journal} {New Journal of Physics}\ } (\bibinfo {year} {2020})}\BibitemShut
  {NoStop}%
\bibitem [{\citenamefont {Serniak}\ \emph {et~al.}(2018)\citenamefont
  {Serniak}, \citenamefont {Hays}, \citenamefont {de~Lange}, \citenamefont
  {Diamond}, \citenamefont {Shankar}, \citenamefont {Burkhart}, \citenamefont
  {Frunzio}, \citenamefont {Houzet},\ and\ \citenamefont
  {Devoret}}]{Serniak2018}%
  \BibitemOpen
  \bibfield  {author} {\bibinfo {author} {\bibnamefont {Serniak}, \bibfnamefont
  {K.}}, \bibinfo {author} {\bibnamefont {Hays}, \bibfnamefont {M.}}, \bibinfo
  {author} {\bibnamefont {de~Lange}, \bibfnamefont {G.}}, \bibinfo {author}
  {\bibnamefont {Diamond}, \bibfnamefont {S.}}, \bibinfo {author} {\bibnamefont
  {Shankar}, \bibfnamefont {S.}}, \bibinfo {author} {\bibnamefont {Burkhart},
  \bibfnamefont {L.}}, \bibinfo {author} {\bibnamefont {Frunzio}, \bibfnamefont
  {L.}}, \bibinfo {author} {\bibnamefont {Houzet}, \bibfnamefont {M.}}, \ and\
  \bibinfo {author} {\bibnamefont {Devoret}, \bibfnamefont {M.}},\ }\bibfield
  {title} {\enquote {\bibinfo {title} {Hot nonequilibrium quasiparticles in
  transmon qubits},}\ }\href@noop {} {\bibfield  {journal} {\bibinfo  {journal}
  {Phys. Rev. Lett.}\ }\textbf {\bibinfo {volume} {121}},\ \bibinfo {pages}
  {157701} (\bibinfo {year} {2018})}\BibitemShut {NoStop}%
\bibitem [{\citenamefont {Shearrow}\ \emph {et~al.}(2018)\citenamefont
  {Shearrow}, \citenamefont {Koolstra}, \citenamefont {Whiteley}, \citenamefont
  {Earnest}, \citenamefont {Barry}, \citenamefont {Heremans}, \citenamefont
  {Awschalom}, \citenamefont {Shirokoff},\ and\ \citenamefont
  {Schuster}}]{Shearrow2018}%
  \BibitemOpen
  \bibfield  {author} {\bibinfo {author} {\bibnamefont {Shearrow},
  \bibfnamefont {A.}}, \bibinfo {author} {\bibnamefont {Koolstra},
  \bibfnamefont {G.}}, \bibinfo {author} {\bibnamefont {Whiteley},
  \bibfnamefont {S.~J.}}, \bibinfo {author} {\bibnamefont {Earnest},
  \bibfnamefont {N.}}, \bibinfo {author} {\bibnamefont {Barry}, \bibfnamefont
  {P.~S.}}, \bibinfo {author} {\bibnamefont {Heremans}, \bibfnamefont {F.~J.}},
  \bibinfo {author} {\bibnamefont {Awschalom}, \bibfnamefont {D.~D.}}, \bibinfo
  {author} {\bibnamefont {Shirokoff}, \bibfnamefont {E.}}, \ and\ \bibinfo
  {author} {\bibnamefont {Schuster}, \bibfnamefont {D.~I.}},\ }\bibfield
  {title} {\enquote {\bibinfo {title} {{Atomic layer deposition of titanium
  nitride for quantum circuits}},}\ }\href@noop {} {\bibfield  {journal}
  {\bibinfo  {journal} {Appl. Phys. Lett.}\ }\textbf {\bibinfo {volume}
  {113}},\ \bibinfo {pages} {212601} (\bibinfo {year} {2018})}\BibitemShut
  {NoStop}%
\bibitem [{\citenamefont {Sheldon}\ \emph {et~al.}(2017)\citenamefont
  {Sheldon}, \citenamefont {Sandberg}, \citenamefont {Paik}, \citenamefont
  {Abdo}, \citenamefont {Chow}, \citenamefont {Steffen},\ and\ \citenamefont
  {Gambetta}}]{Sheldon2017}%
  \BibitemOpen
  \bibfield  {author} {\bibinfo {author} {\bibnamefont {Sheldon}, \bibfnamefont
  {S.}}, \bibinfo {author} {\bibnamefont {Sandberg}, \bibfnamefont {M.}},
  \bibinfo {author} {\bibnamefont {Paik}, \bibfnamefont {H.}}, \bibinfo
  {author} {\bibnamefont {Abdo}, \bibfnamefont {B.}}, \bibinfo {author}
  {\bibnamefont {Chow}, \bibfnamefont {J.~M.}}, \bibinfo {author} {\bibnamefont
  {Steffen}, \bibfnamefont {M.}}, \ and\ \bibinfo {author} {\bibnamefont
  {Gambetta}, \bibfnamefont {J.~M.}},\ }\bibfield  {title} {\enquote {\bibinfo
  {title} {Characterization of hidden modes in networks of superconducting
  qubits},}\ }\href {\doibase 10.1063/1.4990033} {\bibfield  {journal}
  {\bibinfo  {journal} {Appl. Phys. Lett.}\ }\textbf {\bibinfo {volume}
  {111}},\ \bibinfo {pages} {222601} (\bibinfo {year} {2017})},\ \Eprint
  {http://arxiv.org/abs/https://doi.org/10.1063/1.4990033}
  {https://doi.org/10.1063/1.4990033} \BibitemShut {NoStop}%
\bibitem [{\citenamefont {Sillanp{\"a}{\"a}}, \citenamefont {Park},\ and\
  \citenamefont {Simmonds}(2007)}]{Sillanpaa2007}%
  \BibitemOpen
  \bibfield  {author} {\bibinfo {author} {\bibnamefont {Sillanp{\"a}{\"a}},
  \bibfnamefont {M.~A.}}, \bibinfo {author} {\bibnamefont {Park}, \bibfnamefont
  {J.~I.}}, \ and\ \bibinfo {author} {\bibnamefont {Simmonds}, \bibfnamefont
  {R.~W.}},\ }\bibfield  {title} {\enquote {\bibinfo {title} {Coherent quantum
  state storage and transfer between two phase qubits via a resonant cavity},}\
  }\href@noop {} {\bibfield  {journal} {\bibinfo  {journal} {Nature}\ }\textbf
  {\bibinfo {volume} {449}},\ \bibinfo {pages} {438--442} (\bibinfo {year}
  {2007})}\BibitemShut {NoStop}%
\bibitem [{\citenamefont {Skacel}\ \emph {et~al.}(2015)\citenamefont {Skacel},
  \citenamefont {Kaiser}, \citenamefont {Wuensch}, \citenamefont {Rotzinger},
  \citenamefont {Lukashenko}, \citenamefont {Jerger}, \citenamefont {Weiss},
  \citenamefont {Siegel},\ and\ \citenamefont {Ustinov}}]{Skacel2015}%
  \BibitemOpen
  \bibfield  {author} {\bibinfo {author} {\bibnamefont {Skacel}, \bibfnamefont
  {S.~T.}}, \bibinfo {author} {\bibnamefont {Kaiser}, \bibfnamefont {C.}},
  \bibinfo {author} {\bibnamefont {Wuensch}, \bibfnamefont {S.}}, \bibinfo
  {author} {\bibnamefont {Rotzinger}, \bibfnamefont {H.}}, \bibinfo {author}
  {\bibnamefont {Lukashenko}, \bibfnamefont {A.}}, \bibinfo {author}
  {\bibnamefont {Jerger}, \bibfnamefont {M.}}, \bibinfo {author} {\bibnamefont
  {Weiss}, \bibfnamefont {G.}}, \bibinfo {author} {\bibnamefont {Siegel},
  \bibfnamefont {M.}}, \ and\ \bibinfo {author} {\bibnamefont {Ustinov},
  \bibfnamefont {A.~V.}},\ }\bibfield  {title} {\enquote {\bibinfo {title}
  {Probing the density of states of two-level tunneling systems in silicon
  oxide films using superconducting lumped element resonators},}\ }\href@noop
  {} {\bibfield  {journal} {\bibinfo  {journal} {Appl. Phys. Lett.}\ }\textbf
  {\bibinfo {volume} {106}},\ \bibinfo {pages} {022603} (\bibinfo {year}
  {2015})}\BibitemShut {NoStop}%
\bibitem [{\citenamefont {Song}\ \emph
  {et~al.}(2009{\natexlab{a}})\citenamefont {Song}, \citenamefont {DeFeo},
  \citenamefont {Yu},\ and\ \citenamefont {Plourde}}]{song2009reducing}%
  \BibitemOpen
  \bibfield  {author} {\bibinfo {author} {\bibnamefont {Song}, \bibfnamefont
  {C.}}, \bibinfo {author} {\bibnamefont {DeFeo}, \bibfnamefont {M.~P.}},
  \bibinfo {author} {\bibnamefont {Yu}, \bibfnamefont {K.}}, \ and\ \bibinfo
  {author} {\bibnamefont {Plourde}, \bibfnamefont {B.~L.}},\ }\bibfield
  {title} {\enquote {\bibinfo {title} {Reducing microwave loss in
  superconducting resonators due to trapped vortices},}\ }\href@noop {}
  {\bibfield  {journal} {\bibinfo  {journal} {Appl. Phys. Lett.}\ }\textbf
  {\bibinfo {volume} {95}},\ \bibinfo {pages} {232501} (\bibinfo {year}
  {2009}{\natexlab{a}})}\BibitemShut {NoStop}%
\bibitem [{\citenamefont {Song}\ \emph
  {et~al.}(2009{\natexlab{b}})\citenamefont {Song}, \citenamefont {Heitmann},
  \citenamefont {DeFeo}, \citenamefont {Yu}, \citenamefont {McDermott},
  \citenamefont {Neeley}, \citenamefont {Martinis},\ and\ \citenamefont
  {Plourde}}]{Song2009}%
  \BibitemOpen
  \bibfield  {author} {\bibinfo {author} {\bibnamefont {Song}, \bibfnamefont
  {C.}}, \bibinfo {author} {\bibnamefont {Heitmann}, \bibfnamefont {T.~W.}},
  \bibinfo {author} {\bibnamefont {DeFeo}, \bibfnamefont {M.~P.}}, \bibinfo
  {author} {\bibnamefont {Yu}, \bibfnamefont {K.}}, \bibinfo {author}
  {\bibnamefont {McDermott}, \bibfnamefont {R.}}, \bibinfo {author}
  {\bibnamefont {Neeley}, \bibfnamefont {M.}}, \bibinfo {author} {\bibnamefont
  {Martinis}, \bibfnamefont {J.~M.}}, \ and\ \bibinfo {author} {\bibnamefont
  {Plourde}, \bibfnamefont {B.~L.}},\ }\bibfield  {title} {\enquote {\bibinfo
  {title} {{Microwave response of vortices in superconducting thin films of Re
  and Al}},}\ }\href@noop {} {\bibfield  {journal} {\bibinfo  {journal} {Phys.
  Rev. B}\ }\textbf {\bibinfo {volume} {79}},\ \bibinfo {pages} {174512}
  (\bibinfo {year} {2009}{\natexlab{b}})}\BibitemShut {NoStop}%
\bibitem [{\citenamefont {Stan}, \citenamefont {Field},\ and\ \citenamefont
  {Martinis}(2004)}]{Stan2004}%
  \BibitemOpen
  \bibfield  {author} {\bibinfo {author} {\bibnamefont {Stan}, \bibfnamefont
  {G.}}, \bibinfo {author} {\bibnamefont {Field}, \bibfnamefont {S.~B.}}, \
  and\ \bibinfo {author} {\bibnamefont {Martinis}, \bibfnamefont {J.~M.}},\
  }\bibfield  {title} {\enquote {\bibinfo {title} {Critical field for complete
  vortex expulsion from narrow superconducting strips},}\ }\href@noop {}
  {\bibfield  {journal} {\bibinfo  {journal} {Phys. Rev. Lett.}\ }\textbf
  {\bibinfo {volume} {92}},\ \bibinfo {pages} {097003} (\bibinfo {year}
  {2004})}\BibitemShut {NoStop}%
\bibitem [{\citenamefont {Stoutimore}\ \emph {et~al.}(2012)\citenamefont
  {Stoutimore}, \citenamefont {Khalil}, \citenamefont {Lobb},\ and\
  \citenamefont {Osborn}}]{Stoutimore2012}%
  \BibitemOpen
  \bibfield  {author} {\bibinfo {author} {\bibnamefont {Stoutimore},
  \bibfnamefont {M.}}, \bibinfo {author} {\bibnamefont {Khalil}, \bibfnamefont
  {M.}}, \bibinfo {author} {\bibnamefont {Lobb}, \bibfnamefont {C.}}, \ and\
  \bibinfo {author} {\bibnamefont {Osborn}, \bibfnamefont {K.}},\ }\bibfield
  {title} {\enquote {\bibinfo {title} {A josephson junction defect spectrometer
  for measuring two-level systems},}\ }\href@noop {} {\bibfield  {journal}
  {\bibinfo  {journal} {Appl. Phys. Lett.}\ }\textbf {\bibinfo {volume}
  {101}},\ \bibinfo {pages} {062602} (\bibinfo {year} {2012})}\BibitemShut
  {NoStop}%
\bibitem [{\citenamefont {Vahidpour}\ \emph {et~al.}(2017)\citenamefont
  {Vahidpour}, \citenamefont {O'Brien}, \citenamefont {Whyland}, \citenamefont
  {Angeles}, \citenamefont {Marshall}, \citenamefont {Scarabelli},
  \citenamefont {Crossman}, \citenamefont {Yadav}, \citenamefont {Mohan},
  \citenamefont {Bui}, \citenamefont {Rawat}, \citenamefont {Renzas},
  \citenamefont {Vodrahalli}, \citenamefont {Bestwick},\ and\ \citenamefont
  {Rigetti}}]{Vahidpour2017}%
  \BibitemOpen
  \bibfield  {author} {\bibinfo {author} {\bibnamefont {Vahidpour},
  \bibfnamefont {M.}}, \bibinfo {author} {\bibnamefont {O'Brien}, \bibfnamefont
  {W.}}, \bibinfo {author} {\bibnamefont {Whyland}, \bibfnamefont {J.~T.}},
  \bibinfo {author} {\bibnamefont {Angeles}, \bibfnamefont {J.}}, \bibinfo
  {author} {\bibnamefont {Marshall}, \bibfnamefont {J.}}, \bibinfo {author}
  {\bibnamefont {Scarabelli}, \bibfnamefont {D.}}, \bibinfo {author}
  {\bibnamefont {Crossman}, \bibfnamefont {G.}}, \bibinfo {author}
  {\bibnamefont {Yadav}, \bibfnamefont {K.}}, \bibinfo {author} {\bibnamefont
  {Mohan}, \bibfnamefont {Y.}}, \bibinfo {author} {\bibnamefont {Bui},
  \bibfnamefont {C.}}, \bibinfo {author} {\bibnamefont {Rawat}, \bibfnamefont
  {V.}}, \bibinfo {author} {\bibnamefont {Renzas}, \bibfnamefont {R.}},
  \bibinfo {author} {\bibnamefont {Vodrahalli}, \bibfnamefont {N.}}, \bibinfo
  {author} {\bibnamefont {Bestwick}, \bibfnamefont {A.}}, \ and\ \bibinfo
  {author} {\bibnamefont {Rigetti}, \bibfnamefont {C.}},\ }\bibfield  {title}
  {\enquote {\bibinfo {title} {{Superconducting Through-Silicon Vias for
  Quantum Integrated Circuits}},}\ }\href@noop {} {\  (\bibinfo {year}
  {2017})},\ \Eprint {http://arxiv.org/abs/1708.02226} {arXiv:1708.02226}
  \BibitemShut {NoStop}%
\bibitem [{\citenamefont {Vepsäläinen}\ \emph {et~al.}(2020)\citenamefont
  {Vepsäläinen}, \citenamefont {Karamlou}, \citenamefont {Orrell},
  \citenamefont {Dogra}, \citenamefont {Loer}, \citenamefont {Vasconcelos},
  \citenamefont {Kim}, \citenamefont {Melville}, \citenamefont {Niedzielski},
  \citenamefont {Yoder}, \citenamefont {Gustavsson}, \citenamefont {Formaggio},
  \citenamefont {VanDevender},\ and\ \citenamefont {Oliver}}]{vepslinen2020}%
  \BibitemOpen
  \bibfield  {author} {\bibinfo {author} {\bibnamefont {Vepsäläinen},
  \bibfnamefont {A.}}, \bibinfo {author} {\bibnamefont {Karamlou},
  \bibfnamefont {A.~H.}}, \bibinfo {author} {\bibnamefont {Orrell},
  \bibfnamefont {J.~L.}}, \bibinfo {author} {\bibnamefont {Dogra},
  \bibfnamefont {A.~S.}}, \bibinfo {author} {\bibnamefont {Loer}, \bibfnamefont
  {B.}}, \bibinfo {author} {\bibnamefont {Vasconcelos}, \bibfnamefont {F.}},
  \bibinfo {author} {\bibnamefont {Kim}, \bibfnamefont {D.~K.}}, \bibinfo
  {author} {\bibnamefont {Melville}, \bibfnamefont {A.~J.}}, \bibinfo {author}
  {\bibnamefont {Niedzielski}, \bibfnamefont {B.~M.}}, \bibinfo {author}
  {\bibnamefont {Yoder}, \bibfnamefont {J.~L.}}, \bibinfo {author}
  {\bibnamefont {Gustavsson}, \bibfnamefont {S.}}, \bibinfo {author}
  {\bibnamefont {Formaggio}, \bibfnamefont {J.~A.}}, \bibinfo {author}
  {\bibnamefont {VanDevender}, \bibfnamefont {B.~A.}}, \ and\ \bibinfo {author}
  {\bibnamefont {Oliver}, \bibfnamefont {W.~D.}},\ }\href@noop {} {\enquote
  {\bibinfo {title} {Impact of ionizing radiation on superconducting qubit
  coherence},}\ } (\bibinfo {year} {2020}),\ \Eprint
  {http://arxiv.org/abs/2001.09190} {arXiv:2001.09190 [quant-ph]} \BibitemShut
  {NoStop}%
\bibitem [{\citenamefont {Vissers}\ \emph {et~al.}(2010)\citenamefont
  {Vissers}, \citenamefont {Gao}, \citenamefont {Wisbey}, \citenamefont {Hite},
  \citenamefont {Tsuei}, \citenamefont {Corcoles}, \citenamefont {Steffen},\
  and\ \citenamefont {Pappas}}]{Vissers2010}%
  \BibitemOpen
  \bibfield  {author} {\bibinfo {author} {\bibnamefont {Vissers}, \bibfnamefont
  {M.~R.}}, \bibinfo {author} {\bibnamefont {Gao}, \bibfnamefont {J.}},
  \bibinfo {author} {\bibnamefont {Wisbey}, \bibfnamefont {D.~S.}}, \bibinfo
  {author} {\bibnamefont {Hite}, \bibfnamefont {D.~A.}}, \bibinfo {author}
  {\bibnamefont {Tsuei}, \bibfnamefont {C.~C.}}, \bibinfo {author}
  {\bibnamefont {Corcoles}, \bibfnamefont {A.~D.}}, \bibinfo {author}
  {\bibnamefont {Steffen}, \bibfnamefont {M.}}, \ and\ \bibinfo {author}
  {\bibnamefont {Pappas}, \bibfnamefont {D.~P.}},\ }\bibfield  {title}
  {\enquote {\bibinfo {title} {Low loss superconducting titanium nitride
  coplanar waveguide resonators},}\ }\href@noop {} {\bibfield  {journal}
  {\bibinfo  {journal} {Appl. Phys. Lett.}\ }\textbf {\bibinfo {volume} {97}},\
  \bibinfo {pages} {232509} (\bibinfo {year} {2010})}\BibitemShut {NoStop}%
\bibitem [{\citenamefont {Vissers}\ \emph
  {et~al.}(2012{\natexlab{a}})\citenamefont {Vissers}, \citenamefont {Kline},
  \citenamefont {Gao}, \citenamefont {Wisbey},\ and\ \citenamefont
  {Pappas}}]{vissers2012reduced}%
  \BibitemOpen
  \bibfield  {author} {\bibinfo {author} {\bibnamefont {Vissers}, \bibfnamefont
  {M.~R.}}, \bibinfo {author} {\bibnamefont {Kline}, \bibfnamefont {J.~S.}},
  \bibinfo {author} {\bibnamefont {Gao}, \bibfnamefont {J.}}, \bibinfo {author}
  {\bibnamefont {Wisbey}, \bibfnamefont {D.~S.}}, \ and\ \bibinfo {author}
  {\bibnamefont {Pappas}, \bibfnamefont {D.~P.}},\ }\bibfield  {title}
  {\enquote {\bibinfo {title} {Reduced microwave loss in trenched
  superconducting coplanar waveguides},}\ }\href@noop {} {\bibfield  {journal}
  {\bibinfo  {journal} {Appl. Phys. Lett.}\ }\textbf {\bibinfo {volume}
  {100}},\ \bibinfo {pages} {082602} (\bibinfo {year}
  {2012}{\natexlab{a}})}\BibitemShut {NoStop}%
\bibitem [{\citenamefont {Vissers}\ \emph
  {et~al.}(2012{\natexlab{b}})\citenamefont {Vissers}, \citenamefont {Weides},
  \citenamefont {Kline}, \citenamefont {Sandberg},\ and\ \citenamefont
  {Pappas}}]{Vissers2012}%
  \BibitemOpen
  \bibfield  {author} {\bibinfo {author} {\bibnamefont {Vissers}, \bibfnamefont
  {M.~R.}}, \bibinfo {author} {\bibnamefont {Weides}, \bibfnamefont {M.~P.}},
  \bibinfo {author} {\bibnamefont {Kline}, \bibfnamefont {J.~S.}}, \bibinfo
  {author} {\bibnamefont {Sandberg}, \bibfnamefont {M.}}, \ and\ \bibinfo
  {author} {\bibnamefont {Pappas}, \bibfnamefont {D.~P.}},\ }\bibfield  {title}
  {\enquote {\bibinfo {title} {Identifying capacitive and inductive loss in
  lumped element superconducting hybrid titanium nitride/aluminum
  resonators},}\ }\href@noop {} {\bibfield  {journal} {\bibinfo  {journal}
  {Appl. Phys. Lett.}\ }\textbf {\bibinfo {volume} {101}},\ \bibinfo {pages}
  {022601} (\bibinfo {year} {2012}{\natexlab{b}})}\BibitemShut {NoStop}%
\bibitem [{\citenamefont {{Walker}}\ and\ \citenamefont
  {{Williams}}(1998)}]{SOLTvsTRL}%
  \BibitemOpen
  \bibfield  {author} {\bibinfo {author} {\bibnamefont {{Walker}},
  \bibfnamefont {D.~K.}}\ and\ \bibinfo {author} {\bibnamefont {{Williams}},
  \bibfnamefont {D.~F.}},\ }\bibfield  {title} {\enquote {\bibinfo {title}
  {Comparison of solr and trl calibrations},}\ }\href@noop {} {\bibfield
  {journal} {\bibinfo  {journal} {51st ARFTG Conference Digest}\ }\textbf
  {\bibinfo {volume} {33}},\ \bibinfo {pages} {83--87} (\bibinfo {year}
  {1998})}\BibitemShut {NoStop}%
\bibitem [{\citenamefont {Wallraff}\ \emph {et~al.}(2004)\citenamefont
  {Wallraff}, \citenamefont {Schuster}, \citenamefont {Blais}, \citenamefont
  {Frunzio}, \citenamefont {Huang}, \citenamefont {Majer}, \citenamefont
  {Kumar}, \citenamefont {Girvin},\ and\ \citenamefont
  {Schoelkopf}}]{Wallraff2004}%
  \BibitemOpen
  \bibfield  {author} {\bibinfo {author} {\bibnamefont {Wallraff},
  \bibfnamefont {A.}}, \bibinfo {author} {\bibnamefont {Schuster},
  \bibfnamefont {D.~I.}}, \bibinfo {author} {\bibnamefont {Blais},
  \bibfnamefont {A.}}, \bibinfo {author} {\bibnamefont {Frunzio}, \bibfnamefont
  {L.}}, \bibinfo {author} {\bibnamefont {Huang}, \bibfnamefont {R.-S.}},
  \bibinfo {author} {\bibnamefont {Majer}, \bibfnamefont {J.}}, \bibinfo
  {author} {\bibnamefont {Kumar}, \bibfnamefont {S.}}, \bibinfo {author}
  {\bibnamefont {Girvin}, \bibfnamefont {S.~M.}}, \ and\ \bibinfo {author}
  {\bibnamefont {Schoelkopf}, \bibfnamefont {R.~J.}},\ }\bibfield  {title}
  {\enquote {\bibinfo {title} {Strong coupling of a single photon to a
  superconducting qubit using circuit quantum electrodynamics},}\ }\href@noop
  {} {\bibfield  {journal} {\bibinfo  {journal} {Nature}\ }\textbf {\bibinfo
  {volume} {431}},\ \bibinfo {pages} {162--167} (\bibinfo {year}
  {2004})}\BibitemShut {NoStop}%
\bibitem [{\citenamefont {Wang}\ \emph {et~al.}(2015)\citenamefont {Wang},
  \citenamefont {Axline}, \citenamefont {Gao}, \citenamefont {Brecht},
  \citenamefont {Chu}, \citenamefont {Frunzio}, \citenamefont {Devoret},\ and\
  \citenamefont {Schoelkopf}}]{wang2015}%
  \BibitemOpen
  \bibfield  {author} {\bibinfo {author} {\bibnamefont {Wang}, \bibfnamefont
  {C.}}, \bibinfo {author} {\bibnamefont {Axline}, \bibfnamefont {C.}},
  \bibinfo {author} {\bibnamefont {Gao}, \bibfnamefont {Y.~Y.}}, \bibinfo
  {author} {\bibnamefont {Brecht}, \bibfnamefont {T.}}, \bibinfo {author}
  {\bibnamefont {Chu}, \bibfnamefont {Y.}}, \bibinfo {author} {\bibnamefont
  {Frunzio}, \bibfnamefont {L.}}, \bibinfo {author} {\bibnamefont {Devoret},
  \bibfnamefont {M.}}, \ and\ \bibinfo {author} {\bibnamefont {Schoelkopf},
  \bibfnamefont {R.~J.}},\ }\bibfield  {title} {\enquote {\bibinfo {title}
  {Surface participation and dielectric loss in superconducting qubits},}\
  }\href@noop {} {\bibfield  {journal} {\bibinfo  {journal} {Appl. Phys.
  Lett.}\ }\textbf {\bibinfo {volume} {107}},\ \bibinfo {pages} {162601}
  (\bibinfo {year} {2015})}\BibitemShut {NoStop}%
\bibitem [{\citenamefont {Wang}\ \emph {et~al.}(2014)\citenamefont {Wang},
  \citenamefont {Gao}, \citenamefont {Pop}, \citenamefont {Vool}, \citenamefont
  {Axline}, \citenamefont {Brecht}, \citenamefont {Heeres}, \citenamefont
  {Frunzio}, \citenamefont {Devoret}, \citenamefont {Catelani} \emph
  {et~al.}}]{Wang2014}%
  \BibitemOpen
  \bibfield  {author} {\bibinfo {author} {\bibnamefont {Wang}, \bibfnamefont
  {C.}}, \bibinfo {author} {\bibnamefont {Gao}, \bibfnamefont {Y.~Y.}},
  \bibinfo {author} {\bibnamefont {Pop}, \bibfnamefont {I.~M.}}, \bibinfo
  {author} {\bibnamefont {Vool}, \bibfnamefont {U.}}, \bibinfo {author}
  {\bibnamefont {Axline}, \bibfnamefont {C.}}, \bibinfo {author} {\bibnamefont
  {Brecht}, \bibfnamefont {T.}}, \bibinfo {author} {\bibnamefont {Heeres},
  \bibfnamefont {R.~W.}}, \bibinfo {author} {\bibnamefont {Frunzio},
  \bibfnamefont {L.}}, \bibinfo {author} {\bibnamefont {Devoret}, \bibfnamefont
  {M.~H.}}, \bibinfo {author} {\bibnamefont {Catelani}, \bibfnamefont {G.}},
  \emph {et~al.},\ }\bibfield  {title} {\enquote {\bibinfo {title} {Measurement
  and control of quasiparticle dynamics in a superconducting qubit},}\
  }\href@noop {} {\bibfield  {journal} {\bibinfo  {journal} {Nat. Commun.}\
  }\textbf {\bibinfo {volume} {5}},\ \bibinfo {pages} {1--7} (\bibinfo {year}
  {2014})}\BibitemShut {NoStop}%
\bibitem [{\citenamefont {Wang}\ \emph {et~al.}(2009)\citenamefont {Wang},
  \citenamefont {Hofheinz}, \citenamefont {Wenner}, \citenamefont {Ansmann},
  \citenamefont {Bialczak}, \citenamefont {Lenander}, \citenamefont {Lucero},
  \citenamefont {Neeley}, \citenamefont {O’Connell}, \citenamefont {Sank}
  \emph {et~al.}}]{Wang2009}%
  \BibitemOpen
  \bibfield  {author} {\bibinfo {author} {\bibnamefont {Wang}, \bibfnamefont
  {H.}}, \bibinfo {author} {\bibnamefont {Hofheinz}, \bibfnamefont {M.}},
  \bibinfo {author} {\bibnamefont {Wenner}, \bibfnamefont {J.}}, \bibinfo
  {author} {\bibnamefont {Ansmann}, \bibfnamefont {M.}}, \bibinfo {author}
  {\bibnamefont {Bialczak}, \bibfnamefont {R.}}, \bibinfo {author}
  {\bibnamefont {Lenander}, \bibfnamefont {M.}}, \bibinfo {author}
  {\bibnamefont {Lucero}, \bibfnamefont {E.}}, \bibinfo {author} {\bibnamefont
  {Neeley}, \bibfnamefont {M.}}, \bibinfo {author} {\bibnamefont {O’Connell},
  \bibfnamefont {A.}}, \bibinfo {author} {\bibnamefont {Sank}, \bibfnamefont
  {D.}},  \emph {et~al.},\ }\bibfield  {title} {\enquote {\bibinfo {title}
  {Improving the coherence time of superconducting coplanar resonators},}\
  }\href@noop {} {\bibfield  {journal} {\bibinfo  {journal} {Appl. Phys.
  Lett.}\ }\textbf {\bibinfo {volume} {95}},\ \bibinfo {pages} {233508}
  (\bibinfo {year} {2009})}\BibitemShut {NoStop}%
\bibitem [{\citenamefont {Wang}\ \emph {et~al.}(2019)\citenamefont {Wang},
  \citenamefont {Shankar}, \citenamefont {Minev}, \citenamefont
  {Campagne-Ibarcq}, \citenamefont {Narla},\ and\ \citenamefont
  {Devoret}}]{Wang2019}%
  \BibitemOpen
  \bibfield  {author} {\bibinfo {author} {\bibnamefont {Wang}, \bibfnamefont
  {Z.}}, \bibinfo {author} {\bibnamefont {Shankar}, \bibfnamefont {S.}},
  \bibinfo {author} {\bibnamefont {Minev}, \bibfnamefont {Z.}}, \bibinfo
  {author} {\bibnamefont {Campagne-Ibarcq}, \bibfnamefont {P.}}, \bibinfo
  {author} {\bibnamefont {Narla}, \bibfnamefont {A.}}, \ and\ \bibinfo {author}
  {\bibnamefont {Devoret}, \bibfnamefont {M.~H.}},\ }\bibfield  {title}
  {\enquote {\bibinfo {title} {Cavity attenuators for superconducting
  qubits},}\ }\href@noop {} {\bibfield  {journal} {\bibinfo  {journal} {Phys.
  Rev. Applied}\ }\textbf {\bibinfo {volume} {11}},\ \bibinfo {pages} {014031}
  (\bibinfo {year} {2019})}\BibitemShut {NoStop}%
\bibitem [{\citenamefont {Weides}\ \emph {et~al.}(2011)\citenamefont {Weides},
  \citenamefont {Kline}, \citenamefont {Vissers}, \citenamefont {Sandberg},
  \citenamefont {Wisbey}, \citenamefont {Johnson}, \citenamefont {Ohki},\ and\
  \citenamefont {Pappas}}]{weides2011_epiAl2O3}%
  \BibitemOpen
  \bibfield  {author} {\bibinfo {author} {\bibnamefont {Weides}, \bibfnamefont
  {M.~P.}}, \bibinfo {author} {\bibnamefont {Kline}, \bibfnamefont {J.~S.}},
  \bibinfo {author} {\bibnamefont {Vissers}, \bibfnamefont {M.~R.}}, \bibinfo
  {author} {\bibnamefont {Sandberg}, \bibfnamefont {M.~O.}}, \bibinfo {author}
  {\bibnamefont {Wisbey}, \bibfnamefont {D.~S.}}, \bibinfo {author}
  {\bibnamefont {Johnson}, \bibfnamefont {B.~R.}}, \bibinfo {author}
  {\bibnamefont {Ohki}, \bibfnamefont {T.~A.}}, \ and\ \bibinfo {author}
  {\bibnamefont {Pappas}, \bibfnamefont {D.~P.}},\ }\bibfield  {title}
  {\enquote {\bibinfo {title} {Coherence in a transmon qubit with epitaxial
  tunnel junctions},}\ }\href@noop {} {\bibfield  {journal} {\bibinfo
  {journal} {Appl. Phys. Lett.}\ }\textbf {\bibinfo {volume} {99}},\ \bibinfo
  {pages} {262502} (\bibinfo {year} {2011})}\BibitemShut {NoStop}%
\bibitem [{\citenamefont {Wenner}\ \emph {et~al.}(2011)\citenamefont {Wenner},
  \citenamefont {Neeley}, \citenamefont {Bialczak}, \citenamefont {Lenander},
  \citenamefont {Lucero}, \citenamefont {O’Connell}, \citenamefont {Sank},
  \citenamefont {Wang}, \citenamefont {Weides}, \citenamefont {Cleland} \emph
  {et~al.}}]{Wenner2011}%
  \BibitemOpen
  \bibfield  {author} {\bibinfo {author} {\bibnamefont {Wenner}, \bibfnamefont
  {J.}}, \bibinfo {author} {\bibnamefont {Neeley}, \bibfnamefont {M.}},
  \bibinfo {author} {\bibnamefont {Bialczak}, \bibfnamefont {R.~C.}}, \bibinfo
  {author} {\bibnamefont {Lenander}, \bibfnamefont {M.}}, \bibinfo {author}
  {\bibnamefont {Lucero}, \bibfnamefont {E.}}, \bibinfo {author} {\bibnamefont
  {O’Connell}, \bibfnamefont {A.~D.}}, \bibinfo {author} {\bibnamefont
  {Sank}, \bibfnamefont {D.}}, \bibinfo {author} {\bibnamefont {Wang},
  \bibfnamefont {H.}}, \bibinfo {author} {\bibnamefont {Weides}, \bibfnamefont
  {M.}}, \bibinfo {author} {\bibnamefont {Cleland}, \bibfnamefont {A.~N.}},
  \emph {et~al.},\ }\bibfield  {title} {\enquote {\bibinfo {title} {Wirebond
  crosstalk and cavity modes in large chip mounts for superconducting
  qubits},}\ }\href@noop {} {\bibfield  {journal} {\bibinfo  {journal}
  {Supercond. Sci. Technol.}\ }\textbf {\bibinfo {volume} {24}},\ \bibinfo
  {pages} {065001} (\bibinfo {year} {2011})}\BibitemShut {NoStop}%
\bibitem [{\citenamefont {White}\ \emph {et~al.}(2015)\citenamefont {White},
  \citenamefont {Mutus}, \citenamefont {Hoi}, \citenamefont {Barends},
  \citenamefont {Campbell}, \citenamefont {Chen}, \citenamefont {Chen},
  \citenamefont {Chiaro}, \citenamefont {Dunsworth}, \citenamefont {Jeffrey}
  \emph {et~al.}}]{white2015traveling}%
  \BibitemOpen
  \bibfield  {author} {\bibinfo {author} {\bibnamefont {White}, \bibfnamefont
  {T.}}, \bibinfo {author} {\bibnamefont {Mutus}, \bibfnamefont {J.}}, \bibinfo
  {author} {\bibnamefont {Hoi}, \bibfnamefont {I.-C.}}, \bibinfo {author}
  {\bibnamefont {Barends}, \bibfnamefont {R.}}, \bibinfo {author} {\bibnamefont
  {Campbell}, \bibfnamefont {B.}}, \bibinfo {author} {\bibnamefont {Chen},
  \bibfnamefont {Y.}}, \bibinfo {author} {\bibnamefont {Chen}, \bibfnamefont
  {Z.}}, \bibinfo {author} {\bibnamefont {Chiaro}, \bibfnamefont {B.}},
  \bibinfo {author} {\bibnamefont {Dunsworth}, \bibfnamefont {A.}}, \bibinfo
  {author} {\bibnamefont {Jeffrey}, \bibfnamefont {E.}},  \emph {et~al.},\
  }\bibfield  {title} {\enquote {\bibinfo {title} {Traveling wave parametric
  amplifier with josephson junctions using minimal resonator phase matching},}\
  }\href@noop {} {\bibfield  {journal} {\bibinfo  {journal} {Appl. Phys.
  Lett.}\ }\textbf {\bibinfo {volume} {106}},\ \bibinfo {pages} {242601}
  (\bibinfo {year} {2015})}\BibitemShut {NoStop}%
\bibitem [{\citenamefont {Wisbey}\ \emph {et~al.}(2014)\citenamefont {Wisbey},
  \citenamefont {Martin}, \citenamefont {Reinisch},\ and\ \citenamefont
  {Gao}}]{wisbey2014_Qc}%
  \BibitemOpen
  \bibfield  {author} {\bibinfo {author} {\bibnamefont {Wisbey}, \bibfnamefont
  {D.}}, \bibinfo {author} {\bibnamefont {Martin}, \bibfnamefont {A.}},
  \bibinfo {author} {\bibnamefont {Reinisch}, \bibfnamefont {A.}}, \ and\
  \bibinfo {author} {\bibnamefont {Gao}, \bibfnamefont {J.}},\ }\bibfield
  {title} {\enquote {\bibinfo {title} {New method for determining the quality
  factor and resonance frequency of superconducting micro-resonators from
  sonnet simulations},}\ }\href@noop {} {\bibfield  {journal} {\bibinfo
  {journal} {J. Low Temp. Phys.}\ }\textbf {\bibinfo {volume} {176}},\ \bibinfo
  {pages} {538--544} (\bibinfo {year} {2014})}\BibitemShut {NoStop}%
\bibitem [{\citenamefont {Wisbey}\ \emph {et~al.}(2010)\citenamefont {Wisbey},
  \citenamefont {Gao}, \citenamefont {Vissers}, \citenamefont {da~Silva},
  \citenamefont {Kline}, \citenamefont {Vale},\ and\ \citenamefont
  {Pappas}}]{Wisbey2010}%
  \BibitemOpen
  \bibfield  {author} {\bibinfo {author} {\bibnamefont {Wisbey}, \bibfnamefont
  {D.~S.}}, \bibinfo {author} {\bibnamefont {Gao}, \bibfnamefont {J.}},
  \bibinfo {author} {\bibnamefont {Vissers}, \bibfnamefont {M.~R.}}, \bibinfo
  {author} {\bibnamefont {da~Silva}, \bibfnamefont {F.~C.}}, \bibinfo {author}
  {\bibnamefont {Kline}, \bibfnamefont {J.~S.}}, \bibinfo {author}
  {\bibnamefont {Vale}, \bibfnamefont {L.}}, \ and\ \bibinfo {author}
  {\bibnamefont {Pappas}, \bibfnamefont {D.~P.}},\ }\bibfield  {title}
  {\enquote {\bibinfo {title} {Effect of metal/substrate interfaces on
  radio-frequency loss in superconducting coplanar waveguides},}\ }\href@noop
  {} {\bibfield  {journal} {\bibinfo  {journal} {J. Appl. Phys.}\ }\textbf
  {\bibinfo {volume} {108}},\ \bibinfo {pages} {093918} (\bibinfo {year}
  {2010})}\BibitemShut {NoStop}%
\bibitem [{\citenamefont {Wisbey}\ \emph {et~al.}(2019)\citenamefont {Wisbey},
  \citenamefont {Vissers}, \citenamefont {Gao}, \citenamefont {Kline},
  \citenamefont {Sandberg}, \citenamefont {Weides}, \citenamefont {Paquette},
  \citenamefont {Karki}, \citenamefont {Brewster}, \citenamefont {Alameri}
  \emph {et~al.}}]{wisbey2019_boron}%
  \BibitemOpen
  \bibfield  {author} {\bibinfo {author} {\bibnamefont {Wisbey}, \bibfnamefont
  {D.~S.}}, \bibinfo {author} {\bibnamefont {Vissers}, \bibfnamefont {M.~R.}},
  \bibinfo {author} {\bibnamefont {Gao}, \bibfnamefont {J.}}, \bibinfo {author}
  {\bibnamefont {Kline}, \bibfnamefont {J.~S.}}, \bibinfo {author}
  {\bibnamefont {Sandberg}, \bibfnamefont {M.~O.}}, \bibinfo {author}
  {\bibnamefont {Weides}, \bibfnamefont {M.~P.}}, \bibinfo {author}
  {\bibnamefont {Paquette}, \bibfnamefont {M.}}, \bibinfo {author}
  {\bibnamefont {Karki}, \bibfnamefont {S.}}, \bibinfo {author} {\bibnamefont
  {Brewster}, \bibfnamefont {J.}}, \bibinfo {author} {\bibnamefont {Alameri},
  \bibfnamefont {D.}},  \emph {et~al.},\ }\bibfield  {title} {\enquote
  {\bibinfo {title} {Dielectric loss of boron-based dielectrics on niobium
  resonators},}\ }\href@noop {} {\bibfield  {journal} {\bibinfo  {journal} {J.
  Low Temp. Phys.}\ }\textbf {\bibinfo {volume} {195}},\ \bibinfo {pages}
  {474--486} (\bibinfo {year} {2019})}\BibitemShut {NoStop}%
\bibitem [{\citenamefont {Woods}\ \emph {et~al.}(2019)\citenamefont {Woods},
  \citenamefont {Calusine}, \citenamefont {Melville}, \citenamefont {Sevi},
  \citenamefont {Golden}, \citenamefont {Kim}, \citenamefont {Rosenberg},
  \citenamefont {Yoder},\ and\ \citenamefont {Oliver}}]{woods2019_interface}%
  \BibitemOpen
  \bibfield  {author} {\bibinfo {author} {\bibnamefont {Woods}, \bibfnamefont
  {W.}}, \bibinfo {author} {\bibnamefont {Calusine}, \bibfnamefont {G.}},
  \bibinfo {author} {\bibnamefont {Melville}, \bibfnamefont {A.}}, \bibinfo
  {author} {\bibnamefont {Sevi}, \bibfnamefont {A.}}, \bibinfo {author}
  {\bibnamefont {Golden}, \bibfnamefont {E.}}, \bibinfo {author} {\bibnamefont
  {Kim}, \bibfnamefont {D.~K.}}, \bibinfo {author} {\bibnamefont {Rosenberg},
  \bibfnamefont {D.}}, \bibinfo {author} {\bibnamefont {Yoder}, \bibfnamefont
  {J.~L.}}, \ and\ \bibinfo {author} {\bibnamefont {Oliver}, \bibfnamefont
  {W.~D.}},\ }\bibfield  {title} {\enquote {\bibinfo {title} {Determining
  interface dielectric losses in superconducting coplanar-waveguide
  resonators},}\ }\href@noop {} {\bibfield  {journal} {\bibinfo  {journal}
  {Phys. Rev. Applied}\ }\textbf {\bibinfo {volume} {12}},\ \bibinfo {pages}
  {014012} (\bibinfo {year} {2019})}\BibitemShut {NoStop}%
\bibitem [{\citenamefont {Yeh}\ \emph {et~al.}(2017)\citenamefont {Yeh},
  \citenamefont {LeFebvre}, \citenamefont {Premaratne}, \citenamefont
  {Wellstood},\ and\ \citenamefont {Palmer}}]{Yeh2017}%
  \BibitemOpen
  \bibfield  {author} {\bibinfo {author} {\bibnamefont {Yeh}, \bibfnamefont
  {J.-H.}}, \bibinfo {author} {\bibnamefont {LeFebvre}, \bibfnamefont {J.}},
  \bibinfo {author} {\bibnamefont {Premaratne}, \bibfnamefont {S.}}, \bibinfo
  {author} {\bibnamefont {Wellstood}, \bibfnamefont {F.}}, \ and\ \bibinfo
  {author} {\bibnamefont {Palmer}, \bibfnamefont {B.}},\ }\bibfield  {title}
  {\enquote {\bibinfo {title} {Microwave attenuators for use with quantum
  devices below 100 mk},}\ }\href@noop {} {\bibfield  {journal} {\bibinfo
  {journal} {J. Appl. Phys.}\ }\textbf {\bibinfo {volume} {121}},\ \bibinfo
  {pages} {224501} (\bibinfo {year} {2017})}\BibitemShut {NoStop}%
\bibitem [{\citenamefont {Zmuidzinas}(2012)}]{Zmuidzinas2012}%
  \BibitemOpen
  \bibfield  {author} {\bibinfo {author} {\bibnamefont {Zmuidzinas},
  \bibfnamefont {J.}},\ }\bibfield  {title} {\enquote {\bibinfo {title}
  {Superconducting microresonators: Physics and applications},}\ }\href@noop {}
  {\bibfield  {journal} {\bibinfo  {journal} {Annu. Rev. Condens. Matter
  Phys.}\ }\textbf {\bibinfo {volume} {3}},\ \bibinfo {pages} {169--214}
  (\bibinfo {year} {2012})}\BibitemShut {NoStop}%
\end{thebibliography}%

\end{document}